\documentclass[usenatbib,iop,numberedappendix]{aeb_emulateapj_2010}
\usepackage{xspace}
\usepackage{color}

\usepackage{ifthen}
\ifx\pdftexversion\undefined
\usepackage[dvips]{graphicx}
\else
\usepackage{epstopdf}
\usepackage{thumbpdf}
\fi

\def\del#1{{}}

\defcitealias{BBPS1}{BBPS1}
\defcitealias{BBPS2}{BBPS2}
\defcitealias{BBPS3}{BBPS3}
\defcitealias{BBPS4}{BBPS4}
\defcitealias{BBPS5}{BBPS5}

\SetSymbolFont{symbols}{bold}{OMS}{cmsy}{b}{n}
\DeclareSymbolFont{bmisymbols}{OML}{cmm}{b}{it}
\DeclareMathSymbol{\balpha}{0}{bmisymbols}{"0B}
\DeclareMathSymbol{\bbeta}{0}{bmisymbols}{"0C}
\DeclareMathSymbol{\bgamma}{0}{bmisymbols}{"0D}
\DeclareMathSymbol{\bdelta}{0}{bmisymbols}{"0E}
\DeclareMathSymbol{\bepsilon}{0}{bmisymbols}{"0F}
\DeclareMathSymbol{\bzeta}{0}{bmisymbols}{"10}
\DeclareMathSymbol{\boldeta}{0}{bmisymbols}{"11}
\DeclareMathSymbol{\btheta}{0}{bmisymbols}{"12}
\DeclareMathSymbol{\biota}{0}{bmisymbols}{"13}
\DeclareMathSymbol{\bkappa}{0}{bmisymbols}{"14}
\DeclareMathSymbol{\blambda}{0}{bmisymbols}{"15}
\DeclareMathSymbol{\bmu}{0}{bmisymbols}{"16}
\DeclareMathSymbol{\bnu}{0}{bmisymbols}{"17}
\DeclareMathSymbol{\bxi}{0}{bmisymbols}{"18}
\DeclareMathSymbol{\bpi}{0}{bmisymbols}{"19}
\DeclareMathSymbol{\brho}{0}{bmisymbols}{"1A}
\DeclareMathSymbol{\bsigma}{0}{bmisymbols}{"1B}
\DeclareMathSymbol{\btau}{0}{bmisymbols}{"1C}
\DeclareMathSymbol{\bupsilon}{0}{bmisymbols}{"1D}
\DeclareMathSymbol{\bphi}{0}{bmisymbols}{"1E}
\DeclareMathSymbol{\bchi}{0}{bmisymbols}{"1F}
\DeclareMathSymbol{\bpsi}{0}{bmisymbols}{"20}
\DeclareMathSymbol{\bomega}{0}{bmisymbols}{"21}
\DeclareMathSymbol{\bvarepsilon}{0}{bmisymbols}{"22}
\DeclareMathSymbol{\bvartheta}{0}{bmisymbols}{"23}
\DeclareMathSymbol{\bvarpi}{0}{bmisymbols}{"24}
\DeclareMathSymbol{\bvarrho}{0}{bmisymbols}{"25}
\DeclareMathSymbol{\bvarsigma}{0}{bmisymbols}{"26}
\DeclareMathSymbol{\bvarphi}{0}{bmisymbols}{"27}

\newcommand{\mathbfit}[1]{\textbf{\textit{#1}}}

\newcommand{\bra}{\langle}
\newcommand{\ket}{\rangle}

\newcommand{\rmn}{\mathrm}
\newcommand{\dd}{\mathrm{d}}
\newcommand{\vecbf}{\mathbfit}

\newcommand{\bvel}{\bupsilon}

\newcommand{\KU}{K/U}

\slugcomment{Submitted to ApJ}

\shorttitle{Density and Pressure Clumping in Galaxy Clusters}
\shortauthors{Battaglia et al.}

\begin{document}

\title{On the Cluster Physics of Sunyaev-Zel'dovich and X-ray Surveys IV:\\
Characterizing Density and Pressure Clumping due to Infalling Substructures}

\author{N. Battaglia\altaffilmark{1}, J. R. Bond\altaffilmark{2}, C. Pfrommer\altaffilmark{3},  J. L. Sievers\altaffilmark{4,5}}

\altaffiltext{1}{McWilliams Center for Cosmology, Carnegie Mellon University,
Department of Physics, 5000 Forbes Ave., Pittsburgh PA, USA, 15213; nbattaglia@cmu.edu}
\altaffiltext{2}{Canadian Institute for Theoretical Astrophysics, 60 St George, Toronto ON, Canada, M5S 3H8}
\altaffiltext{3}{Heidelberg Institute for Theoretical Studies, Schloss-Wolfsbrunnenweg 35, D-69118 Heidelberg, Germany; christoph.pfrommer@h-its.org}
\altaffiltext{4}{Joseph Henry Laboratories of Physics, Jadwin Hall, Princeton University, Princeton NJ, USA, 08544}
\altaffiltext{5}{Department of Chemistry and Physics, University of Kwa-Zulu Natal, University Road, Westville, KZN, South Africa}

\begin{abstract}
  Understanding the outskirts of galaxy clusters at the virial radius
  ($R_{200}$) and beyond is critical for an accurate determination of cluster
  masses, structure growth, and to ensure unbiased cosmological parameter
  estimates from cluster surveys. This problem has drawn renewed interest due to
  recent determinations of gas mass fractions beyond $R_{200}$, which appear to
  be considerably larger than the cosmic mean, and because the clusters' total
  Sunyaev-Zel'dovich flux receives a significant contribution from these
  regions. Here, we use a large suite of cosmological hydrodynamical simulations
  to study the clumpiness of density and pressure and employ different variants
  of simulated physics, including radiative gas physics and thermal feedback by
  active galactic nuclei. We find that density and pressure clumping closely
  trace each other as a function of radius, but the bias on density remains on
  average $<20\%$ within the virial radius $R_{200}$ (if we only account for the
  X-ray emitting gas with $T\gtrsim10^6$~K). At larger radius, clumping
  increases steeply due to the continuous infall of coherent structures that
  have not yet passed the accretion shock. Density and pressure clumping
  increase with cluster mass and redshift, which probes on average dynamically
  younger objects that are still in the process of assembling.  The angular
  power spectra of gas density and pressure show that the clumping signal is
  dominated by comparably large substructures with scales $\gtrsim R_{200}/5$,
  signaling the presence of gravitationally-driven ``super-clumping''. In
  contrast, the angular power spectrum of the dark matter (DM) shows an almost
  uniform size distribution due to unimpeded subhalos. The quadrupolar
  anisotropy dominates the signal and correlates well across different radii as
  a result of the prolateness of the DM potential. At smaller angular scales,
  gas density and pressure at different radii lose coherence due to dissipative
  gas physics. We provide a synopsis of the radial dependence of the clusters'
  non-equilibrium measures (kinetic pressure support, ellipticity, and clumping)
  that all increase sharply beyond $R_{200}$.
\end{abstract}


\section{Introduction}

Clusters are the largest and latest objects in the universe that had time to
collapse under the influence of their gravity, sitting atop the mass
hierarchy. They form at sites of constructive interference of long waves in the
primordial density fluctuations, the coherent peak-patches
\citep{1986ApJ...304...15B, 1996ApJS..103....1B}. Since the probability of such
events is exponentially suppressed for larger objects and redshifts, clusters
are rare events. Hence, they are very sensitive tracers of the growth of
structure in the universe and the cosmological parameters governing it \citep[as
demonstrated by recent measurements, e.g., by][]{2008MNRAS.383..879A,
  2009ApJ...692.1060V, 2010MNRAS.406.1759M,HBSBPS2013,Mantz2014}. Most of the cluster mass
is in form of dark matter (DM) while the dominating fraction of cluster baryons
is in the form of a hot ($kT\sim(1-10)\,$~keV) diffuse plasma, the intra-cluster
medium (ICM), with the remaining baryons residing in the clusters' numerous
stars and galaxies.

Recent X-ray observations from {\em Chandra} and {\em XMM-Newton} have witnessed
a rich and complicated picture of the ICM in the interiors of clusters---many
ostensibly ``relaxed'' clusters show substantial small-scale variation in
temperature, metallicity, and surface brightness
\citep[e.g.,][]{2011MNRAS.418.2154F,2012MNRAS.421.1123C}. These effects seem to
by driven by non-gravitational feedback processes in the cluster centers
\citep{2007ARA&A..45..117M} or are remnant features of cluster mergers that are
either caught in the process of dissipating the gravitational binding energy or
at a later relaxation stage that is characterized by sloshing motions. On
increasingly larger scales, the mostly dark-matter (DM) dominated gravity of the
overall cluster potential starts to dominate the dynamics of the hot baryons; so
the hope is that on large enough scales the influence of those non-gravitational
(gastrophysical) effects is minimized. However, when approaching the cluster
outskirts, we encounter other complications that are predicated upon the mass
assembling history of clusters, with an increasing level of kinetic pressure
support \citep[e.g.,][]{1990ApJ...363..349E,2004MNRAS.351..237R,
  2009ApJ...705.1129L, 2009A&A...504...33V,
  2010ApJ...725...91B,Nelson2012,BBPS1}, ellipticity
\citep[e.g.,][]{2005ApJ...629..781K,2008MNRAS.391.1940M,2011ApJ...734...93L,
  BBPS1}, and density clumping
\citep[e.g.,][]{2011ApJ...731L..10N,Zhur2013,Khed2013,Ronc2013,
  2013MNRAS.429..799V,2014arXiv1404.4634A}.

Until recently, we did not have the possibility of directly observing the ICM
outskirts because of their low densities that escaped detections by means of
thermal X-ray bremsstrahlung emission which scales with the square of the gas
density. The high and temporarily variable particle background of {\em Chandra}
and {\em XMM-Newton} also provided a fundamental limit to the sensitivity of
X-ray emission in the ICM outskirts. A complementary probe of the ICM is the
thermal Sunyaev-Zel'dovich (SZ) effect, the Compton up-scattering of cosmic
microwave background (CMB) photons by hot electrons with its unique signature of
a spatially-varying distortion of the CMB spectrum, a decrement in thermodynamic
temperature at frequencies below $\sim 220$~GHz, and an excess above
\citep{1970Ap&SS...7....3S}.  But similarly to X-ray observations, detecting ICM
in cluster outskirts also remained a challenge to SZ experiments despite their
intrinsic advantage to characterize those regions because the signal scales with
the line-of-sight integral of the free electron pressure.  However, SZ
observations of low surface brightness outskirts are predicated upon the ability
to separate the SZ signal from the primary anisotropies of the CMB (as well as
the contribution due to radio and dusty galaxies) and requires comparably
sensitive multi-frequency observations that allow to cleanly separate the SZ
signal owing to its unique spectra signature.

Those outskirts are of particular importance since they connect the cluster
interior to the filamentary cosmic web with its warm-hot intergalactic medium
that is expected to host half of today's baryon budget. Hence, improving our
understanding of the dynamical state of the ICM in cluster outskirts helps
unraveling cluster masses and cosmological structure formation. Precise
characterization of the clumpiness of the ICM density and pressure will be of
crucial importance for current and next generation surveys in the X-rays ({\em
  eROSITA}) and those employing the SZ effect ({\em Atacama Cosmology Telescope}
(ACT, \citealt{2007ApOpt..46.3444F}), {\em South Pole Telescope} (SPT,
\citealt{2011PASP..123..568C}), and {\em Planck} as will be separably laid out in
the following.

\subsection{Density clumping and X-ray inferred gas mass fractions}

The launch of the {\em Suzaku} X-ray observatory opened the window for studies
of cluster outskirts due to its low-Earth orbit that ensures a low and stable
particle background. Recent {\em Suzaku} observations detected cluster X-ray
emission up to and even beyond $R_{200}$ and consistently find decreasing
temperature profiles towards larger radii, in accordance with theoretical
expectations. However, the majority of them observes an excess X-ray emission
for radii $\gtrsim R_{500}$ over what is expected from cosmological
hydrodynamical simulations at different levels of significance
\citep[e.g.,][]{2009MNRAS.395..657G, 2009PASJ...61.1117B, 2009A&A...501..899R,
  2010PASJ...62..371H, 2010ApJ...714..423K, 2011PASJ...63S1019A,
  2011Sci...331.1576S, 2012arXiv1203.0486W, 2012arXiv1203.1700S,
  2014MNRAS.437.3939U}.

This has been interpreted as entropy profiles which rise in the outskirts less
steeply than the predictions of hydrodynamical simulations (almost independent
of the adopted physics) or an increasing profile of the enclosed gas mass
fraction, $f_\rmn{gas} = M_\rmn{gas}/M_\rmn{tot}$, to values above the universal
Hubble-volume-smoothed proportion. While some of that excess can be attributed
to the anisotropic nature of the cosmic web that connects to the outer cluster
parts \citep{BBPS3}, there remain other possible explanations. Those include an
increasing kinetic pressure contribution in the outskirts, which biases
hydrostatic mass estimates of clusters low and gas mass fractions
high. Similarly, it was argued that the presence of substantial density clumping
\citep[e.g.,][]{Eckr2013,Mrnd2014} can bias X-ray inferred densities and, hence,
gas mass fractions high
\citep[e.g.,][]{2011ApJ...731L..10N,Khed2013,Ronc2013,BBPS3} as well as entropy
values low.\footnote{Observational or instrumental systematics for the observed
  excess emission (which is typically a factor of 3-5 below the soft
  extragalactic X-ray background) include fluctuations of unresolved point
  sources, point-spread function leakage from masked point sources, or stray
  light from the bright inner cluster core. This was argued not to be
  responsible for the excess emission \citep[e.g.,][]{2011Sci...331.1576S}.}

\subsection{Pressure clumping and Sunyaev-Zel'dovich observations}

Pressure clumping could have important influences on determinations of SZ
scaling relations and the SZ power spectrum (or its interpretation in terms of
cosmology). There is currently great interest in measuring the SZ power
spectrum, which is an integral of the cluster mass function times the square of
the Fourier transform of the average pressure profile of clusters/groups. Since
radiative physics in combination with non-gravitational energy injection
strongly modifies the pressure profile and since the SZ power spectrum depends
sensitively on the {\it rms} amplitude of the (linear) density power spectrum on
cluster-mass scales, $\sigma_8$, this is potentially a great tool to
simultaneously constrain cluster physics and cosmology. Thus far, {\em ACT} and
{\em SPT} have have detected the SZ effect in the CMB power spectrum
\citep[e.g.,][]{Reic2012,Seiv2013} and used SZ detected clusters to constraints
cosmological parameters \citep[e.g.,][]{Reic2013,Hass2013}.  Additionally, the
{\em Planck Collaboration} has released several papers on SZ science, including
scaling relations \citep[e.g.,][]{2011arXiv1101.2026P, 2011arXiv1101.2024P,
  2011arXiv1101.2043P,2011A&A...536A..11P}, pressure profiles \citep{PlnkP2013},
cosmological constraints from SZ clusters \citep{PlnkSZ2013} and SZ power
spectrum measurements at small multipoles \citep{PlnkY2013}.

Significant pressure clumping would impact the SZ power spectrum on different
angular scales depending on the physical scale of clumping and its strength as
well as the cluster redshift. Pressure clumping at radii $>R_{500}$ manifests
itself on angular multipoles $\ell \lesssim 4000$, which corresponds to angular
scales $\gtrsim 2.5$~arcmin \citep{BBPS2}. For example, at $\ell \simeq 1000$
more than 40\% of the power is contributed from radii $>R_{500}$. However, the
largest impact on the SZ power spectrum originates from clumps within $R_{500}$
that carry a larger SZ signal than their analogues at larger radii due to a
combination of ram-pressure and adiabatic compression of infalling substructures
and an increasing pressure toward the cluster center that amplifies any
deviations from spherical symmetry. This can be indirectly inferred from the
differences between the simulated SZ power spectrum and the model that pastes a
smoothed fit to the sphericalized pressure profile in the simulations; see
Figures 7 and 8 of \citet{BBPS2}.  If semi-analytical SZ power spectrum
templates neglect this contribution from clumping, the resulting cosmological
parameters would be biased, or, if sufficient freedom was left to address these
uncertainties in the cluster physics, this would limit the accuracy of the
cosmological parameter estimation.

More than half of the total SZ flux accumulates from radii $>R_{500}$
\citep{2010ApJ...725...91B}. If there are significant pressure fluctuations
present on these scales, this could change the interpretation of detected signal
strength in the presence of a noisy background, in particular when extrapolating
pseudo-pressure profiles that are informed by X-ray observations of the more
homogeneous inner regions \citep[e.g.,][]{2010A&A...517A..92A}. This could
potentially bias inferred SZ fluxes and resulting scaling relations, especially
in lower angular resolution surveys such as {\em Planck} that rely on accurate
templates for signal detection.

\subsection{Overview}

Those points provide ample motivation for characterizing density and pressure
clumping as a function of cluster mass, radius, and physics as well as
understanding its origin. It seems to be crucial to accurately model its
contribution to get unbiased cosmological constraints in future X-ray and SZ
surveys. To this end it is not sufficient to look at a small sample of simulated
clusters, but rather to use a large statistical sample drawn from comparably
large cosmological volumes to accurately address the cluster-to-cluster
scatter. However, such an undertaking dictates the limiting resolution (for a
given volume and particle number) that is far larger than the desired resolution
needed to follow all the possibly necessary physics. We hence concentrate on
modeling the most important physical agents with physically motivated subgrid
models and carefully check numerical convergence and whether we obtain agreement
with observed integrated quantities such as gas and stellar masses, SZ scaling
relations and SZ power spectra to gain confidence to our effective model. Due
to uncertainties of the subgrid modeling of cluster physics, it appears also
critical to complement such an approach with different model variants that
enable to develop an intuition about the effect of differently modeled physics
on various observables.

This is the forth paper of a series of papers addressing the cluster physics of
SZ and X-ray surveys. In the first two companion papers, we thoroughly
scrutinized the influence of feedback, non-thermal pressure and cluster shapes
on $Y-M$ scaling relations \citep[][hereafter BBPS1]{BBPS1} and the thermal SZ
power spectrum \citep[][hereafter BBPS2]{BBPS2}. The third paper details the
measurement biases and cosmological evolution of gas and stellar mass fractions
\citep[][hereafter BBPS3]{BBPS3}, and the fifth provides an information
theoretic view of clusters and their non-equilibrium entropies
\citep[][hereafter BBPS5]{BBPS5}.

The outline of this paper is as follows. In Section~\ref{sec:sims}, we introduce
our simulations and the subgrid physics models. In Section~\ref{sec:clumping},
we study how density and pressure clumping varies with radius, cluster mass,
redshift, and simulated physics. In Section~\ref{sec:correlations}, we study the
origin of clumping by computing dark matter and gas mass as well as thermal
energy distributions in angular cones and by cross-correlating their
distributions. In Section~\ref{sec:synthesis}, we provide a synthesis of the
different non-equilibrium measures in cluster outskirts, such as kinetic
pressure contribution, ellipticity, and clumping, that all are predicated upon
the mass assembling history of clusters. We conclude in
Section~\ref{sec:conclusions}.

\section{Cosmological simulations and cluster sample}
\label{sec:sims}

We use smoothed particle hydrodynamic (SPH) simulations of large-scale periodic
boxes of cosmological volume, which provide us with large cluster samples to
accurately characterize ICM properties over large ranges of cluster masses and
redshifts. We use a modified version of the {\sc GADGET-2}
\citep{2005MNRAS.364.1105S} code. Our sequence of periodic boxes have lengths of
$165$ and $330\,h^{-1}\,\rmn{Mpc}$, which are filled with $N_\rmn{DM} =
N_{\mathrm{gas}} = 256^3$ and $512^3$ dark matter (DM) and gas particles,
respectively. This choice maintains the same gas particle mass $m_{\mathrm{gas}}
= 3.2\times 10^9\, h^{-1}\,\mathrm{M}_{\sun}$, DM particle mass $m_{\mathrm{DM}}
= 1.54\times 10^{10}\, h^{-1}\,\mathrm{M}_{\sun}$ and a minimum gravitational
smoothing length $\varepsilon_{s}=20\, h^{-1}\,$kpc for different computational
volumes. We compute SPH densities with 32 neighbour particles. For our standard
calculations, we adopt a tilted $\Lambda$CDM cosmology, with total matter
density (in units of the critical) $\Omega_{m}=\Omega_{DM} + \Omega_{b} = 0.25$,
baryon density $\Omega_{b} = 0.043$, cosmological constant
$\Omega_{\Lambda} = 0.75$, Hubble parameter $h = 0.72$ in units of $100 \mbox{
  km s}^{-1} \mbox{ Mpc}^{-1}$, spectral index of the primordial power-spectrum
$n_{s} = 0.96$ and $\sigma_8 = 0.8$.

We show results from simulations employing three different variants of simulated
physics: the first only accounts for gravitational physics such that the gas
entropy can only increase through {\em shock heating}; the second model
additionally accounts for {\em radiative cooling}, star formation, thermal
energy feedback by supernovae, galactic winds, and cosmic-ray physics; and the
last model additionally accounts for thermal {\em feedback by active galactic
  nuclei (AGNs)}. Radiative cooling and heating were computed assuming an
optically thin gas of primordial hydrogen and helium composition in a
time-dependent, spatially uniform ultraviolet background. Star formation and
thermal feedback by supernovae were modelled using the hybrid multiphase model
for the interstellar medium of \citet{2003MNRAS.339..289S}. The CR population is
modelled as a relativistic proton population, which has an isotropic power-law
distribution function in momentum space with a spectral index of $\alpha=2.3$
and only accounts for CR acceleration at cosmological structure formation
shocks, following
\citet{2006MNRAS.367..113P,2007A&A...473...41E,2008A&A...481...33J}. With those
parameters, the CR pressure modifies the SZ effect at most at the percent level,
which is just consistent with the latest cluster gamma-ray limits
\citep{2013arXiv1308.5654A,2012ApJ...757..123A,2012A&A...541A..99A} and causes a
similarly small reduction of the resulting integrated Compton-$y$ parameter
\citep{2007MNRAS.378..385P}. The AGN feedback prescription included in the
simulations \citep[for more details see][]{2010ApJ...725...91B} enables lower
computational resolution and hence can be applied to large-scale structure
simulations. It couples the black hole accretion rate to the global star
formation rate (SFR) of the central region of the cluster, as inspired by a
model of \citet{2005ApJ...630..167T}. The thermal energy is injected into the
ICM such that it is proportional to the stellar mass formed within a given
central spherical region.

We use a two-step algorithm to compute the virial mass of a cluster. First, we
find all clusters in a given snapshot using a friends-of-friends (FOF)
algorithm. Second, using a spherical overdensity method with the FOF values as
starting estimates, we recursively calculate the center of mass, the virial
radius, $R_{\Delta}$, and mass, $M_{\Delta}$, contained within $R_{\Delta}$. The
radially averaged profiles of a given quantity are computed as a function of
cluster-centric radii, which are scaled by $R_{\Delta}$. We define $R_{\Delta}$
as the radius at which the mean interior density equals $\Delta$ times the {\em
  critical density} of the universe, e.g., for $\Delta =200$ or 500. Our cluster
sample (for each simulation type) has $\simeq1300$ clusters with $M_{200} >
7\times 10^{13}\,\rmn{M}_\odot$ and $\simeq800$ clusters with $M_{200} > 1\times
10^{14}\,\rmn{M}_\odot$ at $z=0$.


\section{Clumping factors}
\label{sec:clumping}

\subsection{Definitions and SPH volume bias}

\begin{figure*}[thbp]
  \resizebox{0.5\hsize}{!}{\includegraphics{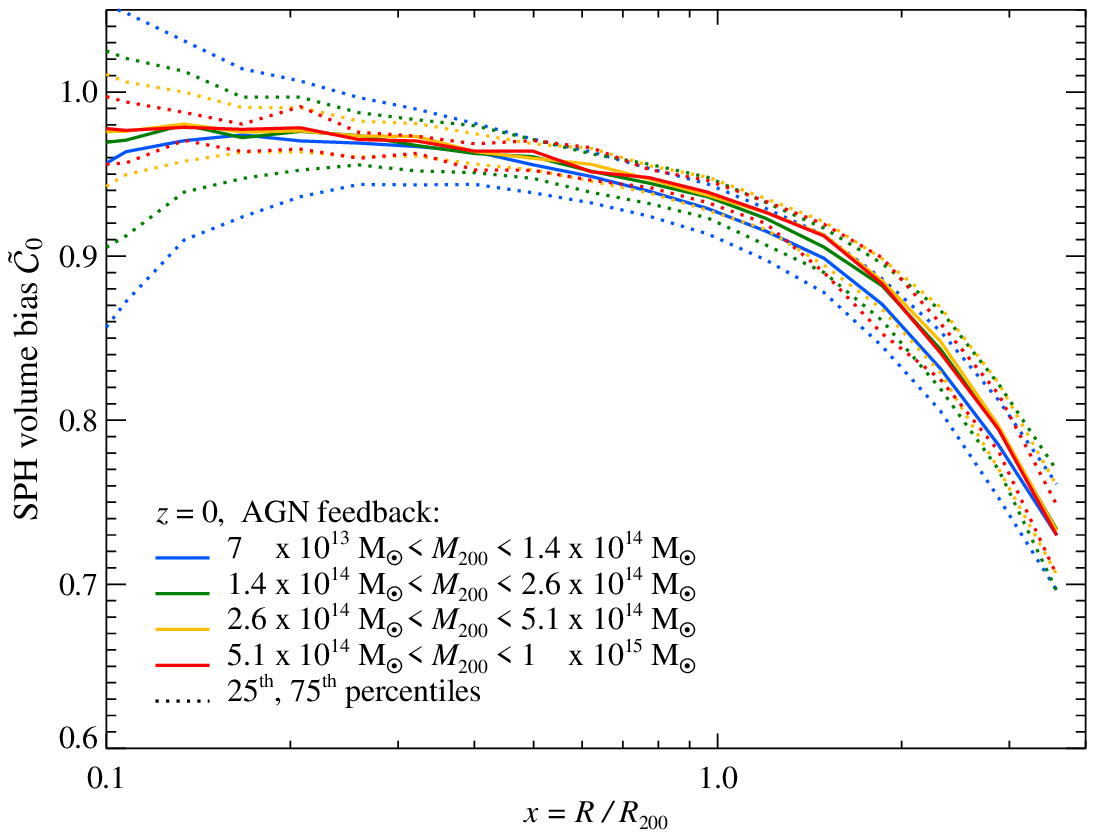}}%
  \resizebox{0.5\hsize}{!}{\includegraphics{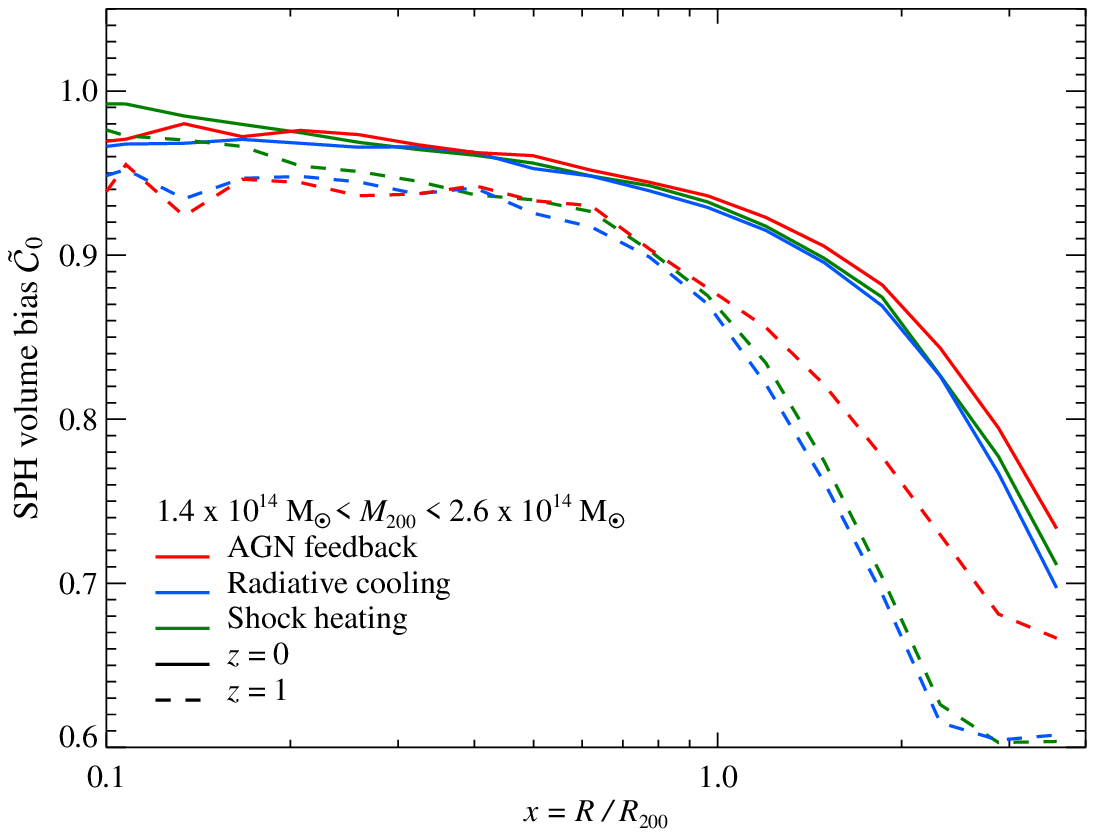}}\\
  \caption{Radial dependence of the SPH volume bias, $\tilde{\mathcal{C}}_{0} =
    \sum_i V_i/V$, which is given by the ratio of the sum of volume elements
    occupied by SPH particles ($V_i$) and the true volume of the concentric
    shell ($V$). Left: median (solid) and percentile profiles (dotted) of
    $\tilde{\mathcal{C}}_{0}$ for different cluster mass ranges in our AGN
    feedback model at $z=0$. Right: median profile of $\tilde{\mathcal{C}}_{0}$
    for different physics models at $z=0$ and 1 for cluster masses $1.4\times
    10^{14}\,\rmn{M}_\odot < M_{200} < 2.6 \times 10^{14}\,\rmn{M}_\odot$. The
    influence of the SPH volume bias becomes progressively greater for larger
    radii and is softened in our AGN feedback model for $R>R_{200}$, in
    particular at $z\gtrsim1$.  }
\label{fig:sph_volume_bias}
\end{figure*}

We define the dimensionless $n$-th order moments of mass density ($\rho$) and
thermal pressure ($P_\rmn{th}$) by
\begin{equation}
\tilde{\mathcal{C}}_{n,\rho} = \frac{\bra \rho^n \ket}{\bra \rho \ket^n}
\qquad\rmn{and}\qquad
\tilde{\mathcal{C}}_{n,P} = \frac{\bra P_\rmn{th}^n \ket}{\bra P_\rmn{th} \ket^n},
\label{eq:Cn}
\end{equation}
where $n\in\mathbb{Z}$. Since mass density and thermal pressure are both
intensive thermodynamic quantities, averages are always volume averages, i.e.,
the $n$-th order moments of mass density and thermal pressure within a radial
shell are defined by
\begin{eqnarray}
\bra \rho^n\ket &=& \frac{1}{V}\,\sum_i\frac{m_{\rmn{gas},i}}{\rho_i}\,\rho_i^n = 
\frac{1}{V}\,\sum_i m_{\rmn{gas},i}\rho_i^{n-1}, \\
\bra P_\rmn{th}^n \ket &=& \frac{1}{V}\,\sum_i\frac{m_{\rmn{gas},i}}{\rho_i}\,P_{\rmn{th},i}^n = 
\frac{1}{V}\,\sum_i m_{\rmn{gas},i}\,\rho_i^{n-1}\,\left(\frac{kT_i}{\mu\,m_p}\right)^n,
\label{eq:P-average}
\end{eqnarray}
where we sum over all SPH particles with mass $m_{\rmn{gas},i}$, mass density
$\rho_i$, and temperature $T_i$ that are within a radial shell of volume $V$,
$\mu$ denotes the mean molecular weight, and $m_p$ is the proton mass. A
particular moment is the zeroth order density moment $\tilde{\mathcal{C}}_{0}$,
which we will call SPH volume bias for reasons that become clear momentarily,
\begin{equation}
\tilde{\mathcal{C}}_{0} = \bra 1 \ket = \frac{1}{V}\,\sum_i \frac{m_{\rmn{gas},i}}{\rho_i}
= \frac{1}{V}\,\sum_i V_i.
\label{eq:C0}
\end{equation}
Clearly, we expect this to be identical to unity everywhere.
Figure~\ref{fig:sph_volume_bias} shows this SPH volume bias as a function of
radius (scaled by $R_{200}$).\footnote{To derive the radial profiles, we choose
  an equidistant spacing in $\log R$ and ensured that the ratio of particle
  smoothing lengths to radial extend of the shells is always smaller than unity
  in our smallest clusters for $R>0.5 R_{200}$.}  While the median of the SPH
volume bias is close to unity at small radii, it deviates progressively for
larger radii, and starts to severely underpredict unity outside $R_{200}$. This
is because of the inability to accurately percolate space that is characterized
by an inhomogeneous density distribution with a Lagrangian spherical density
kernel as realized by SPH.  The relative deviation of the SPH volume bias from
unity almost doubles at $z=1$ in comparison to $z=0$ (with the exception of our
AGN feedback model which shows a smaller increase towards $z=1$). The reason for
this deviation is a progressively larger inhomogeneity of the ICM for larger
radii due to increasing large-scale ellipticity \citepalias{BBPS1} as well as an
increased density clumping \citep[][see also below]{2011ApJ...731L..10N}. The
right-hand side of Figure~\ref{fig:sph_volume_bias} shows that the SPH volume
bias is almost independent of the simulated physics (especially for $z=0$) and
hence is not affected by the presence of the cold interstellar medium (ISM) of
galaxies (see also discussion surrounding Figure~\ref{fig:temp_cuts}).

As a result of this volume bias, we decided to redefine the dimensionless $n$-th
order density and thermal pressure moments and to apply a correction for the SPH
volume bias:
\begin{equation}
\mathcal{C}_{n,\rho} = \tilde{\mathcal{C}}_{n,\rho}\,\tilde{\mathcal{C}}_{0}^{n-1}
\qquad\rmn{and}\qquad
\mathcal{C}_{n,P} = \tilde{\mathcal{C}}_{n,P}\,\tilde{\mathcal{C}}_{0}^{n-1}
\label{eq:C_n,corr}
\end{equation}
where the $\tilde{\mathcal{C}}_{0}$ term has been chosen to cancel each power of the
SPH volume in the averaging procedure. While this corrects the median of the
$n$-th order density or pressure moments (which is our main objective in the
present work), it does not correct their entire distributions. We show in
Appendix~\ref{sec:volume_bias} how this correction impacts clumping factors
(second order density moment) and the negative order density moments, which are
useful for studying underdense regions. In particular for the latter case, the
SPH volume bias correction removes unphysical estimates of the negative order
density moments that are smaller than unity.

Previous work on clumping using SPH simulations \citep{Ronc2013} did not account
for this volume bias. If we neglected our SPH volume-bias correction, our values
for $\mathcal{C}_{2,\rho}$ in the {\em shock heating-only} model, which are not
susceptible to a differing implementation of the sub-grid physics models, agree
with those of \citet{Ronc2013}.  This can be seen by comparing
$\mathcal{C}_{2,\rho}$ for our massive cluster sample with $M_{200} > 5.1\times
10^{14}\,\rmn{M}_\odot$ (red lines in the middle panel of
Figure~\ref{fig:temp_cuts}) to Figure 4 in \citet{Ronc2013}.

\subsection{Density and pressure clumping}
\label{sec:rho_p}
\begin{figure}
\epsscale{1.20}
\plotone{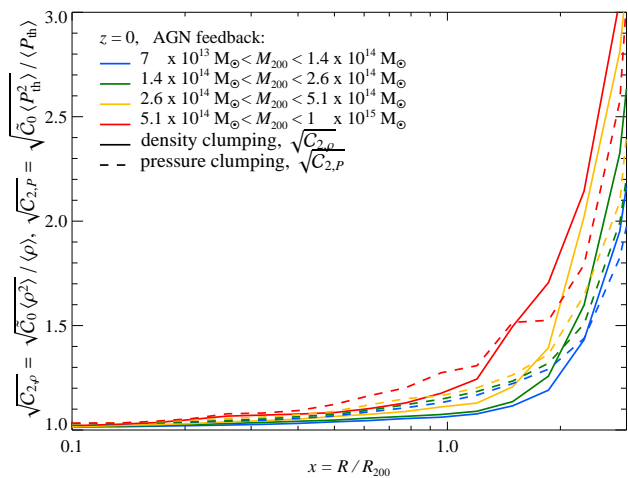}
\caption{Radial profile of the square root of the density clumping factor,
  $\sqrt{\mathcal{C}_{2,\rho}}$ (solid), and the pressure clumping factor,
  $\sqrt{\mathcal{C}_{2,P}}$ (dashed), corrected for the SPH volume bias. Shown
  are the median profiles of the clumping factors for the corresponding cluster
  mass ranges in our AGN feedback model (indicated by different colors). We
  impose a temperature cut to the density clumping factor and only consider SPH
  particles with $T>10^6\,\rmn{K}$ that can be observed by their X-ray
  emission. The X-ray-inferred gas density profile is biased high by
  $\sqrt{\mathcal{C}_{2,\rho}}$ which increases as a function of cluster mass at
  each radius, but remains $<1.2$ for $R<R_{200}$ in all clusters. The pressure
  clumping dominates over the density clumping inside the virial radius, but
  increases at a slower rate outside this radius so that the density clumping
  takes over. }
\label{fig:clumping_linear}
\end{figure}

After introducing our definitions and formalism, we turn to estimating the
clumping factor for density and pressure distribution. In
Figure~\ref{fig:clumping_linear}, we show the radial profile of the square root
of the density clumping factor, $\sqrt{\mathcal{C}_{2,\rho}}$. The emissivity of
thermal bremsstrahlung scales with the square of the gas density. Hence the
X-ray-inferred gas density profile is biased high by
$\sqrt{\mathcal{C}_{2,\rho}}$ which increases as a function of cluster mass at
each radius, but remains $<1.2$ for $R<R_{200}$ in all clusters.\footnote{We
  note that our clumping statistics only accounts for clumping that we can
  numerically resolve. In principle, there could be a spectrum of clumps to much
  smaller masses, but these clumps would have to be non-gravitationally heated
  to X-ray emitting temperatures. In our simulations, we resolve a $2\times
  10^{12} \rmn{M}_{\odot}$ object with 100 resolution elements which has the
  virial temperature $< 10^6$~K.} However, $\sqrt{\mathcal{C}_{2,\rho}}$
increases sharply outside $R_{200}$. There is a clear trend of an increasing
clumping factor for larger clusters, which have on average higher mass
accretion rates, i.e., they are assembling at the current epoch and hence are
dynamically younger.

In Figure~\ref{fig:clumping_linear}, we additionally show the square root of the
pressure clumping factor, $\sqrt{\mathcal{C}_{2,P}}$.  It broadly follows the
trend with radius and mass that we observe for density clumping. However, there
are small but distinct differences. While pressure clumping dominates over
density clumping inside the virial radius, it increases at a slower rate outside
this radius and becomes subdominant in comparison to the density clumping. This
similar radial behavior argues for a common physical origin and we will come
back to this question in Section~\ref{sec:correlations}. In contrast to the
X-rays, the thermal SZ flux scales with the gas pressure, so the Compton-$y$
maps are not biased by the presence of pressure clumping. There will be a
contribution of pressure clumping to the thermal SZ power spectrum, however. The
magnitude of this contribution is set by a balance of the overall signal
strength (that increases towards the cluster core) and the strength of the
clumping signal (that increases for larger radii).

\begin{figure*}[thbp]
  \begin{minipage}[t]{0.33\hsize}
    \centering{\small Density clumping, {\em AGN feedback}:}
  \end{minipage}
  \begin{minipage}[t]{0.33\hsize}
    \centering{\small Density clumping, {\em shock heating}:}
  \end{minipage}
 \begin{minipage}[t]{0.33\hsize}
    \centering{\small Pressure clumping, {\em AGN feedback}:}
  \end{minipage}
  \resizebox{0.33\hsize}{!}{\includegraphics{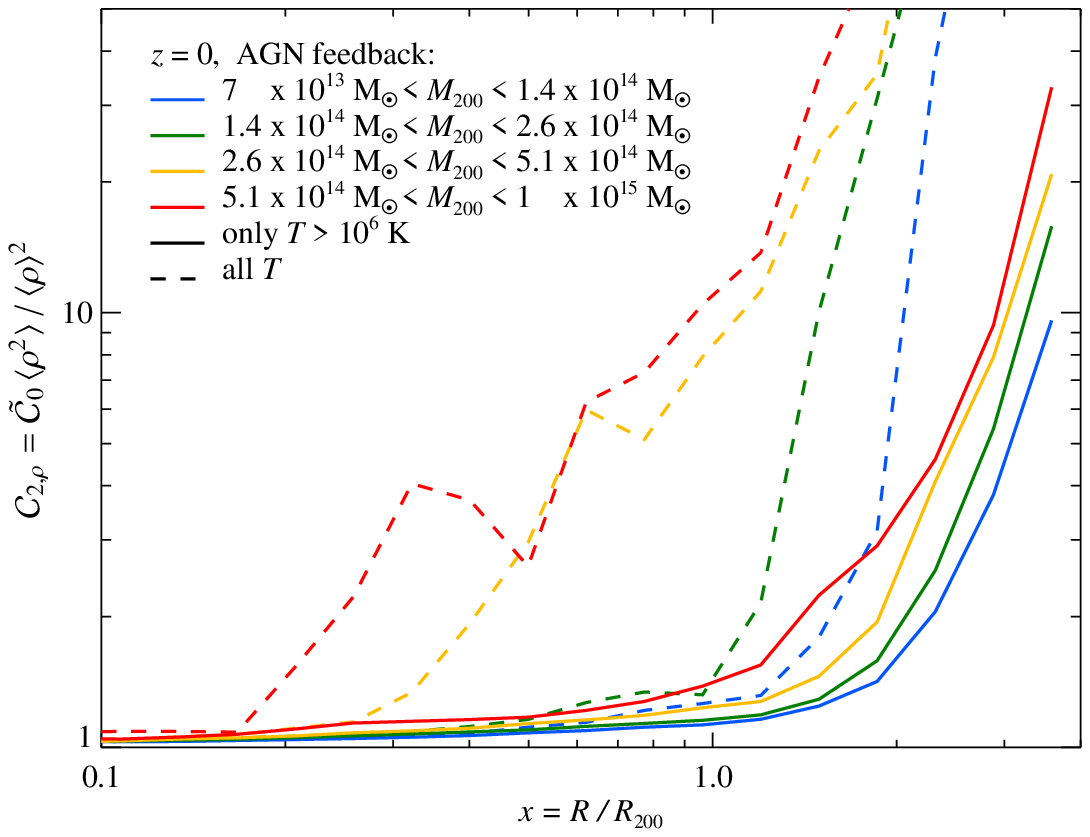}}%
  \resizebox{0.33\hsize}{!}{\includegraphics{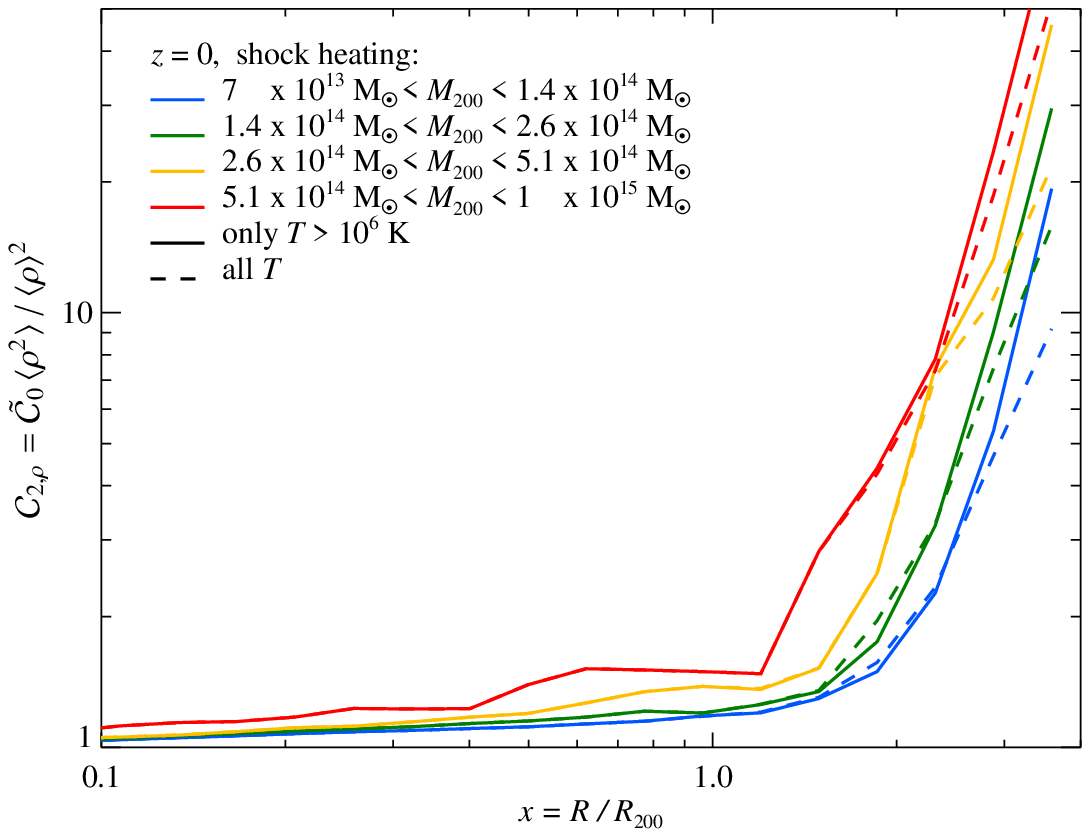}}%
  \resizebox{0.33\hsize}{!}{\includegraphics{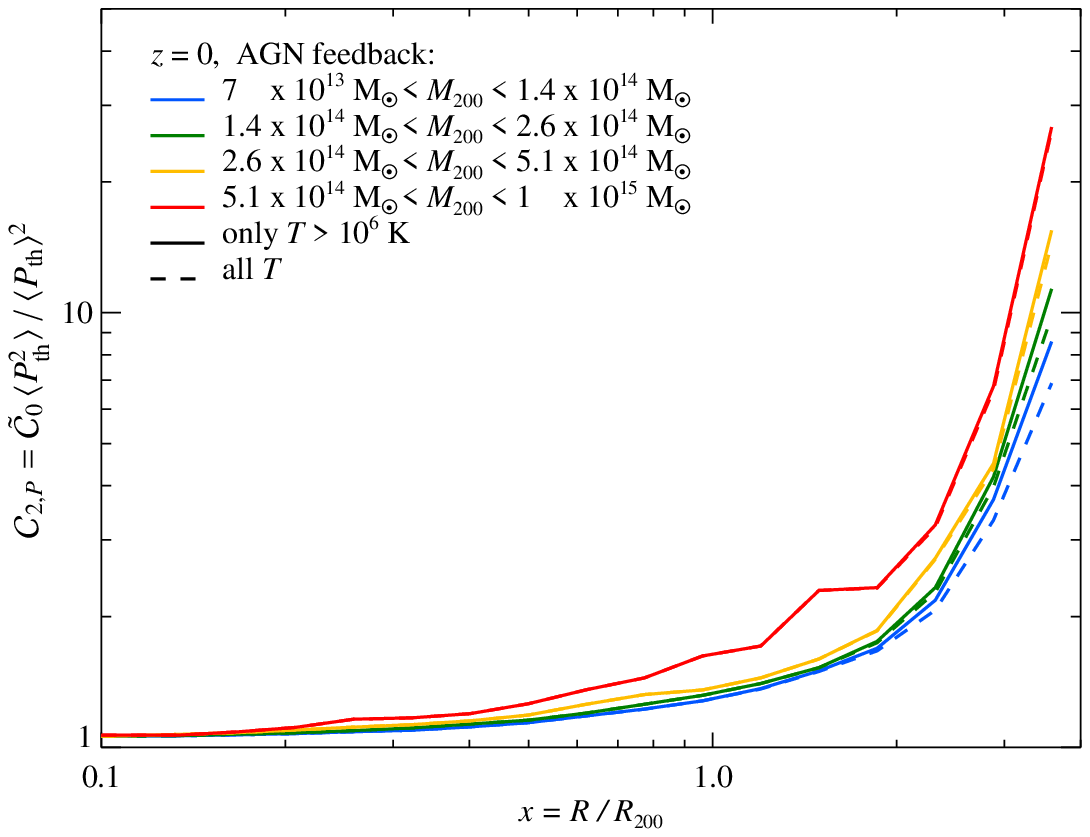}}\\
  \caption{Influence of the temperature cut on the density and pressure clumping
    factors ($\mathcal{C}_{2,\rho}$ and $\mathcal{C}_{2,P}$) at $z=0$. Left:
    density clumping factor in our {\em AGN feedback} model for SPH particles
    with $T>10^6\,\rmn{K}$ that can be observed by their X-ray emission (solid)
    and for all particles, i.e., without temperature restrictions (dashed).  The
    cold ISM of cluster galaxies and ram-pressure stripped dense gas clumps from
    these galaxies bias the density clumping factor high.  Center: to address
    whether the temperature floor removes a physically relevant phase of the ICM,
    we compare the density clumping factor in our {\em shock heating-only} model
    for SPH particles with $T>10^6\,\rmn{K}$ (solid) and for all particles
    (dashed). Clumping with a temperature cut is only slightly enhanced at large
    radii $R\gtrsim2.5 R_{200}$, implying that our temperature cut does not bias
    our clumping results at smaller radii with physically relevant
    substructures.  Right: the pressure clumping factor in our {\em AGN
      feedback} model for SPH particles with $T>10^6\,\rmn{K}$ (solid) and for
    all particles (dashed) are similar, suggesting that the dense, cold gas
    clumps are in pressure equilibrium with the ambient hot ICM.}
\label{fig:temp_cuts}
\end{figure*}

In Figure~\ref{fig:clumping_linear}, we applied an observationally motivated
temperature cut to the density moments and only considered SPH particles with
$T>10^6\,\rmn{K}$ that can be observed by their X-ray emission. We assess the
impact of such a temperature cut on density and pressure clumping in
Figure~\ref{fig:temp_cuts}. Clearly, the density clumping factors are
significantly biased high by the presence of the ISM in cluster galaxies as well
as cold, dense gas clumps that are ram-pressure stripped from galaxies. This
bias needs to be accounted for when comparing to X-ray observations by
employing, e.g., a temperature cut of $T>10^6\,\rmn{K}$, which we will adopt for
all the density clumping calculations hereafter.  We do not include a
volume-selection scheme \citep{Ronc2006} in our analysis since the goal of this
work is to quantify all ICM inhomogeneities, including large-scale features. This
allows us to study the connection to pressure clumping, which is important for
improved analyses of SZ data.

Does this temperature floor remove a physically relevant phase of the ICM
irrespective of the uncertainty of galaxy formation physics and the associated
ISM phase? To answer this, we show the density clumping factors in our
simulations with {\em shock heating-only} in the central panel of
Figure~\ref{fig:temp_cuts}. We find identical values for $\mathcal{C}_{2,\rho}$
within $R_{200}$ irrespective of adopting the temperature cut. However, the
density clumping factors with a temperature cut are slightly enhanced at large
radii $R\gtrsim2.5 R_{200}$: the temperature restriction removes gas with the
lowest density at any given radius that is neither in self-bound substructures
nor has it passed through the final cluster accretion shock. For non-radiative
physics, any temperature enhancement is necessarily accompanied by density
enhancements by either adiabatic compression or shock heating. Hence removing
the temperature and density floor of the warm-hot intergalactic medium leaves
the extrema which also slightly increases the resulting density clumping,
preferably for low-mass objects. We caution that the inclusion of AGN heating
(in addition to radiative cooling and star formation) modifies the pressure and
temperature beyond the virial radius and depending on the sign of this change,
either enhances or suppresses the differences in the clumping factor
\citep{2010ApJ...725...91B}.

In contrast to density clumping, pressure clumping in our {\em AGN feedback}
model (right panel, Figure~\ref{fig:temp_cuts}) depends very little on any
temperature cuts suggesting that these cold gas clumps find themselves in
pressure equilibrium with the ambient hot ICM. Qualitatively similar to density
clumping in our {\em shock heating-only} model, we find a slightly larger
pressure clumping if we adopt a temperature cut of $T>10^6\,\rmn{K}$ in low-mass
systems. This again is due to the colder pre-shock regions of the warm-hot
intergalactic medium that realize $T<10^6$~K. Hence, in order to compute
pressure clumping in the remainder of this work, we do not adopt any temperature
cut. This is again observationally motivated as the Compton-$y$ parameter
represents the line-of-sight integral over all free electrons.

\subsection{Impact of simulated physics on clumping}
\label{sec:clump_phys}

\begin{figure*}[thbp]
  \begin{minipage}[t]{0.5\hsize}
    \centering{\small Density and pressure clumping at $z=0$:}
  \end{minipage}
  \begin{minipage}[t]{0.5\hsize}
    \centering{\small Density and pressure clumping at $z=1$:}
  \end{minipage}
  \resizebox{0.5\hsize}{!}{\includegraphics{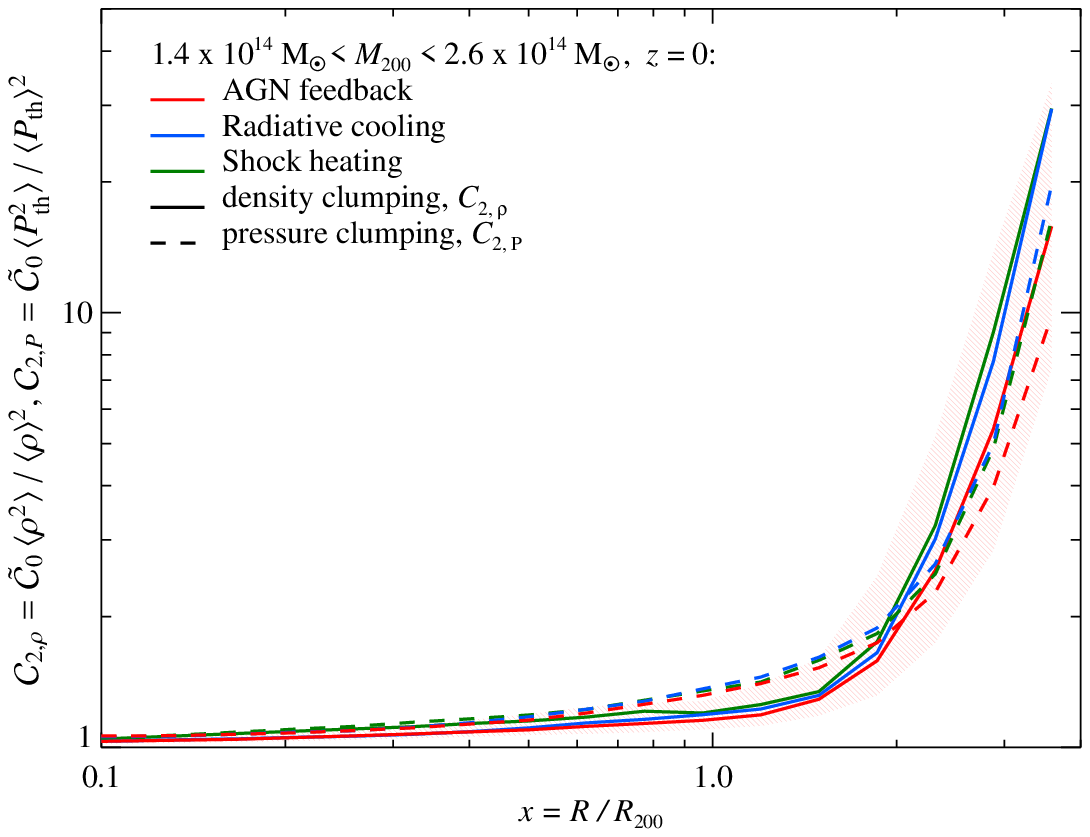}}%
  \resizebox{0.5\hsize}{!}{\includegraphics{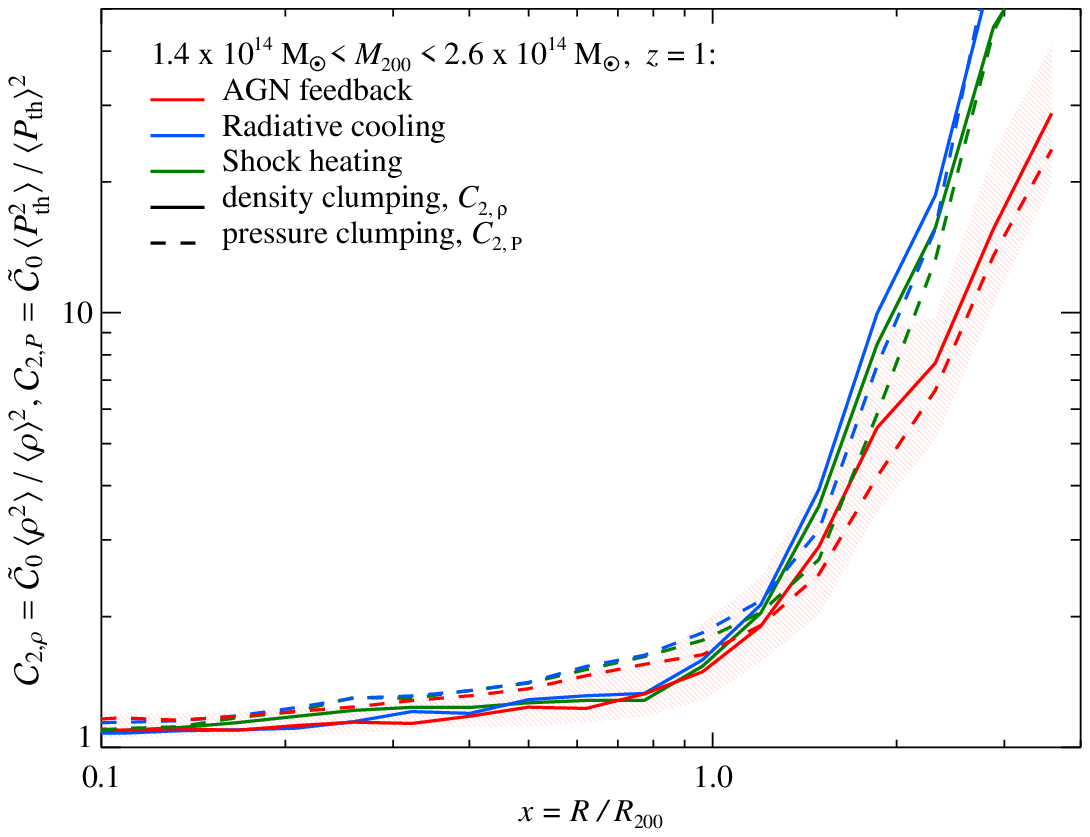}}\\
  \caption{Impact of different physics models on the density (solid) and
    pressure clumping factor (dashed), corrected for the SPH volume bias. We
    compare the median profiles for our {\em AGN feedback} model (red with
    25$^\rmn{th}$ and 75$^\rmn{th}$ percentiles of $\mathcal{C}_{2,\rho}$ shown
    by the error bands), our simulations with {\em radiative cooling} and star
    formation (blue), and with {\em shock heating-only} (green) at $z=0$ (left)
    and $z=1$ (right). Independent of the simulated physics, the slopes of the
    radial profiles of $\mathcal{C}_{2,\rho}$ and $\mathcal{C}_{2,P}$ change at
    $R\sim2 R_{200}$ ($z=0$) and $R\sim R_{200}$ ($z=1$). The AGN feedback model
    shows the lowest clumping in density and pressure at each radius. }
\label{fig:clumping_physics}
\end{figure*}

In Figure~\ref{fig:clumping_physics}, we show how the radial profiles of density
and pressure clumping depend on our different variants of physics. At $z=0$, all
different clumping measures are indistinguishable within the cluster-to-cluster
scatter. The picture is different at $z=1$, where both clumping measures in our
{\em AGN feedback} model fall short by a factor of two at $2\,R_{200}$ in
comparison to the clumping factors in the {\em radiative cooling} and {\em shock
  heating-only} simulation. This difference increases for larger radii due to
the shallower slope of the clumping factors in the {\em AGN feedback} models. This is
due to the central energy injection in this model that smooths out the central
density and pressure values and causes smaller fluctuations with respect to the
floor values \citep{2010ApJ...725...91B}.

\subsection{Clumping changes with cluster mass, redshift and dynamical state}

\begin{figure*}[thbp]
  \begin{minipage}[t]{0.5\hsize}
    \centering{\small Density clumping:}
  \end{minipage}
  \begin{minipage}[t]{0.5\hsize}
    \centering{\small Pressure clumping:}
  \end{minipage}
  \resizebox{0.5\hsize}{!}{\includegraphics{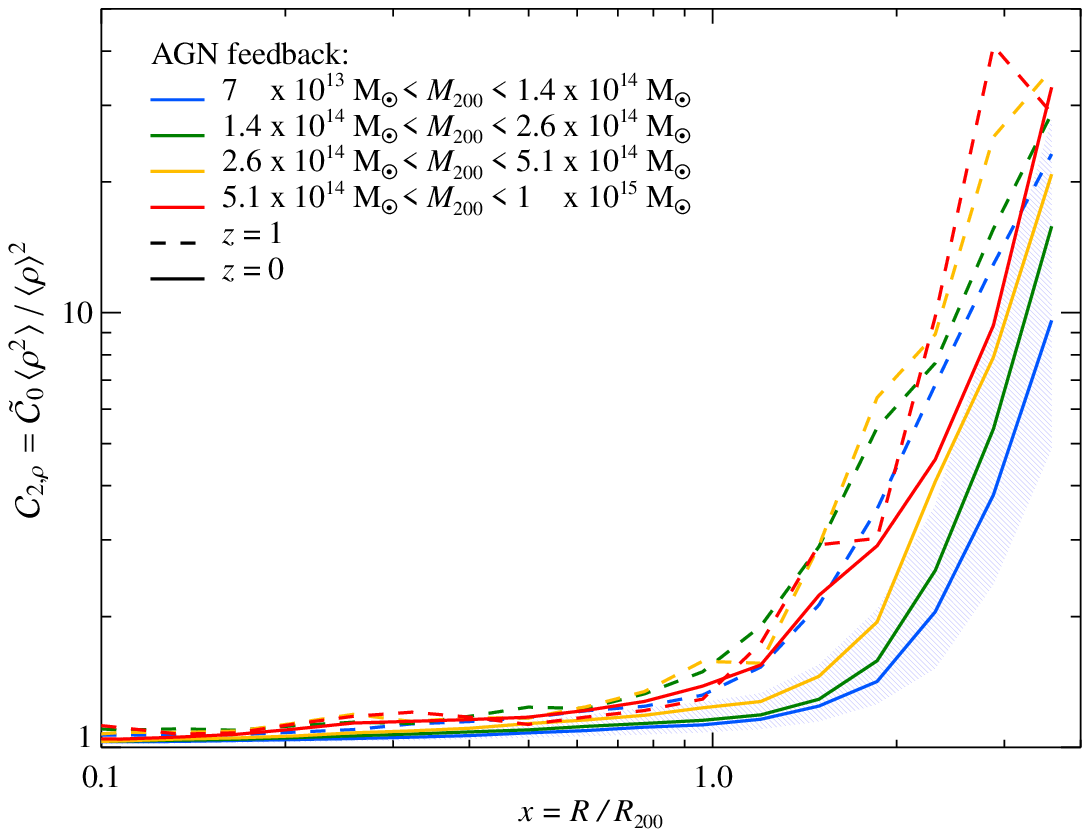}}%
  \resizebox{0.5\hsize}{!}{\includegraphics{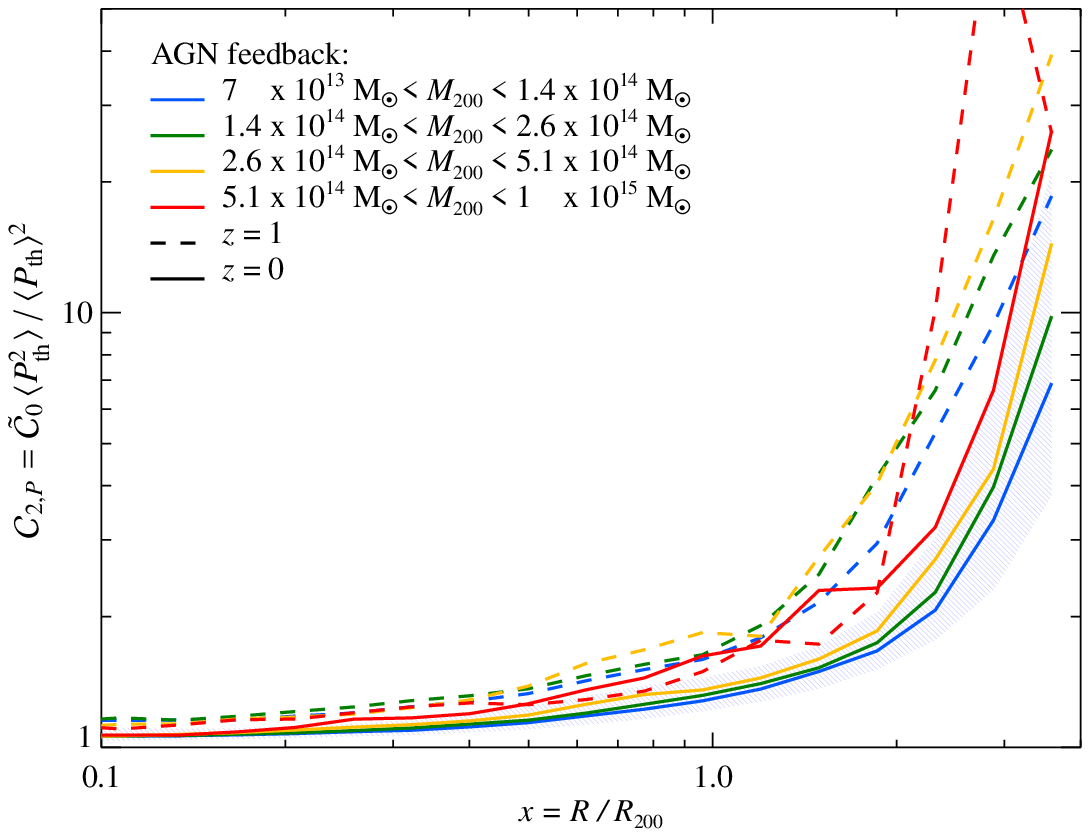}}\\
  \caption{Radial profiles of the density (left) and pressure clumping factor
    (right) for different cluster mass ranges (indicated with different
    colors). Shown are the median profiles at $z=0$ (solid) and $z=1$ (dashed)
    as well as the 25$^\rmn{th}$ and 75$^\rmn{th}$ percentiles for the lowest
    cluster mass range for $z=0$ (error bands).  At each radius, clumping
    increases for heavier clusters which assembled on average more recently.}
\label{fig:clumping_mass}
\end{figure*}

\begin{figure*}[thbp]
  \begin{minipage}[t]{0.5\hsize}
    \centering{\small Redshift evolution:}
  \end{minipage}
  \begin{minipage}[t]{0.5\hsize}
    \centering{\small Dynamical state:}
  \end{minipage}
  \resizebox{0.5\hsize}{!}{\includegraphics{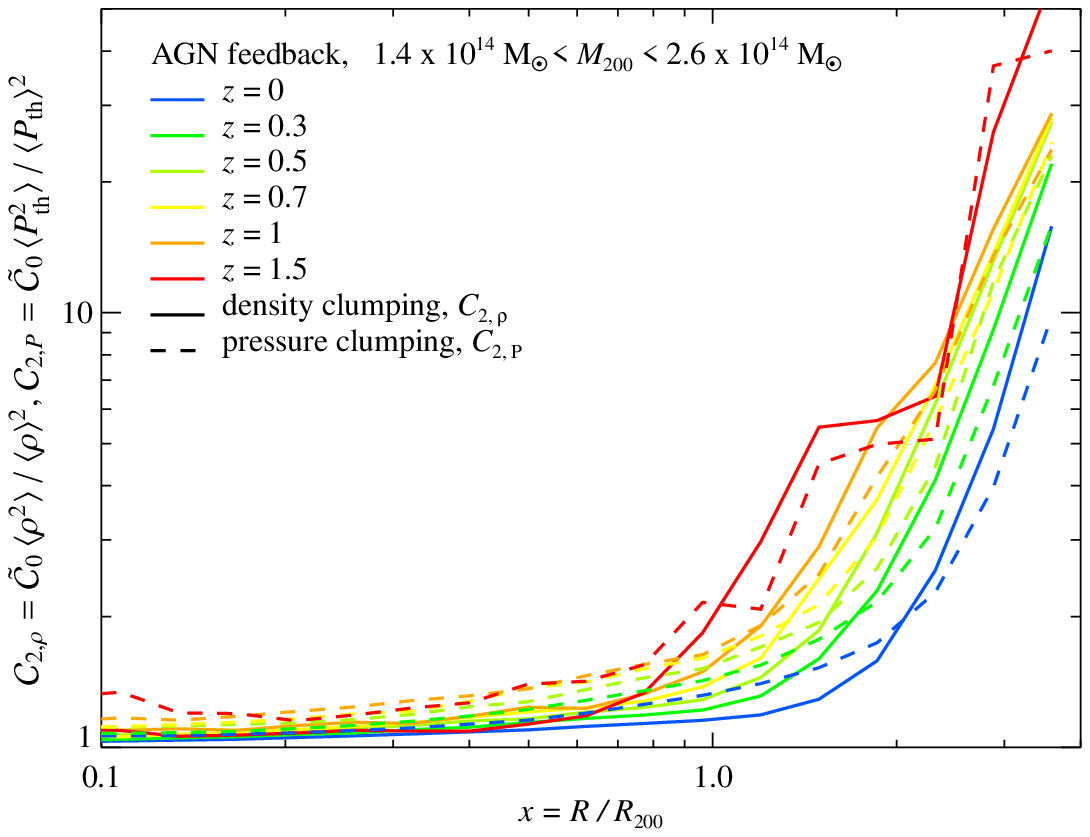}}%
  \resizebox{0.5\hsize}{!}{\includegraphics{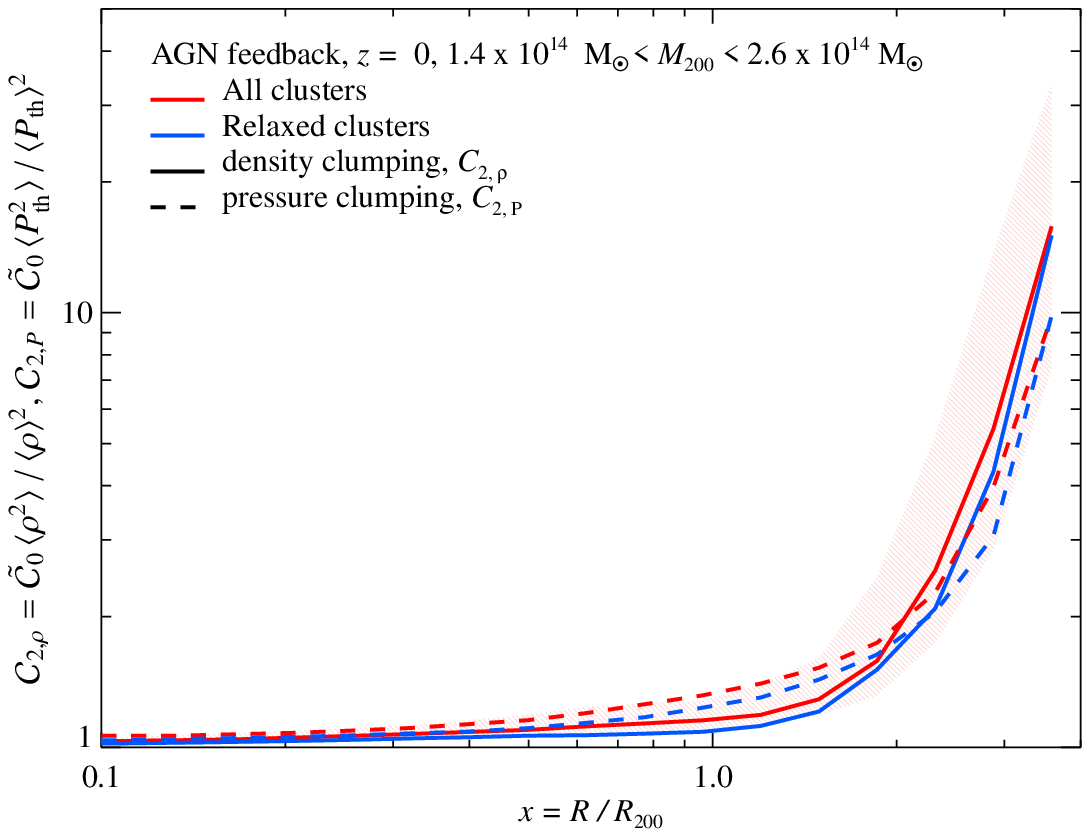}}\\
  \caption{Left: redshift evolution of the clumping factor for density (solid)
    and pressure (dashed). Right: comparison of the density (solid) and pressure
    (dashed) clumping between a subsample of relaxed clusters (blue, with the
    lowest ratio of total kinetic-to-thermal energy, $\KU$) and all clusters
    (red). In both panels, we show the median profiles for our {\em AGN
      feedback} model for cluster masses in the range $1.4\times
    10^{14}\,\rmn{M}_\odot < M_{200} < 2.6 \times 10^{14}\,\rmn{M}_\odot$. The
    red band in the right panel represents the range from the 25$^{\rmn{th}}$ to
    75$^{\rmn{th}}$ percentile of the sample that includes all the clusters.}
\label{fig:clumping_redshift}
\end{figure*}

In Figures~\ref{fig:clumping_mass} and \ref{fig:clumping_redshift}, we
show profiles of density and pressure clumping for different cluster
masses, redshifts, and dynamical states. At each radius, density and
pressure clumping increases for heavier clusters as well as for larger
redshifts (except for the highest cluster mass range of $5.1\times
10^{14}\,\rmn{M}_\odot < M_{200} < 1 \times 10^{15}\,\rmn{M}_\odot$ at
$z=1$ where we do not have sufficient statistics).  If we identify the
clumping with infalling \mbox{(sub-)}structures (see
Section~\ref{sec:correlations}), then we see that those accreting
structures from the cosmic web penetrate to smaller radii in bigger
clusters. 

The dynamical state of a clusters is determined by the ratio of
kinetic-to-thermal energy, $\KU$, within $R_{200}$. In \citetalias{BBPS3} we
show that this tracer correlates with another commonly used tracer known as
centroid shift, $\bra w \ket$ \citep{1993ApJ...413..492M}.  The internal kinetic
energy, $K$, and thermal energy, $U$, of a cluster are defined as
\begin{eqnarray}
\label{eq:KU}
K (<R_{200}) &\equiv& \sum_i \frac{3 m_{\rmn{gas},i}  P_{\mathrm{kin},i}}{2\rho_i}, \\
\label{eq:KU2}
U (<R_{200}) &\equiv& \sum_i \frac{3 m_{\rmn{gas},i}  P_{\mathrm{th},i}} {2\rho_i},
\end{eqnarray}
where $m_{\rmn{gas}}$ and $\rho$ are the gas mass and the SPH density,
respectively for all particles $i$ less than $R_{200}$,
$P_{\mathrm{th}}=kT\rho/(\mu m_p)$ denotes the thermal pressure, and the kinetic
pressure is defined as
\begin{eqnarray}
\label{eq:Pkin}
P_{\mathrm{kin}} &=& \rho \, \delta \bvel^2 / 3, \\
\delta \bvel   &=& a\,\left(\bvel - \bar{\bvel}\right) +
a\,H(a)\left(\vecbf{x}-\bar{\vecbf{x}}\right) .
\end{eqnarray}
\noindent The code uses comoving peculiar velocities, which we convert into the
internal cluster velocities relative to the mean cluster velocity in the
Hubble flow. Here $H(a)$ denotes the Hubble function, $a$ is the scale factor,
$\bvel$ ($=\dd \vecbf{x}\,/\,\dd t$) is the peculiar velocity, and $\vecbf{x}$
is the comoving position of each particle. The averaged cluster bulk flow for
the gas particles is $\bar{\bvel}$ and the center of mass is $\bar{\vecbf{x}}$,
both calculated within $R_{200}$.

The ratio $\KU$ is the volume integrated analog of the ratio
$P_{\mathrm{kin}}/P_{\mathrm{th}}$ and hence is also an indicator of formation
history and substructure. We split our cluster sample equally into lower, middle
and upper bands of $\KU$ and define the sub-sample with the lowest ratio of
$\KU$ as {\em relaxed clusters}.  We find that the clumping factor depends
weakly on the dynamical state of the clusters. Overall, the relaxed clusters
have a lower clumping factor in comparison to the entire
sample. Figure~\ref{fig:clumping_redshift} shows that for small radii
($R\lesssim R_{200}$), the median clumping factor of relaxed clusters is below
the 25$^{\rmn{th}}$ percentile of entire cluster sample.  However, these
differences between the relaxed and full cluster sample become progressively
smaller outside the virial radius.

These trends can be understood by the fact that 1) clusters of a given mass
(range) show a larger degree of morphological disturbance/mass accretion at
higher redshifts which probe on average dynamically younger objects and 2) the
redshift evolution of the velocity anisotropy \citepalias[cf.][]{BBPS5} which
shows that the average location of accretion shocks moves to smaller radii at
larger $z$ (if scaled by $R_{200}$, see Sec.~\ref{sec:synthesis} for more
details). Hence at larger redshifts, also the gas distribution probes the
infall/pre-accretion shock region that is shaped by the tides exerted by the
far-field of clusters.


\section{Characterizing the Structure of Clumping}
\label{sec:correlations}

\begin{figure*}
  \begin{minipage}[t]{0.33\hsize}
    \centering{\small $\delta P$}
    \vspace{1em}
  \end{minipage}
  \begin{minipage}[t]{0.33\hsize}
    \centering{\small $\delta\rho$}
    \vspace{1em}
  \end{minipage}
  \begin{minipage}[t]{0.33\hsize}
    \centering{\small $\delta\rho_\rmn{DM}$}
    \vspace{1em}
  \end{minipage}
    \resizebox{0.33\hsize}{!}{\includegraphics[angle=90,origin=c]{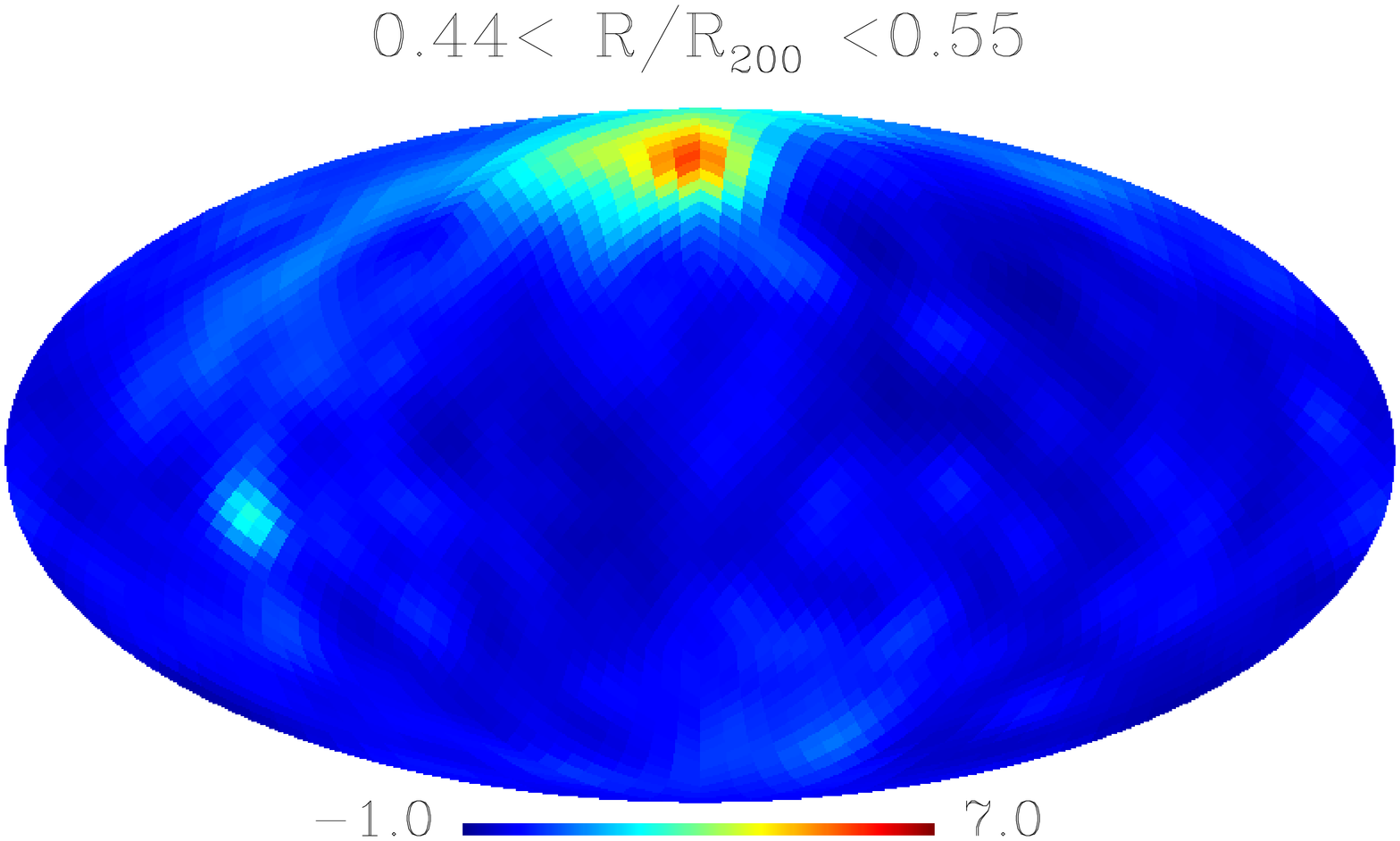}}%
    \resizebox{0.33\hsize}{!}{\includegraphics[angle=90,origin=c]{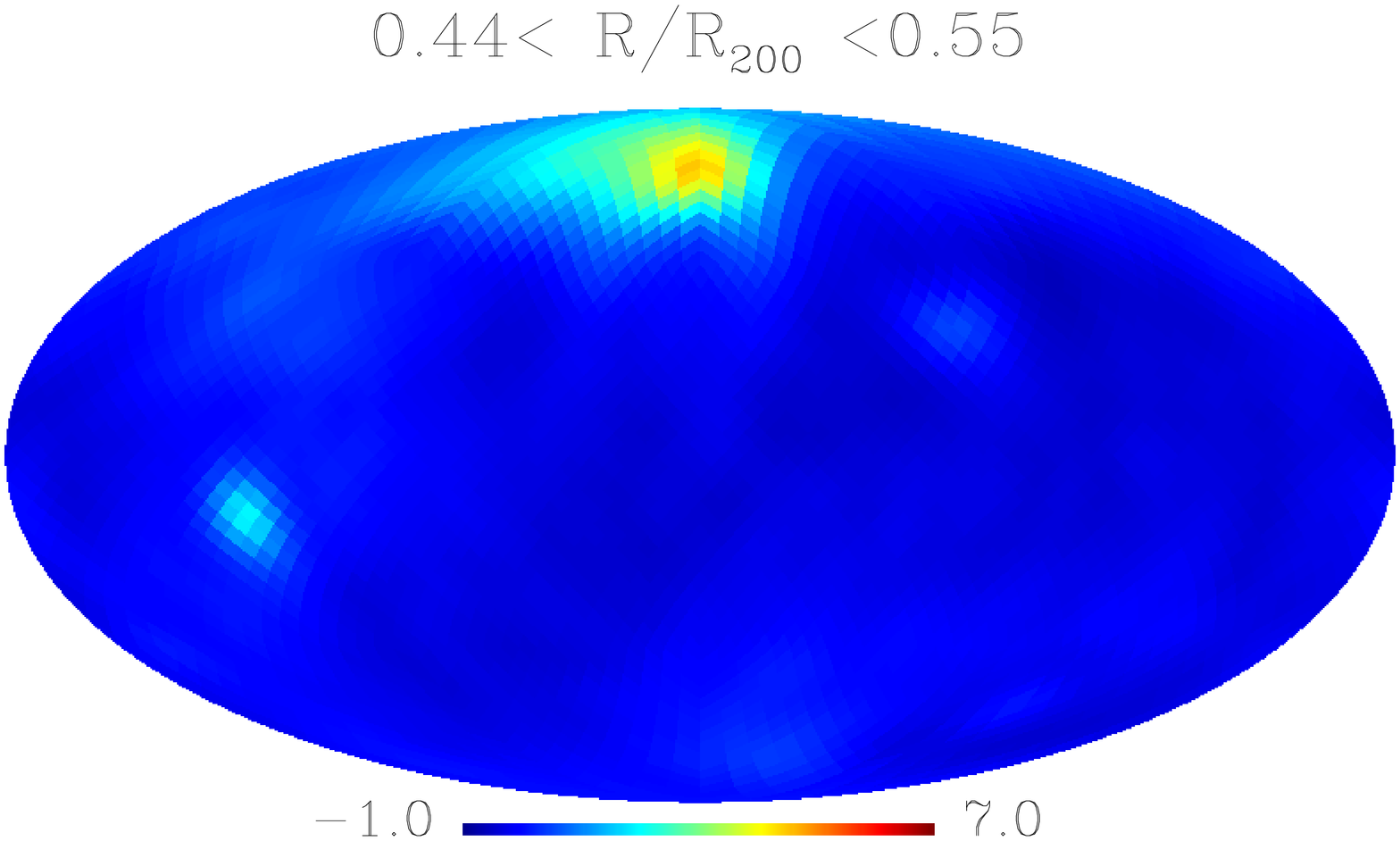}}%
    \resizebox{0.33\hsize}{!}{\includegraphics[angle=90,origin=c]{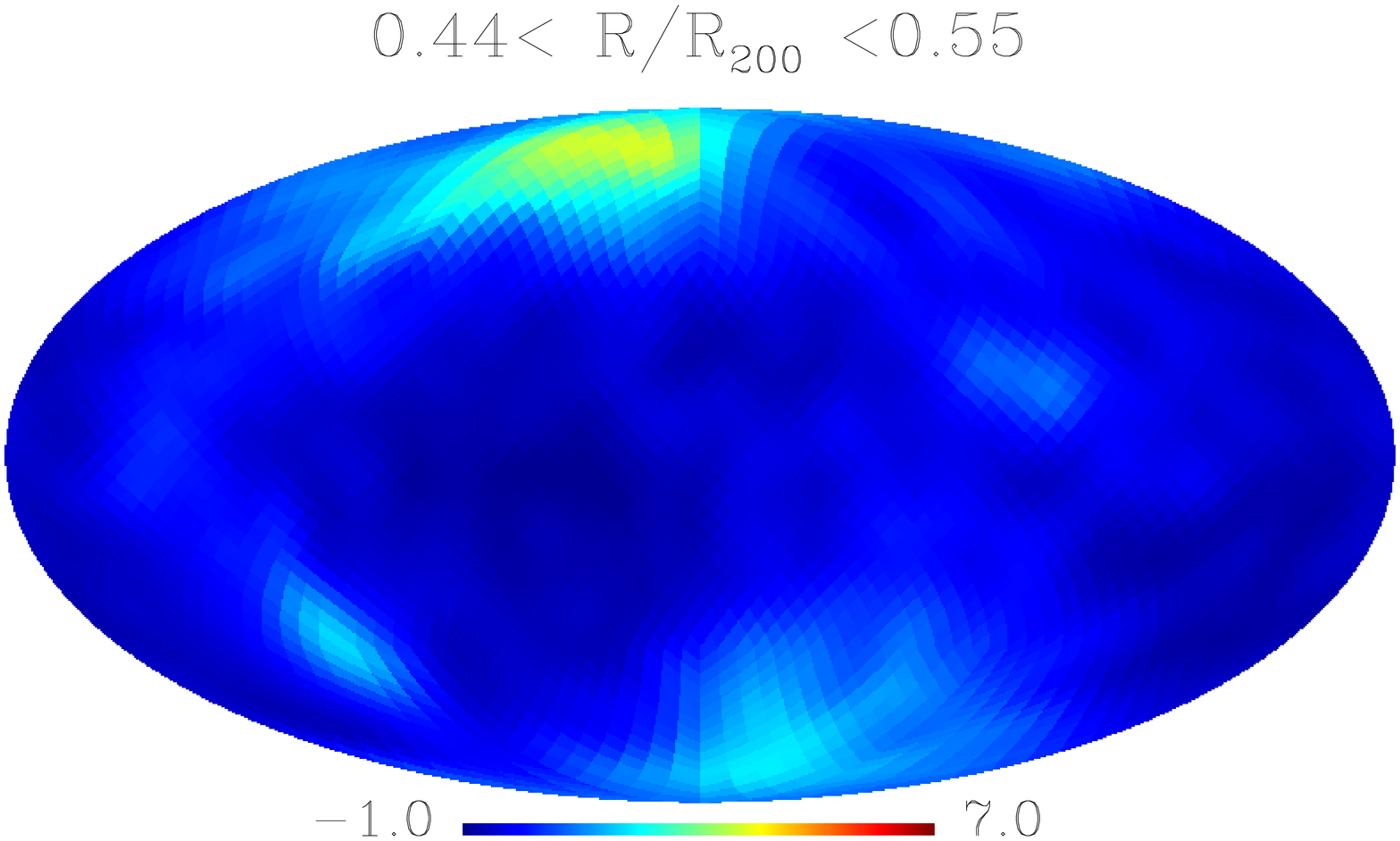}}\\  
    \resizebox{0.33\hsize}{!}{\includegraphics[angle=90,origin=c]{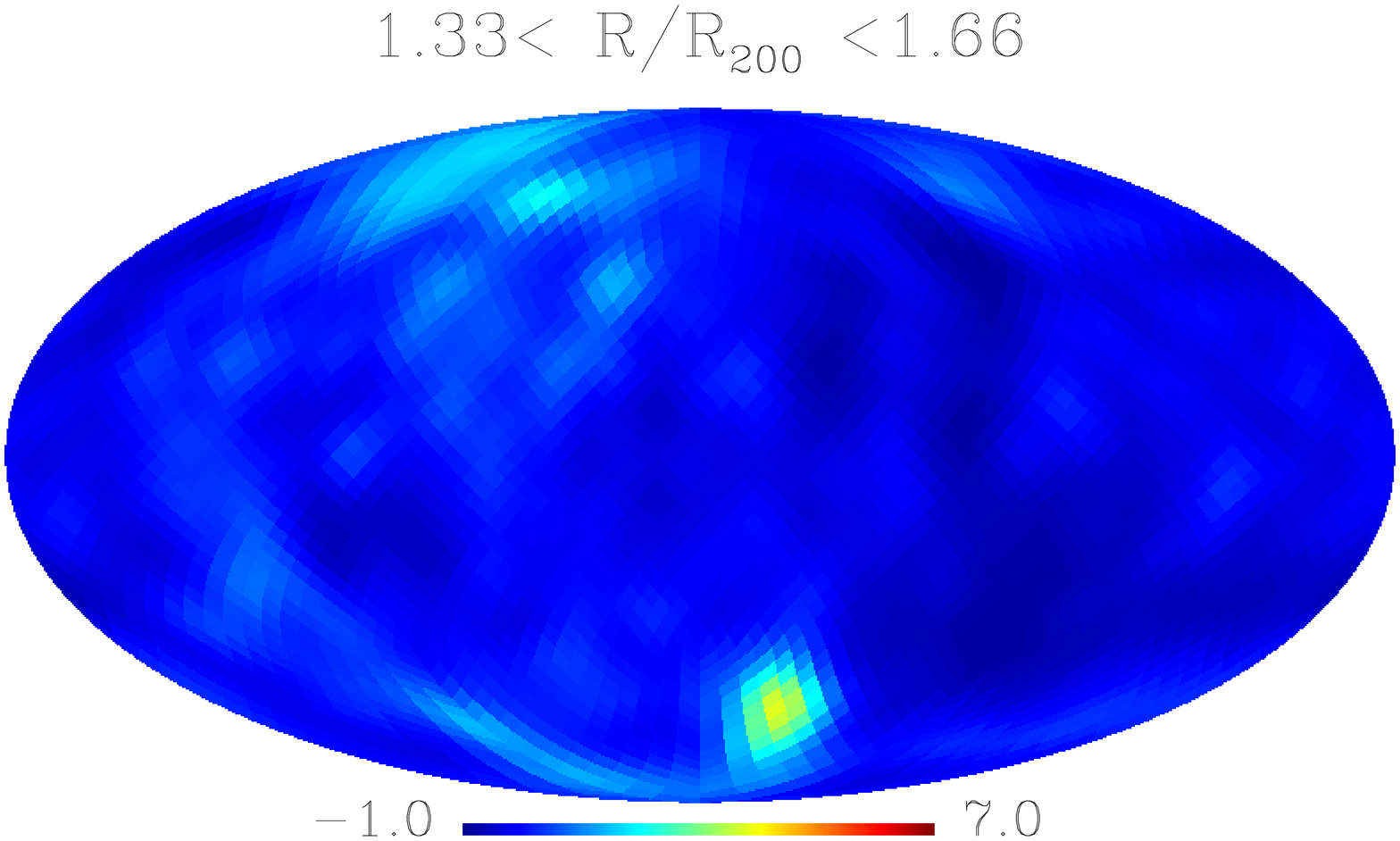}}%
    \resizebox{0.33\hsize}{!}{\includegraphics[angle=90,origin=c]{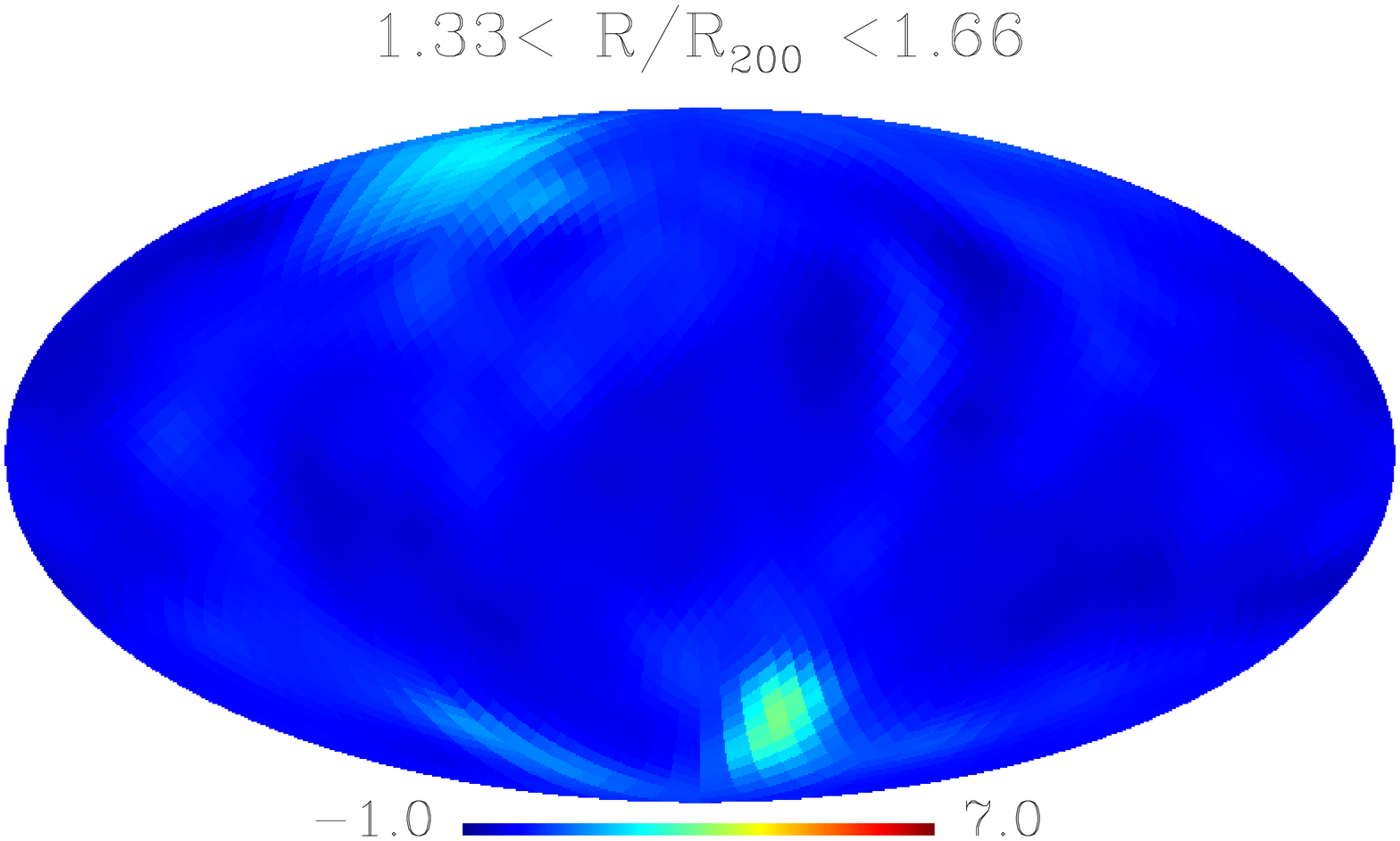}}%
    \resizebox{0.33\hsize}{!}{\includegraphics[angle=90,origin=c]{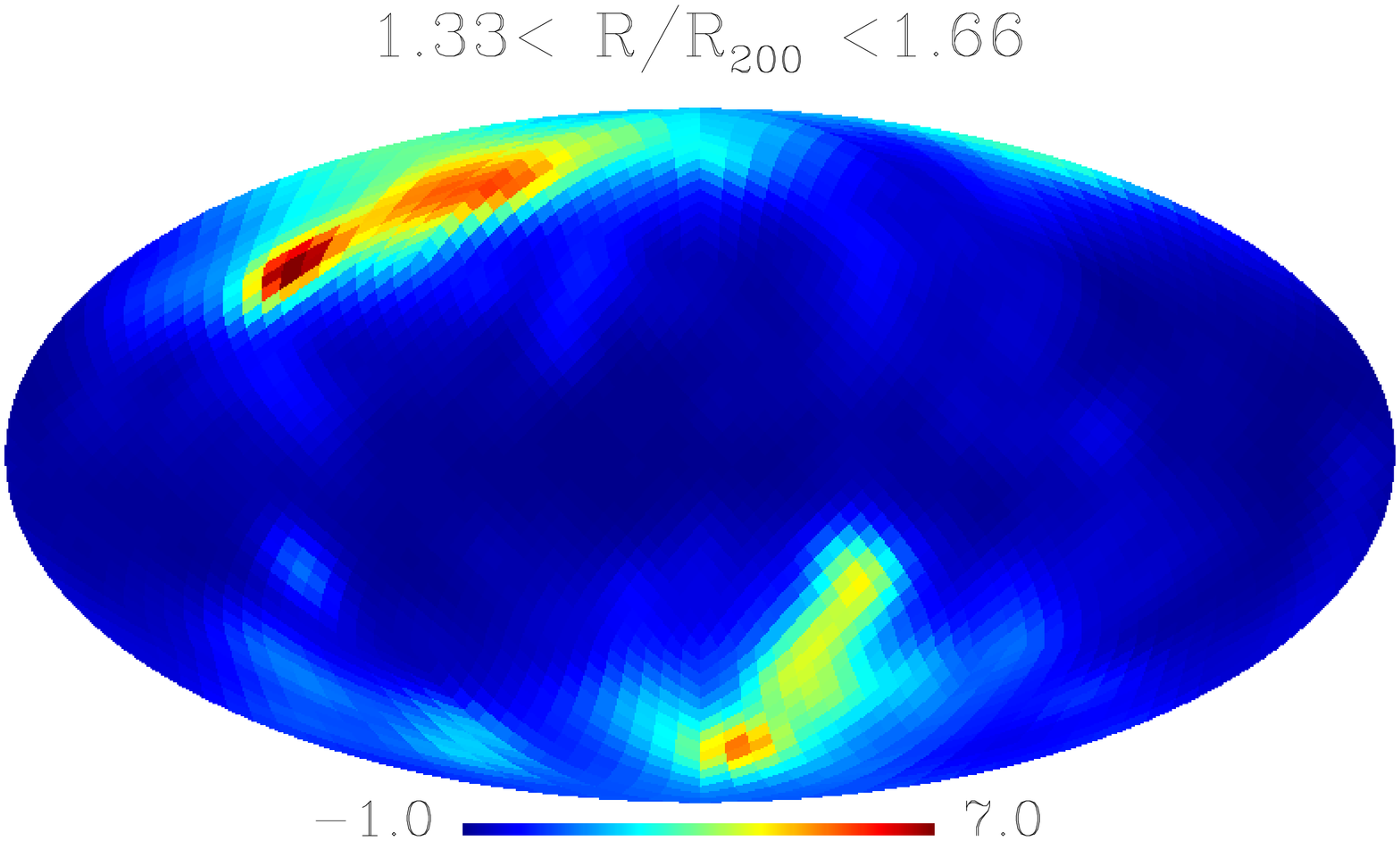}}\\  
    \resizebox{0.33\hsize}{!}{\includegraphics[angle=90,origin=c]{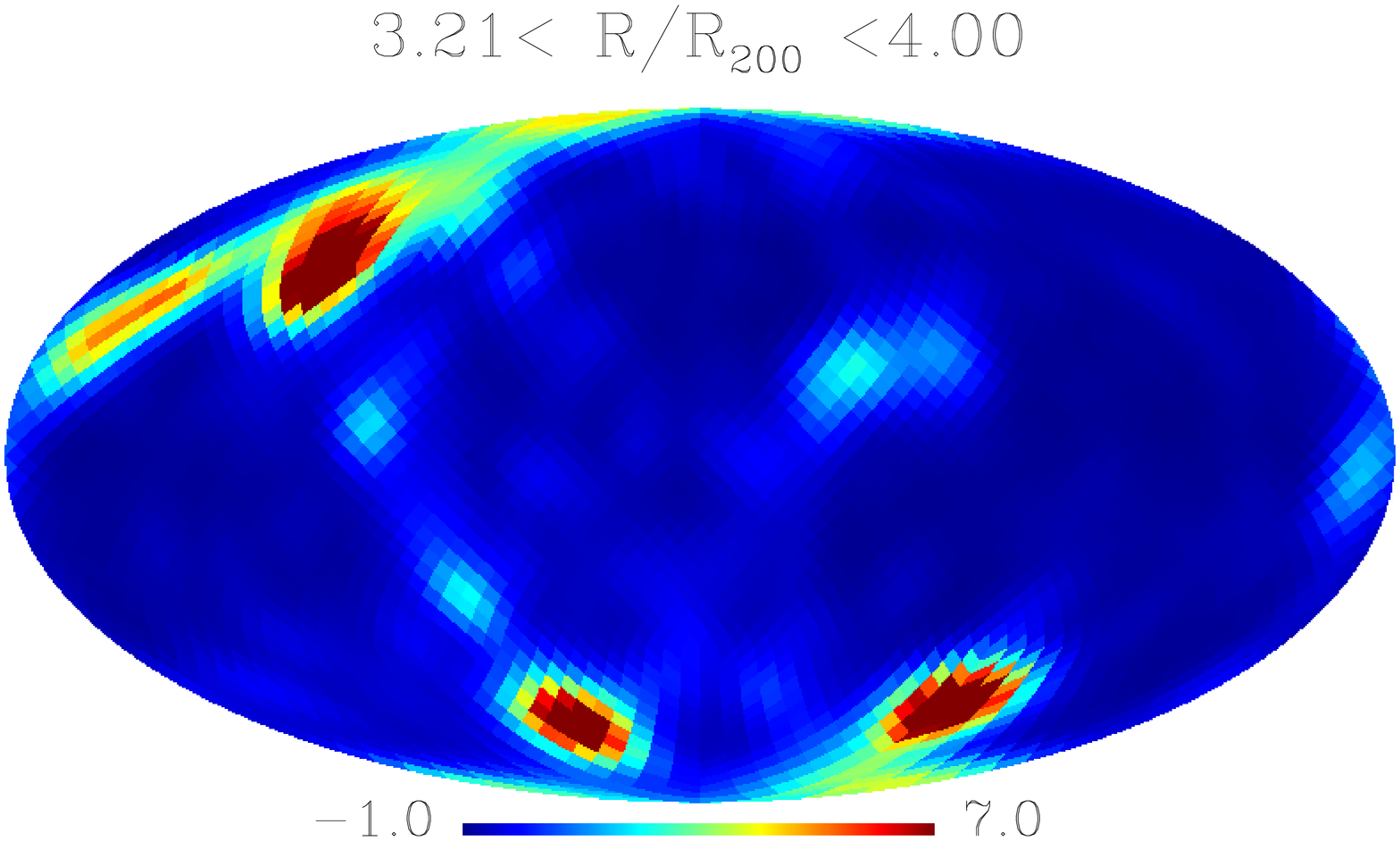}}%
    \resizebox{0.33\hsize}{!}{\includegraphics[angle=90,origin=c]{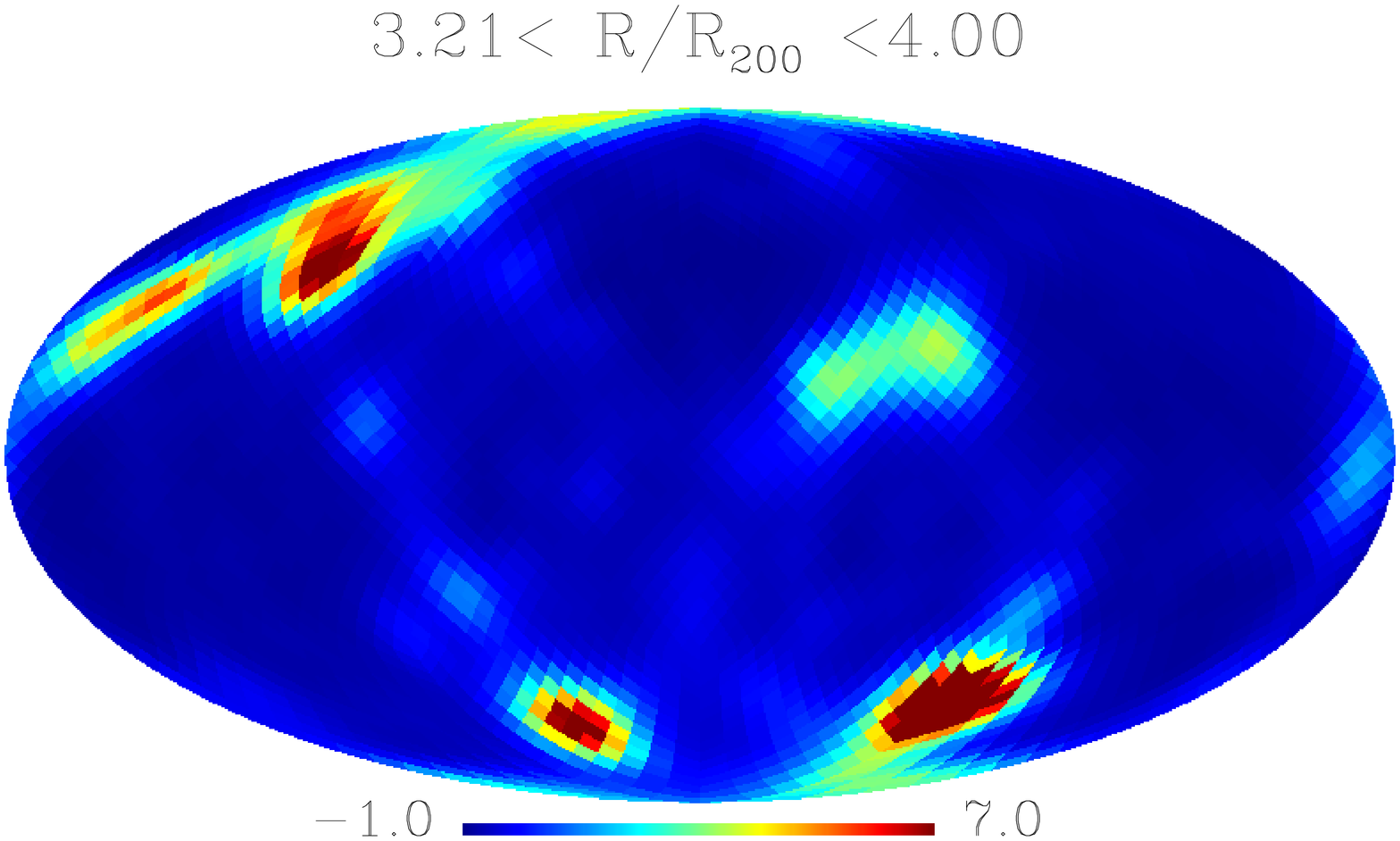}}%
    \resizebox{0.33\hsize}{!}{\includegraphics[angle=90,origin=c]{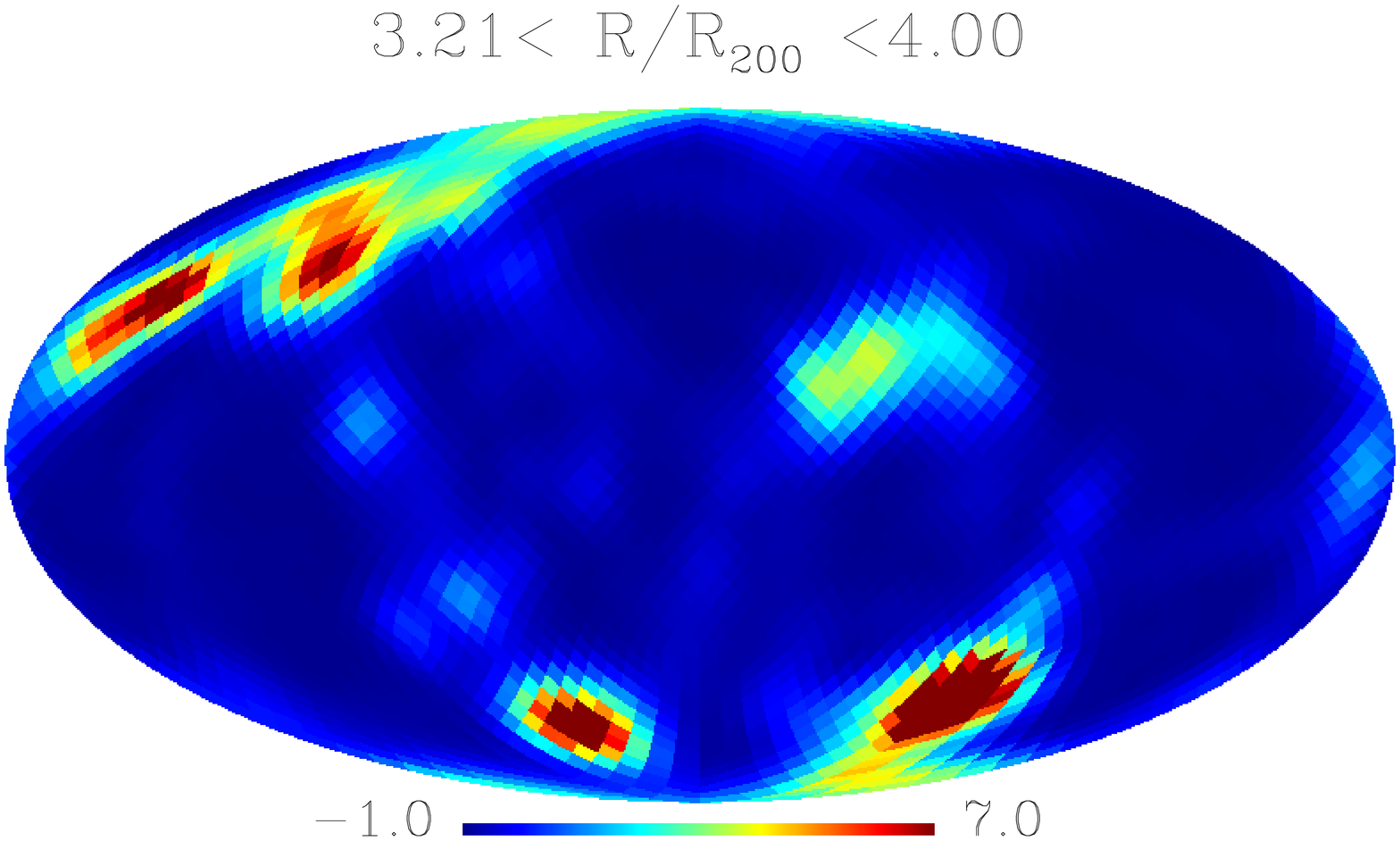}}\\  
    \caption{Mollweide equal area projections of the dimensionless fluctuations
      of the thermal pressure ($\delta P$, left panels), gas density
      ($\delta\rho$, middle panels), and DM density ($\delta\rho_\rmn{DM}$,
      right panels) for a cluster ($M_{200} \simeq 1.15\times 10^{15} M_{\sun}$,
      $z = 0$) in radial shells at $R\sim 0.5\,R_{200}$, $R\sim 1.5\,R_{200}$,
      and $R\sim 4\,R_{200}$ (top to bottom). We project the fields onto 3072
      angular pixels (using HEALPix with Nside = 16) and smooth the maps to
      $12^\circ$ FWHM. The color scale shows dimensionless fluctuations in
      linear units. Outside the virial radius, gas and pressure fluctuations
      show a strong correlation with the DM density, which becomes weaker for
      smaller radii due to dissipative gas effects in the ICM.}
  \label{fig:mole1}
\end{figure*}

\begin{figure*}
  \begin{minipage}[t]{0.33\hsize}
    \centering{\small $\delta P$}
    \vspace{1em}
  \end{minipage}
  \begin{minipage}[t]{0.33\hsize}
    \centering{\small $\delta\rho$}
    \vspace{1em}
  \end{minipage}
  \begin{minipage}[t]{0.33\hsize}
    \centering{\small $\delta\rho_\rmn{DM}$}
    \vspace{1em}
  \end{minipage}
    \resizebox{0.33\hsize}{!}{\includegraphics[angle=90,origin=c]{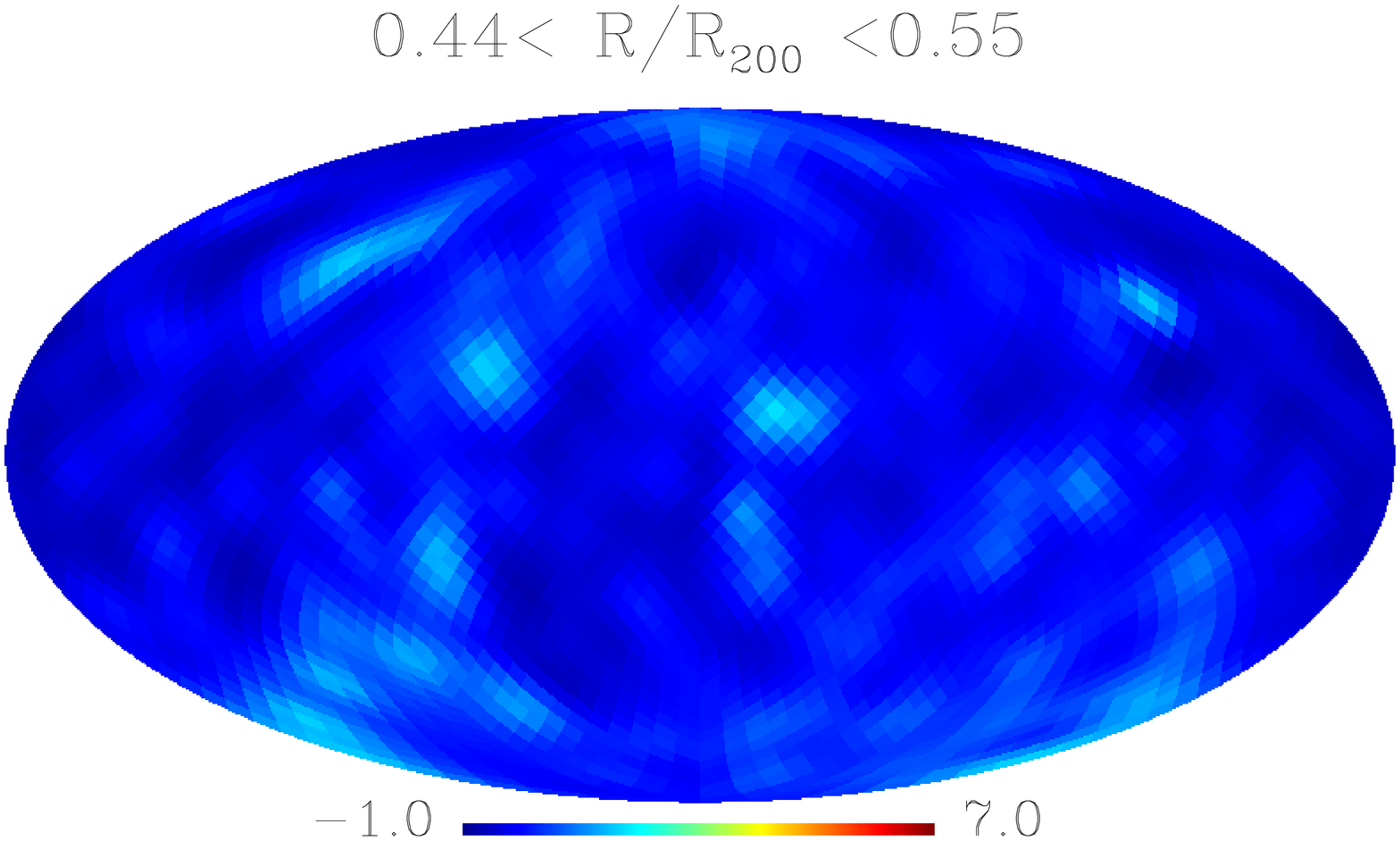}}%
    \resizebox{0.33\hsize}{!}{\includegraphics[angle=90,origin=c]{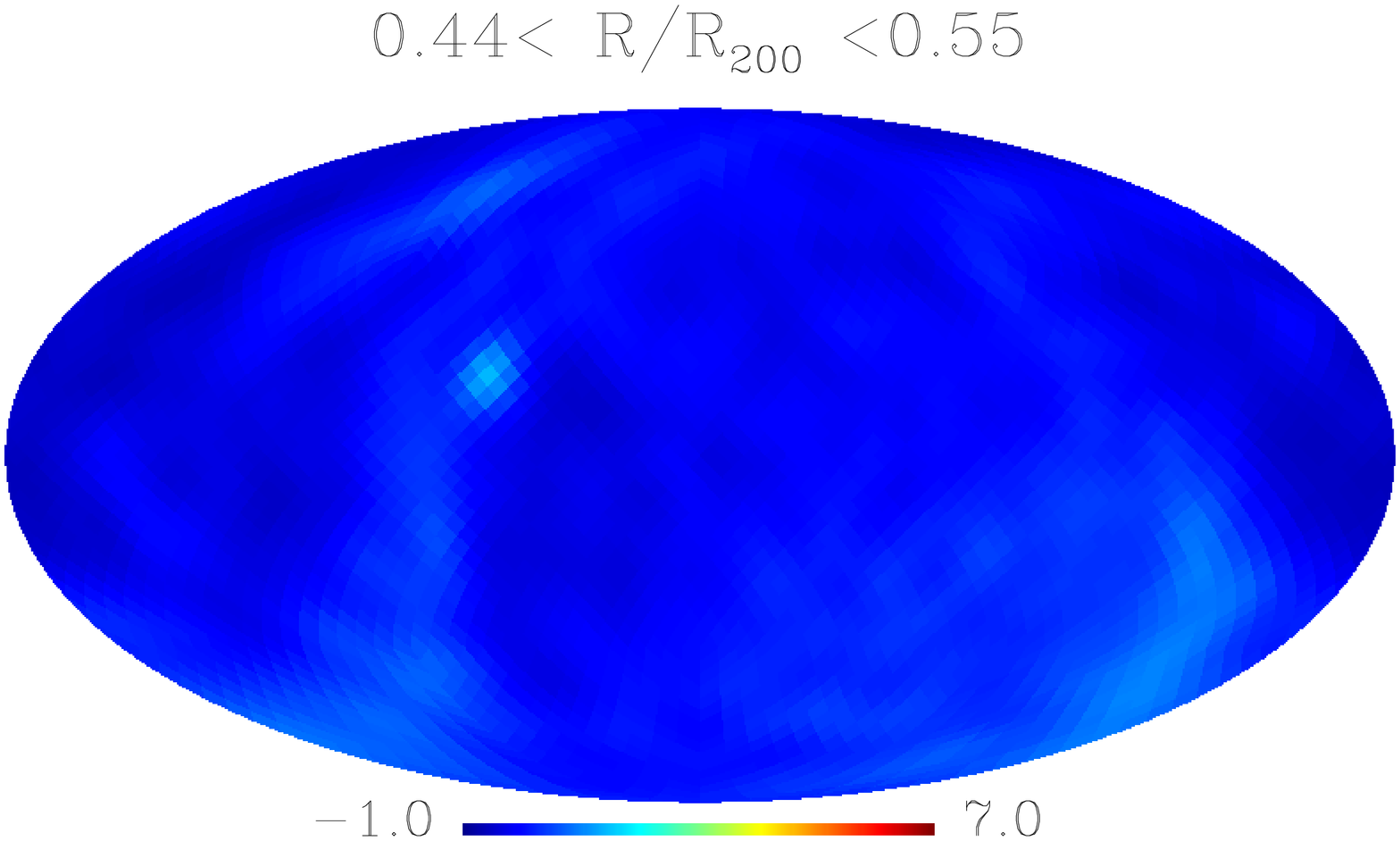}}%
    \resizebox{0.33\hsize}{!}{\includegraphics[angle=90,origin=c]{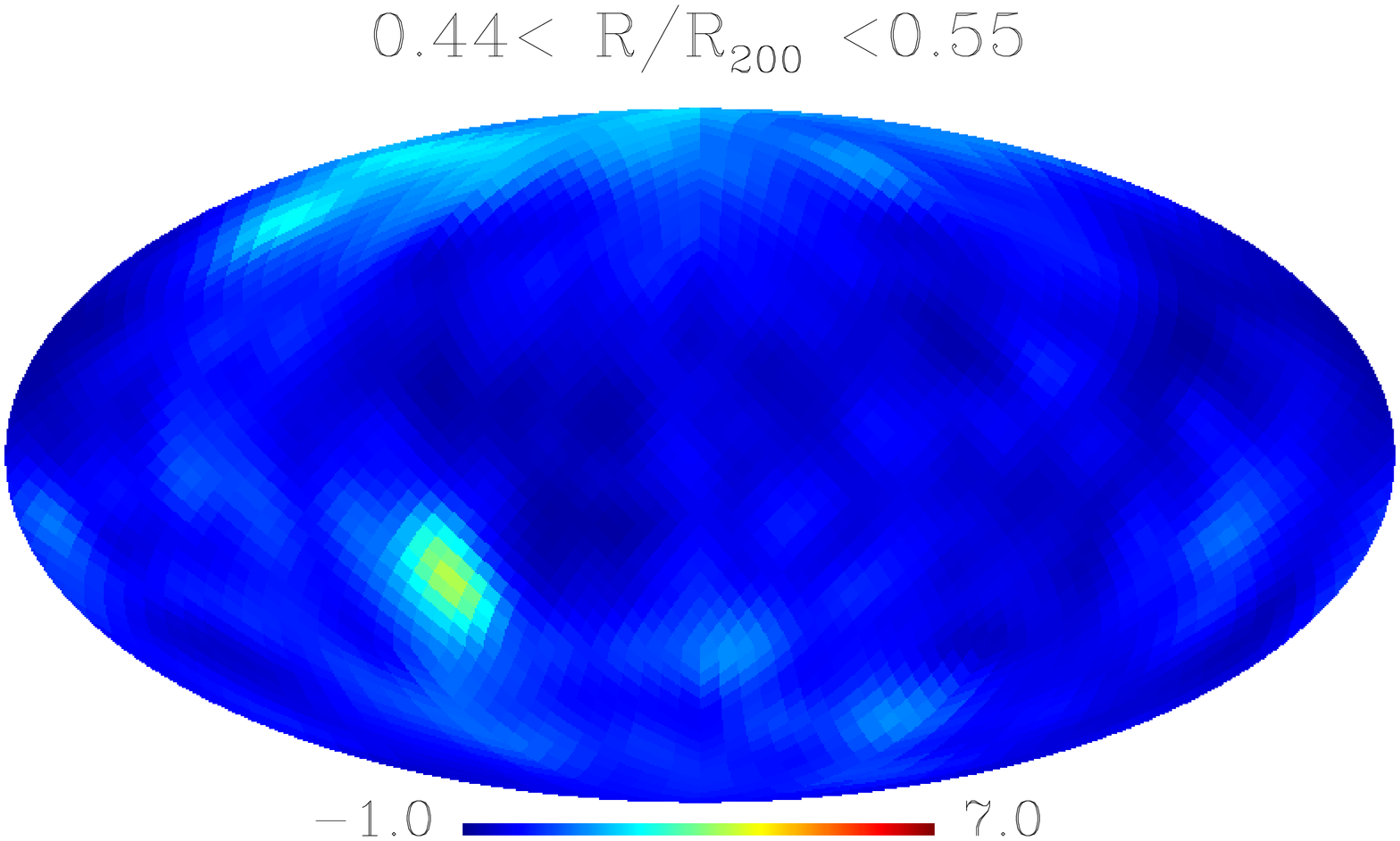}}\\  
    \resizebox{0.33\hsize}{!}{\includegraphics[angle=90,origin=c]{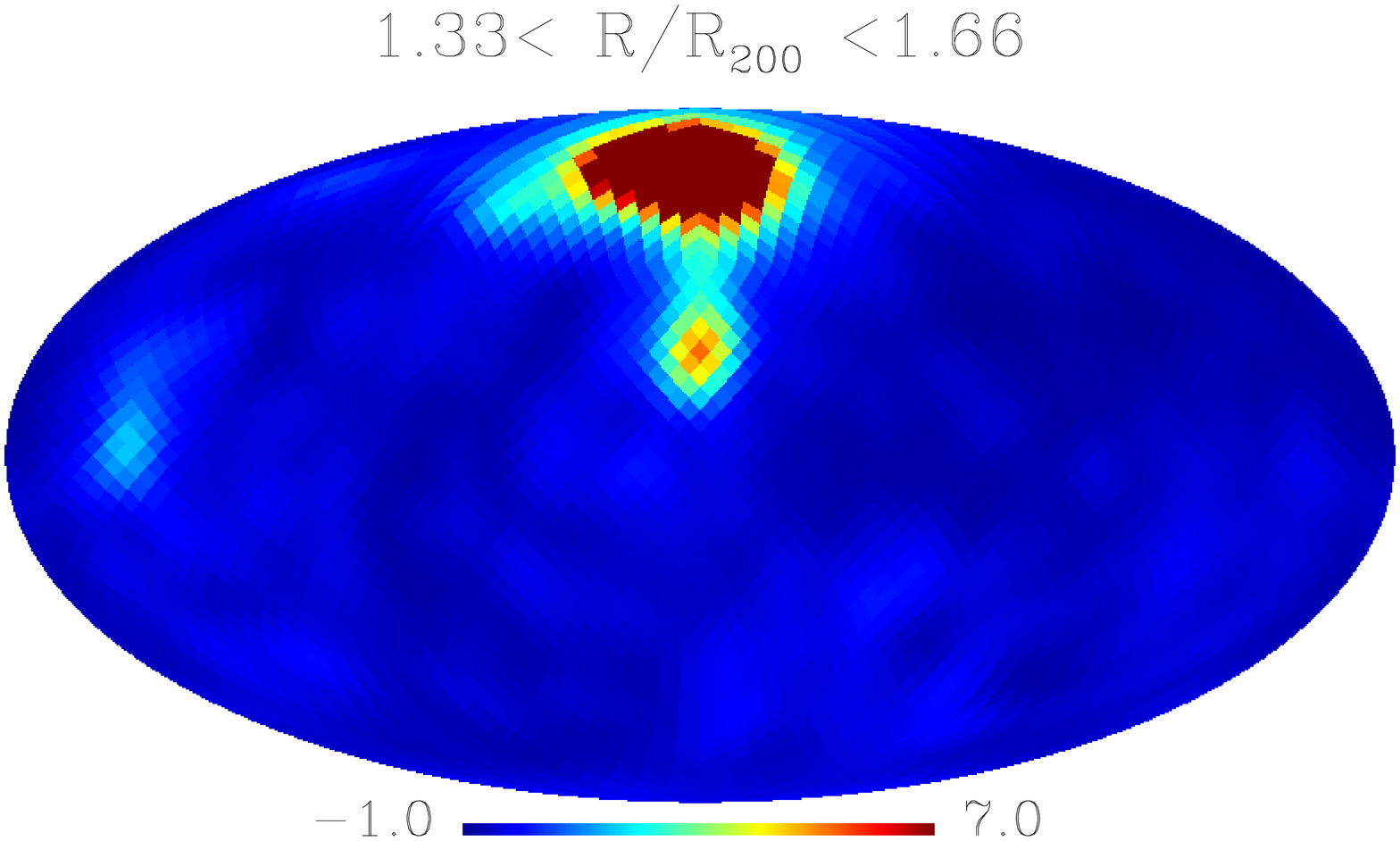}}%
    \resizebox{0.33\hsize}{!}{\includegraphics[angle=90,origin=c]{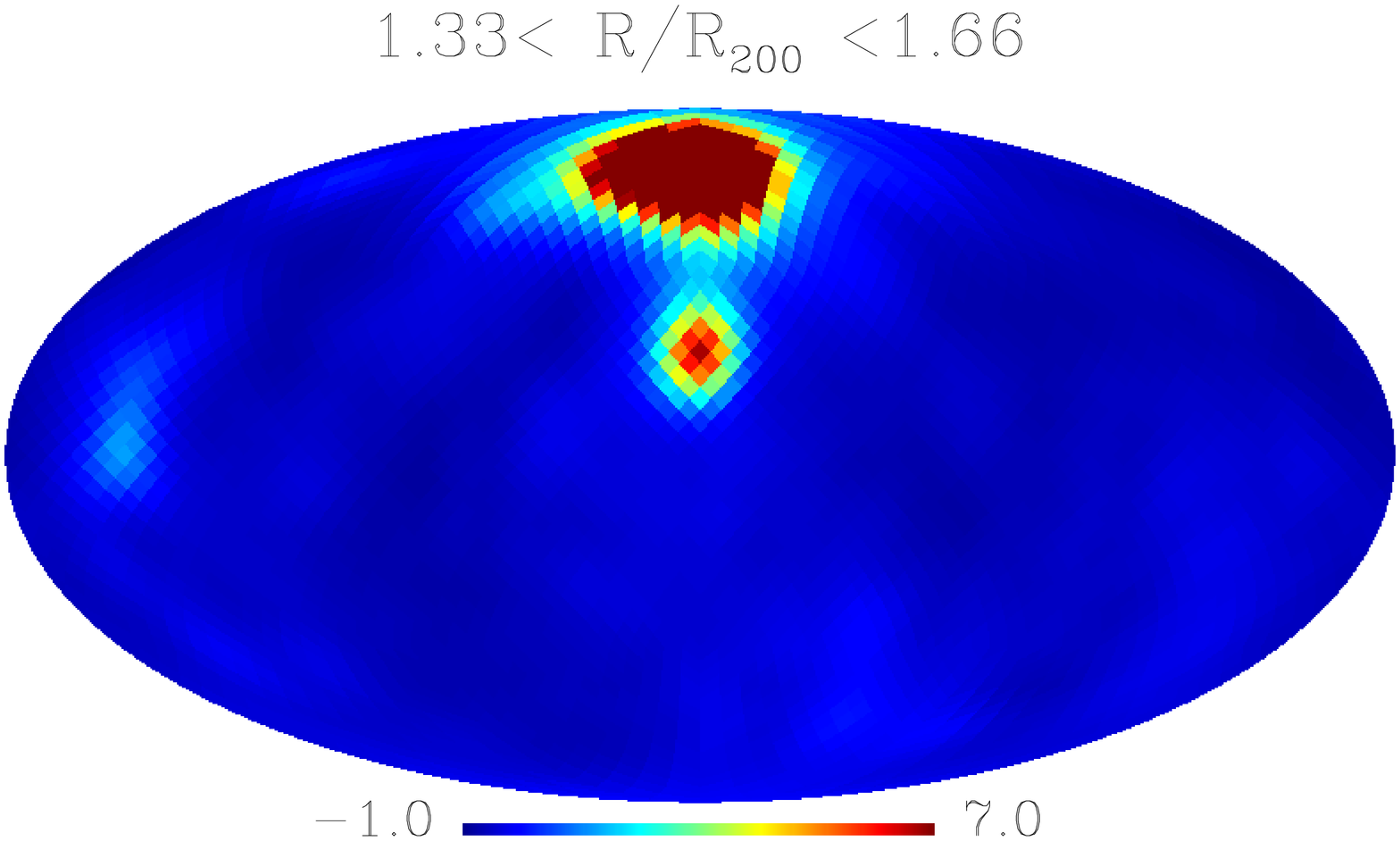}}%
    \resizebox{0.33\hsize}{!}{\includegraphics[angle=90,origin=c]{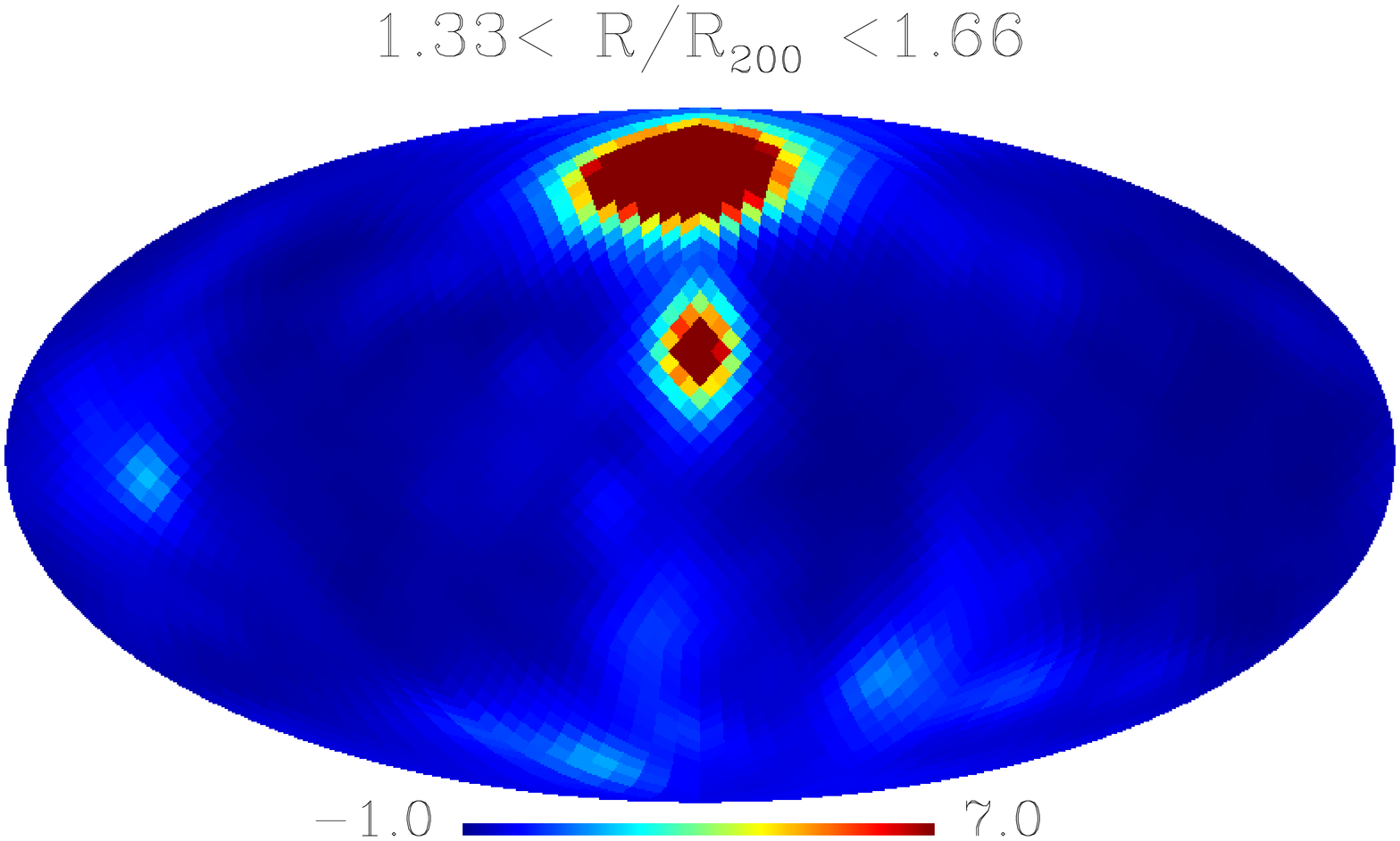}}\\  
    \resizebox{0.33\hsize}{!}{\includegraphics[angle=90,origin=c]{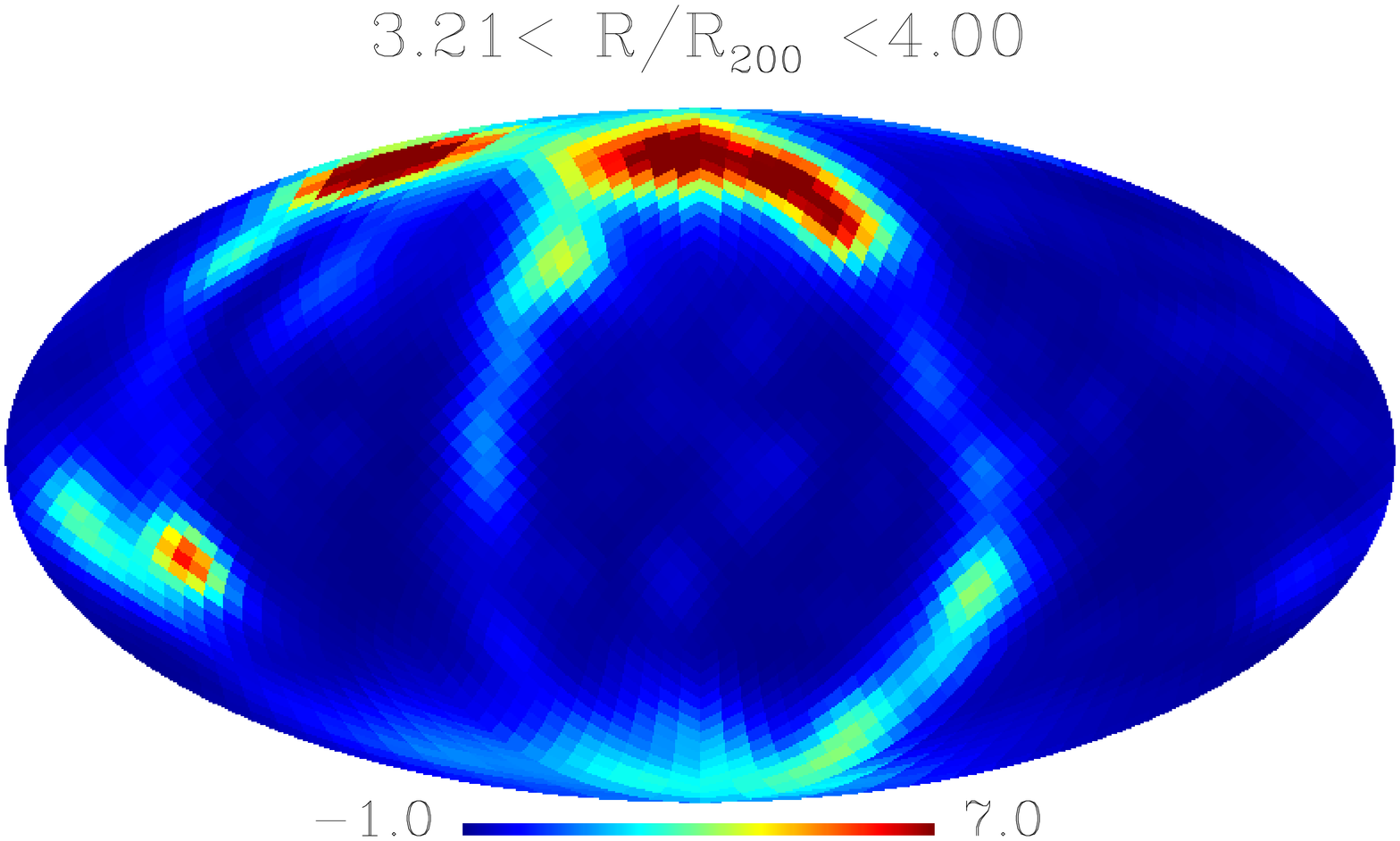}}%
    \resizebox{0.33\hsize}{!}{\includegraphics[angle=90,origin=c]{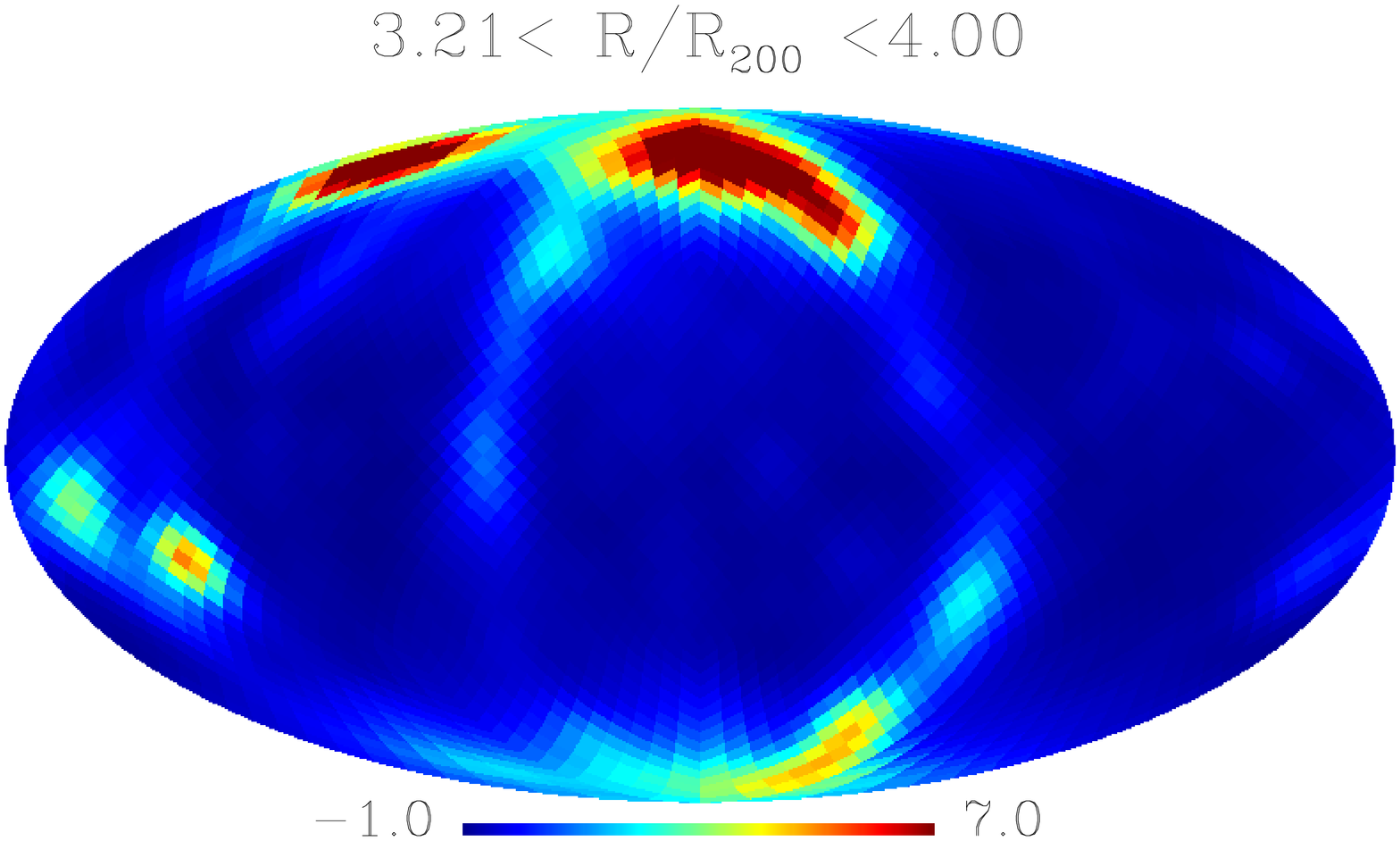}}%
    \resizebox{0.33\hsize}{!}{\includegraphics[angle=90,origin=c]{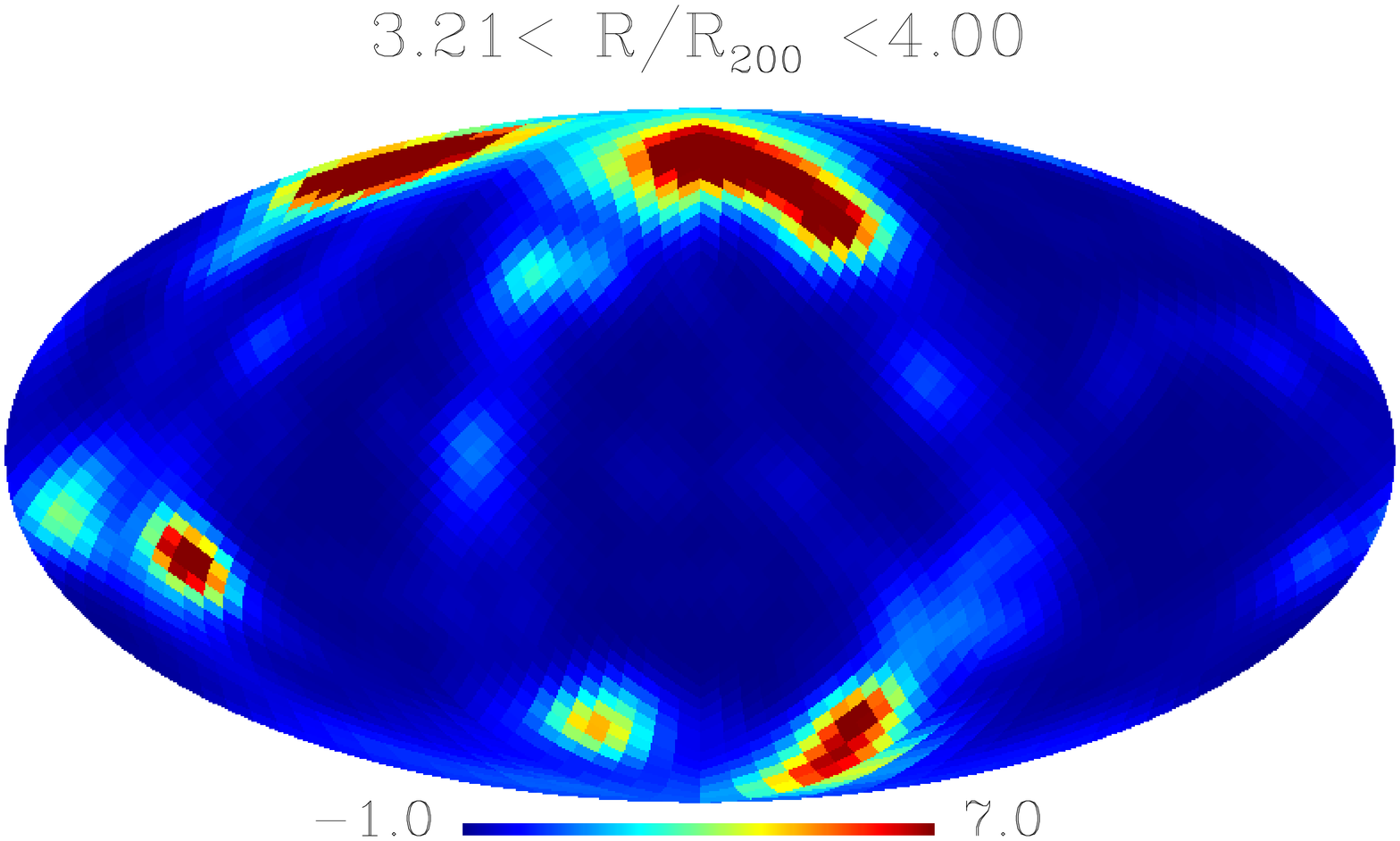}}\\  
  \caption{Same as Figure~\ref{fig:mole1} for a $M_{200} \simeq6.5\times 10^{14}
    M_{\sun}$ cluster at $z = 0$.}
  \label{fig:mole2}
\end{figure*}

In the following, we aim at understanding the origin of the clumping, why there
is a similar radial behavior of the density and pressure clumping, and to
elucidate the relation of gas density and pressure to the underlying DM
distribution. Following a similar procedure as described in \citetalias{BBPS3},
we subdivide the gas density, pressure, and the DM density distributions into
equal area angular cones. We use the HEALPix scheme \citep{gorski}, which uses
equal area pixels on the sphere, with Nside = 16 (3072 pixels). The value of
Nside is a compromise between angular resolution and the rising contribution of
the bin-to-bin variance due to SPH particle shot noise. In our treatment, SPH
particles are projected using the SPH kernel in the radial direction but not in
angular bins in order not to introduce an artificial smoothing of the power
spectrum toward small angular scales, which is associated with the finite
resolution.

\subsection{Projecting density and pressure fields onto the sphere}

In Figures~\ref{fig:mole1} and \ref{fig:mole2}, we show Mollweide equal area
projections of the gas density ($\rho$), pressure ($P$), and the DM density
($\rho_\rmn{DM}$) for two massive, low-redshift clusters in three radial shells
at $R\sim 0.5\,R_{200}$, $R\sim 1.5\,R_{200}$, and $R\sim 4\,R_{200}$. We
project the dimensionless fluctuations of the gas density ($\delta\rho$),
pressure ($\delta P$), and the DM density($\delta\rho_\rmn{DM}$),
\begin{eqnarray}
\delta\rho &=& (\rho - \bar{\rho})/\bar{\rho}, \nonumber \\
\delta P &=& (P - \bar{P})/\bar{P}, \nonumber \\
\delta\rho_\rmn{DM} &=& (\rho_\rmn{DM} - \bar{\rho}_\rmn{DM})/\bar{\rho}_\rmn{DM}.
\end{eqnarray}
\noindent Here $\bar{\rho}$, $\bar{P}$, and $\bar{\rho}_\rmn{DM}$ are the mean
values within each radial bin. We smooth the resulting maps to 12$^\circ$
full-width half-maximum (FWHM) for presentation purposes only. Our analysis is
performed on the unsmoothed maps.

We also conducted the same analyses using the median instead of the mean for
$\bar{\rho}$, $\bar{P}$, and $\bar{\rho}_\rmn{DM}$ and found very similar
results. The construction of these fluctuating fields $\delta{\rho}$,
$\delta{P}$, and $\delta{\rho}_\rmn{DM}$ is similar to subtracting the
underlying smooth density and pressure profiles.  However, by subtracting the
mean in each radial bin we remove any residual radial, bin-to-bin correlations,
which would have remained if we subtracted a best fit global profile. Lastly,
when constructing these fields we adopted the same temperature cuts to
$\delta\rho$ as in Section~\ref{sec:rho_p}, but not to $\delta P$.

Figures \ref{fig:mole1} and \ref{fig:mole2} illustrate that at a given radius
(panels left to right) the gas density, pressure, and DM density are
correlated. This correlation is stronger at $R\sim 4\,R_{200}$ where the gas
closely traces the DM structures while the correlation weakens for small radii
($R\sim 0.5\,R_{200}$) because of dissipational gas effects in the ICM such as
formation shocks and hydrodynamic instabilities. Comparing the maps at different
radii, there is no apparent correlation between the small-scale fluctuations. In
the following subsections we will quantify this.

The Mollweide projections and the observed trends do not change for different
physical simulation models. While global thermodynamic properties change upon
varying the modeled physics \citepalias[see, e.g.,][]{BBPS1}, clumping and
sub-structures remain almost invariant since they are related to halo growth,
which is dominated by the gravitational potential. As a result, clumping factors
and clump properties (such as sizes and shapes) are not affected by baryonic
processes \citep[e.g., Sec.\ref{sec:clump_phys} and][]{Ronc2013}.

\begin{figure*}
  \label{fig:clmole}
  \begin{minipage}[t]{0.33\hsize}
    \centering{\small $\delta P$}
  \end{minipage}
  \begin{minipage}[t]{0.33\hsize}
    \centering{\small $\delta\rho$}
  \end{minipage}
  \begin{minipage}[t]{0.33\hsize}
    \centering{\small $\delta\rho_\rmn{DM}$}
  \end{minipage}
  \resizebox{0.33\hsize}{!}{\includegraphics{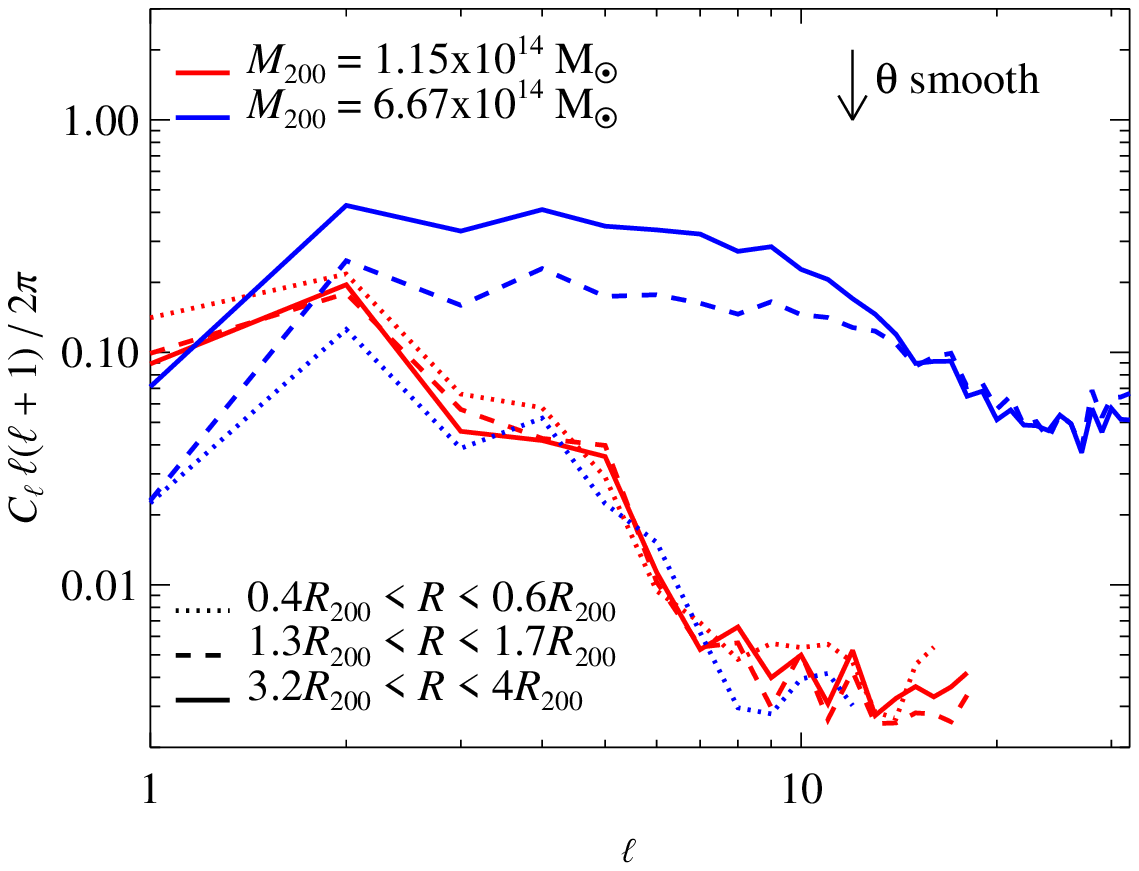}}%
  \resizebox{0.33\hsize}{!}{\includegraphics{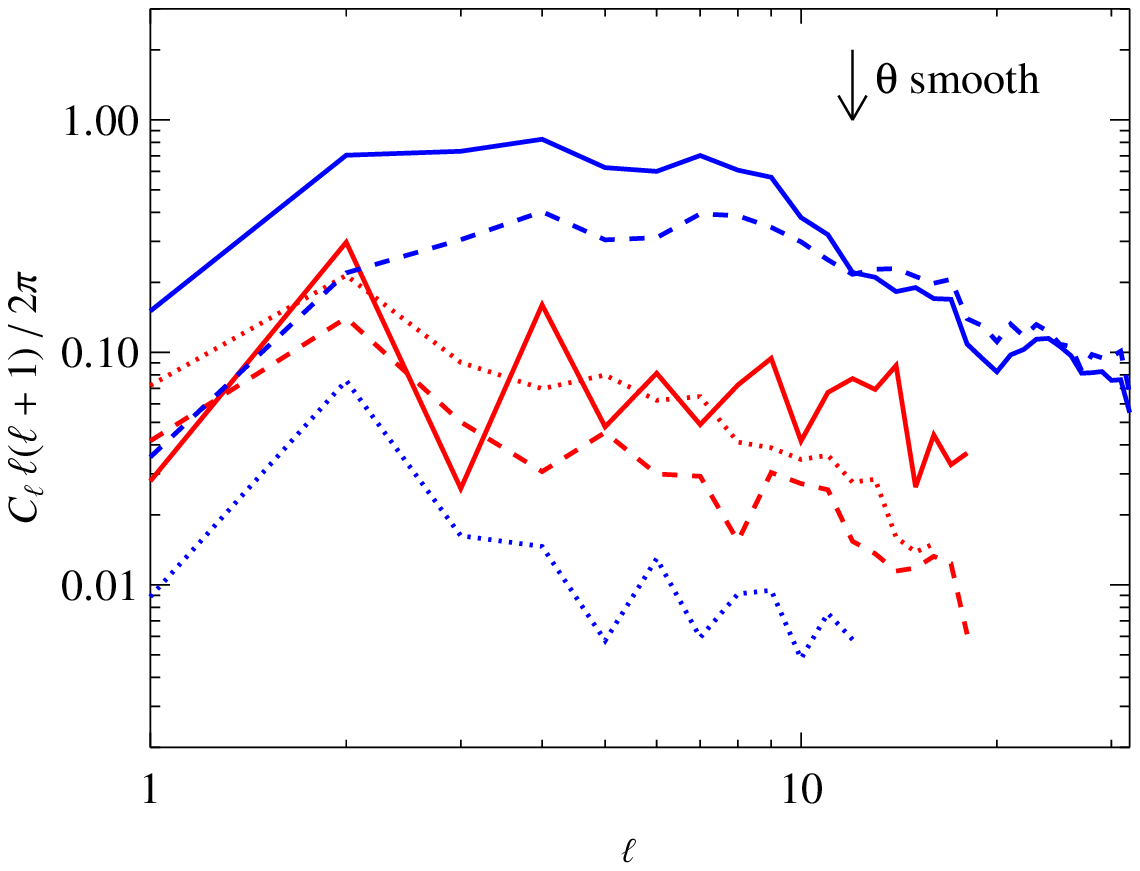}}%
  \resizebox{0.33\hsize}{!}{\includegraphics{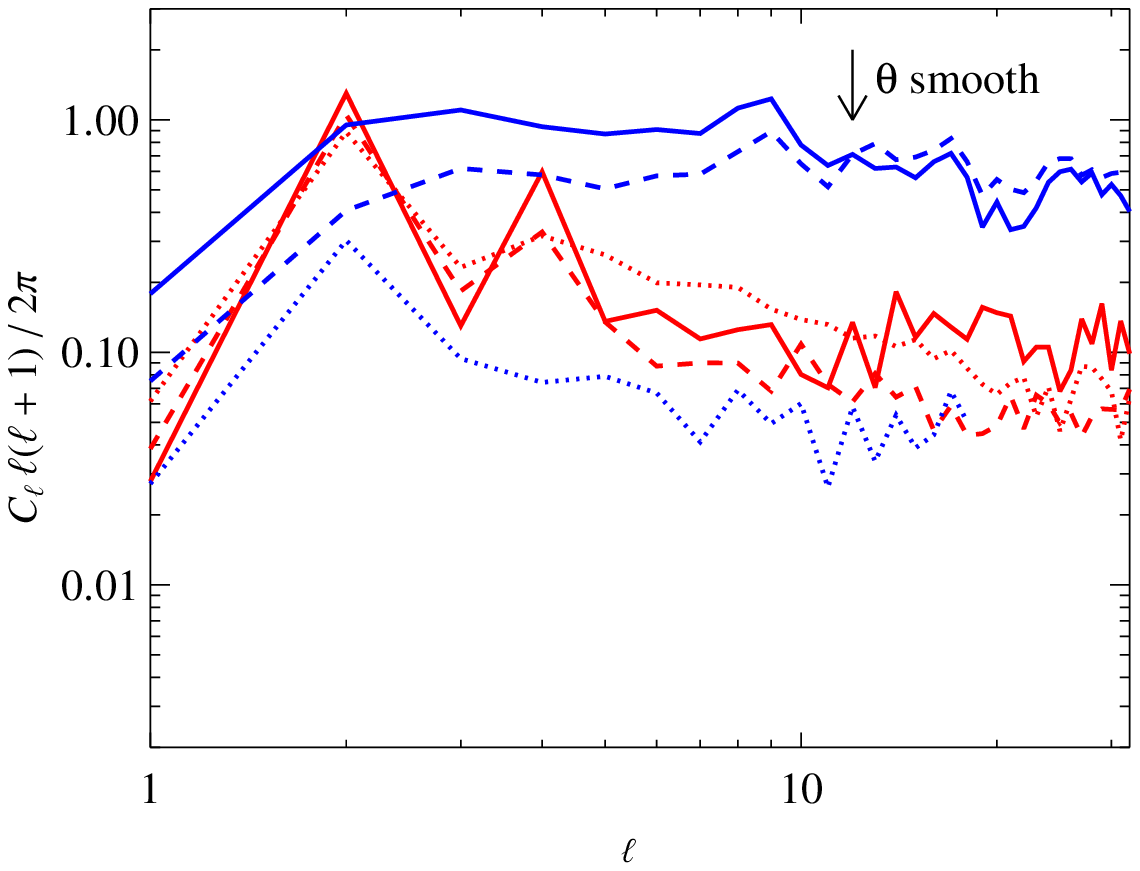}}\\  
  \caption{The angular power spectra ($C_\ell$) of the dimensionless
    fluctuations of the thermal pressure ($\delta P$, left), gas density
    ($\rho$, middle), and DM density ($\rho_\rmn{DM}$, right) for the two
    clusters shown in Figures~\ref{fig:mole1} (red lines) and \ref{fig:mole2}
    (blue lines) at $R\sim 0.5\,R_{200}$, $R\sim 1.5\,R_{200}$, and $R\sim
    4\,R_{200}$. In each panel, we indicate the smoothing scaling that was used
    in Figures~\ref{fig:mole1} and \ref{fig:mole2} by an arrow. The large
    variations of $C_\ell$ among those two clusters indicate the stochasticity
    introduced by the recent formation epoch of these systems.}
  \label{fig:clmole}
\end{figure*}

\subsection{Power-spectrum analysis}

To quantify the angular structures in the projections shown in
Figures~\ref{fig:mole1} and \ref{fig:mole2}, we perform an angular
power-spectrum analysis of each un-smoothed dimensionless fluctuation field on
the sphere.  We use the {\it ANAFAST} module in HEALpix that expands a projected
field $\delta$ into spherical harmonics,
\begin{eqnarray}
\delta(\theta,\phi) &=& \sum_{\ell=0}^\infty\sum_{m=-\ell}^\ell a_{\ell,m}Y_\ell^m(\theta,\phi),\\
a_{\ell,m} &=& \int\dd\Omega(\theta,\phi)\delta(\theta,\phi)Y_\ell^m(\theta,\phi).
\end{eqnarray}
Our analysis is analogous to that of quantifying the acoustic peaks in the
cosmic microwave background with the exception that we obtain a different
surface for each radial shell such that the physical scale of a given angular
mode increases as the radius from the cluster center increases.

For a Gaussian field the statistics are completely described by the
variance of the spherical harmonic eigenfunctions ($a_{\ell,m}$).  The projected
fluctuation fields $\delta{\rho}$, $\delta{P}$, and $\delta{\rho}_\rmn{DM}$ each
have non-Gaussian probability density functions so that there is additional
information contained in their higher-order statistics. Here, we constrain
ourselves to the analog of the two-point correlation in spherical harmonic
space, the angular power spectrum. We calculate the angular auto spectra,
\begin{equation}
C_{\ell} = \frac{1}{2\ell+1}\sum_{m=-\ell}^\ell a_{\ell,m} a_{\ell,m}^*,
\end{equation}
\noindent where the star denotes the complex conjugate, and the cross spectra,
\begin{equation} 
C_{\ell}^{i,j} = \frac{1}{2\ell+1}\sum_{m=-\ell}^\ell a_{\ell,m}^i a_{\ell,m}^{j*}
\end{equation}
\noindent for all radii and projected quantities $i$ and $j$. We use the cross
correlation coefficient,
\begin{equation}
r_{ij} = \frac{C_{\ell}^{i,j}}{\sqrt{C_{\ell}^{i,i} C_{\ell}^{j,j}}},
\label{eq:xcor}
\end{equation}
to quantify the correlations between the different quantities and radii.

In Figure \ref{fig:clmole}, we show the auto spectra of $\delta \rho$, $\delta
P$, and $\delta \rho_\rmn{DM}$, for the Mollweide projections of the two
clusters shown in Figures \ref{fig:mole1} and \ref{fig:mole2}. The large
variations of $C_\ell$ among those two clusters indicate the stochasticity
introduced by the recent formation epoch of these systems, implying an
incomplete relaxation for an ``average'' cluster. In all panels and at all radii
the quadrupole ($\ell = 2$ mode) dominates the normalized power spectrum, which
shows that cluster ellipticity is an important source of clumping. Our results
in \citetalias{BBPS1} demonstrate that clusters are predominately prolate over
the range of radii shown, i.e., that the expansion coefficient of
$Y_2^2(\theta,\phi)$ carries most of the weight in the power-spectrum
average.\footnote{These auto spectra exhibit ringing features in the even
  harmonics, which signal an axis-symmetric mass distribution that cannot be
  fully described by an ellipsoidal shape and exhibits (sub-dominant) higher
  order contributions to the even harmonics.}  However, the auto spectra show a
broad spectrum of modes contributing to the signal and lack of prominent peaks
beyond the quadrupole. Generally, the spectra of the outer radial shells show
more power and are flatter than the spectra at smaller radii. This result is
consistent with our findings in Section~\ref{sec:clumping} \citep[see
also][]{2011ApJ...731L..10N,Zhur2013,Ronc2013}, which shows an increasing
clumping factor at larger radii, particularly for $\delta \rho_\rmn{DM}$.

\begin{figure*}
  \begin{minipage}[t]{0.33\hsize}
    \centering{\small $\delta P$}
  \end{minipage}
  \begin{minipage}[t]{0.33\hsize}
    \centering{\small $\delta\rho$}
  \end{minipage}
  \begin{minipage}[t]{0.33\hsize}
    \centering{\small $\delta\rho_\rmn{DM}$}
  \end{minipage}
  \resizebox{0.33\hsize}{!}{\includegraphics{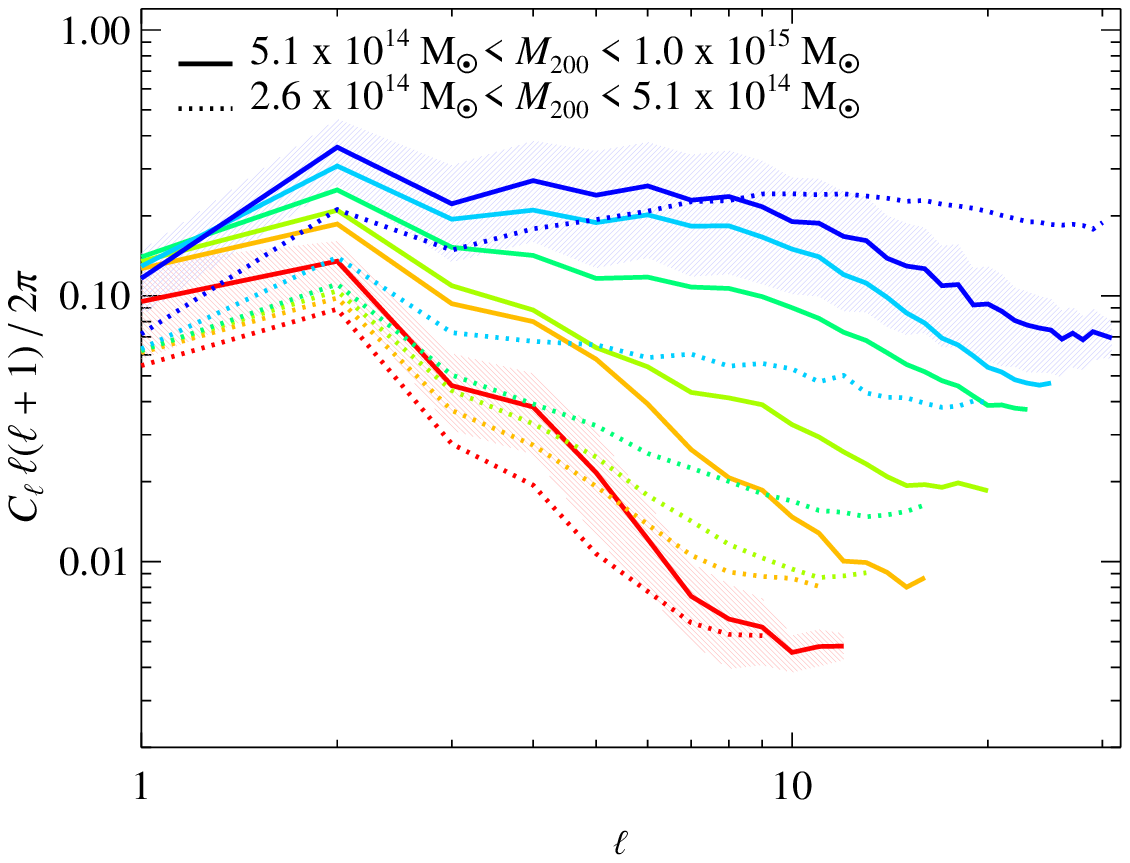}}%
  \resizebox{0.33\hsize}{!}{\includegraphics{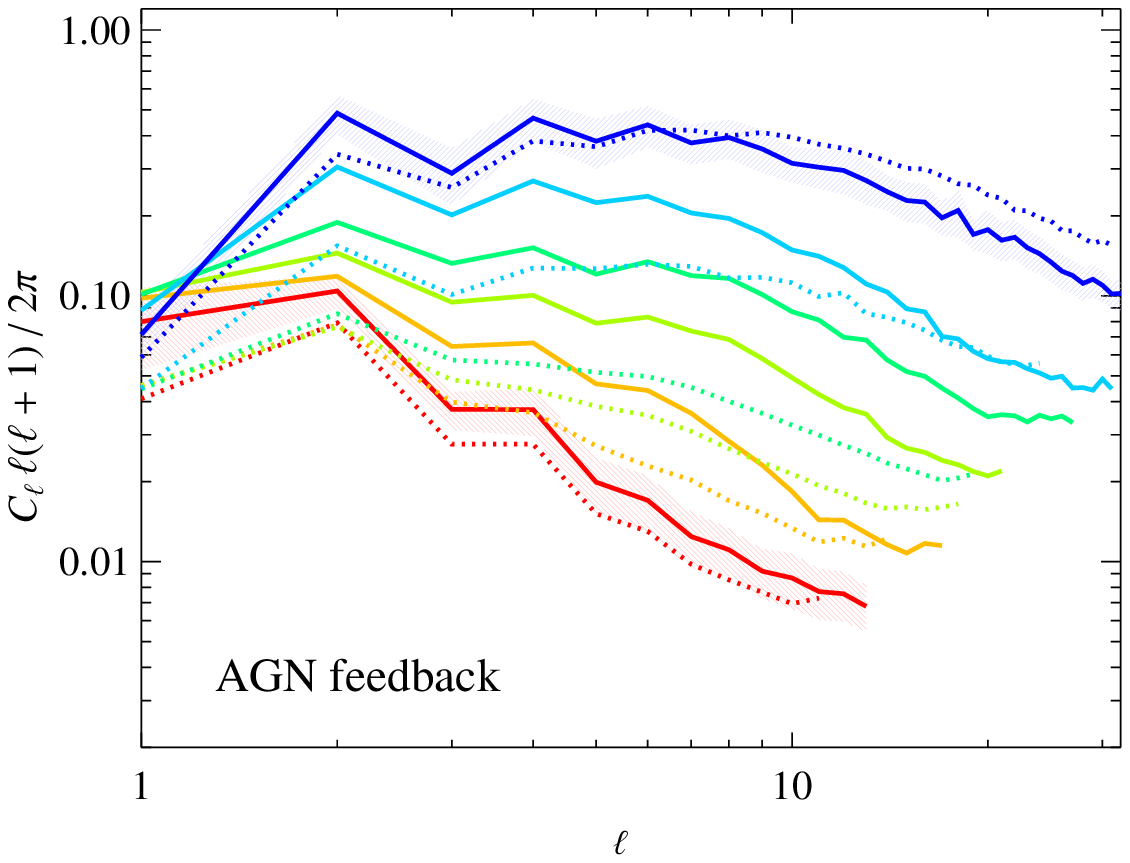}}%
  \resizebox{0.33\hsize}{!}{\includegraphics{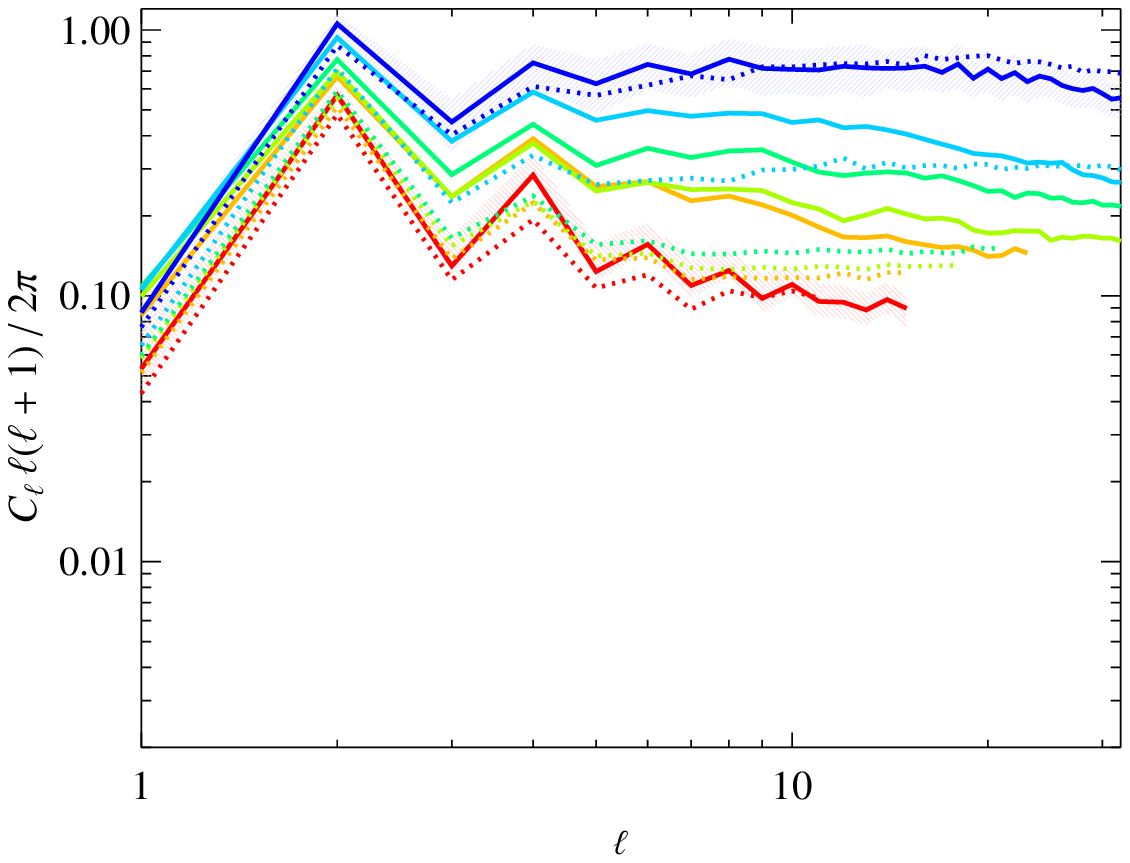}}\\
  \resizebox{0.33\hsize}{!}{\includegraphics{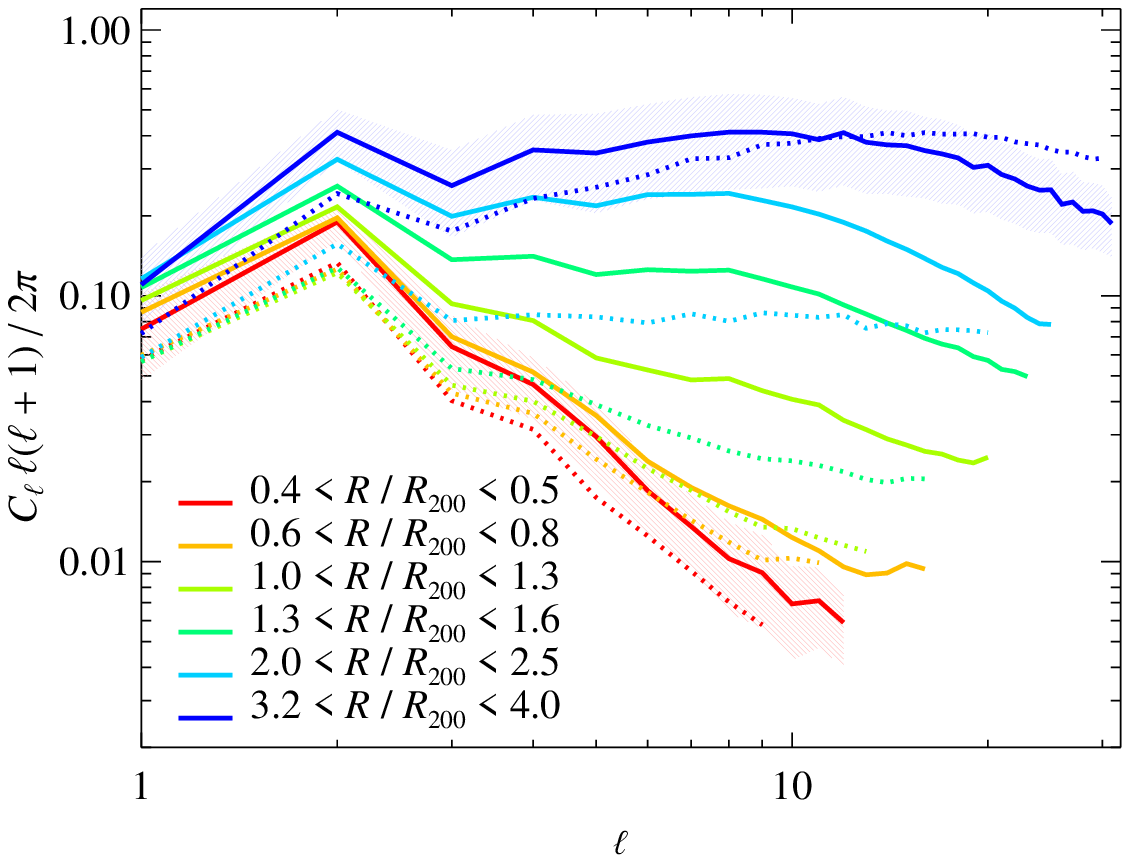}}%
  \resizebox{0.33\hsize}{!}{\includegraphics{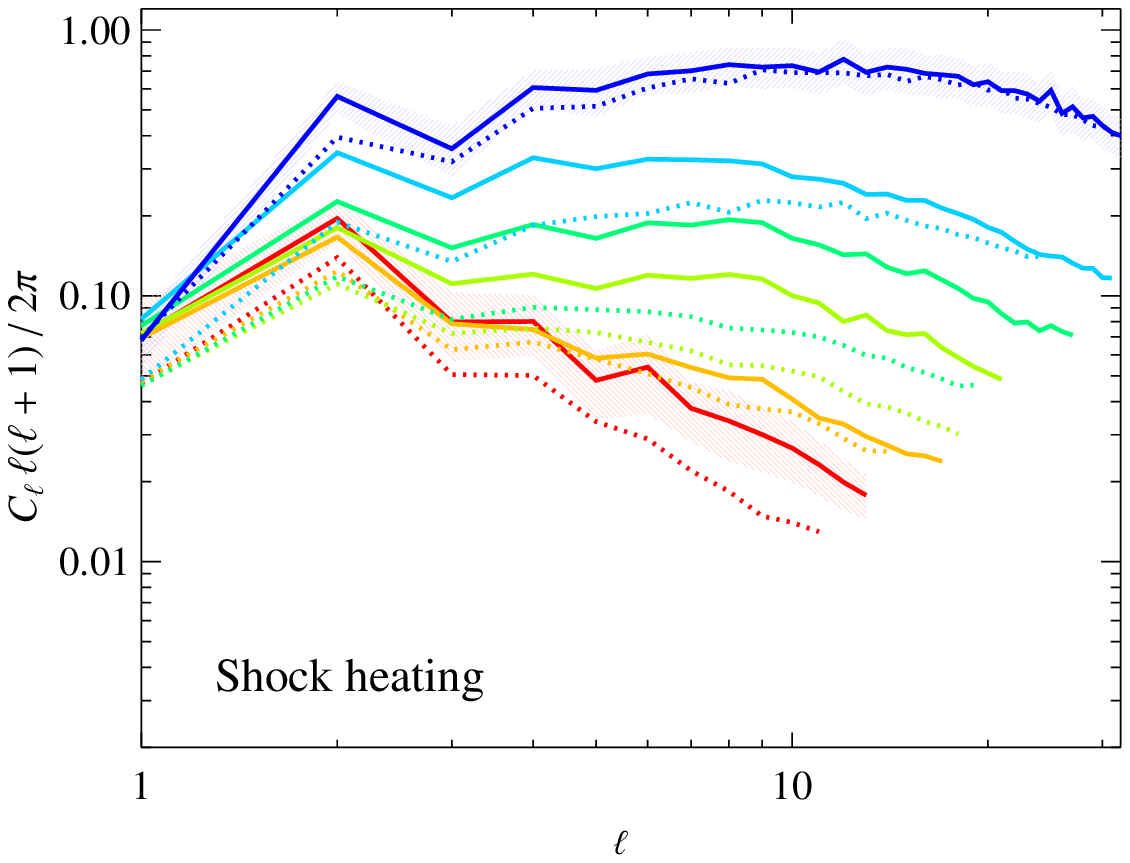}}%
  \resizebox{0.33\hsize}{!}{\includegraphics{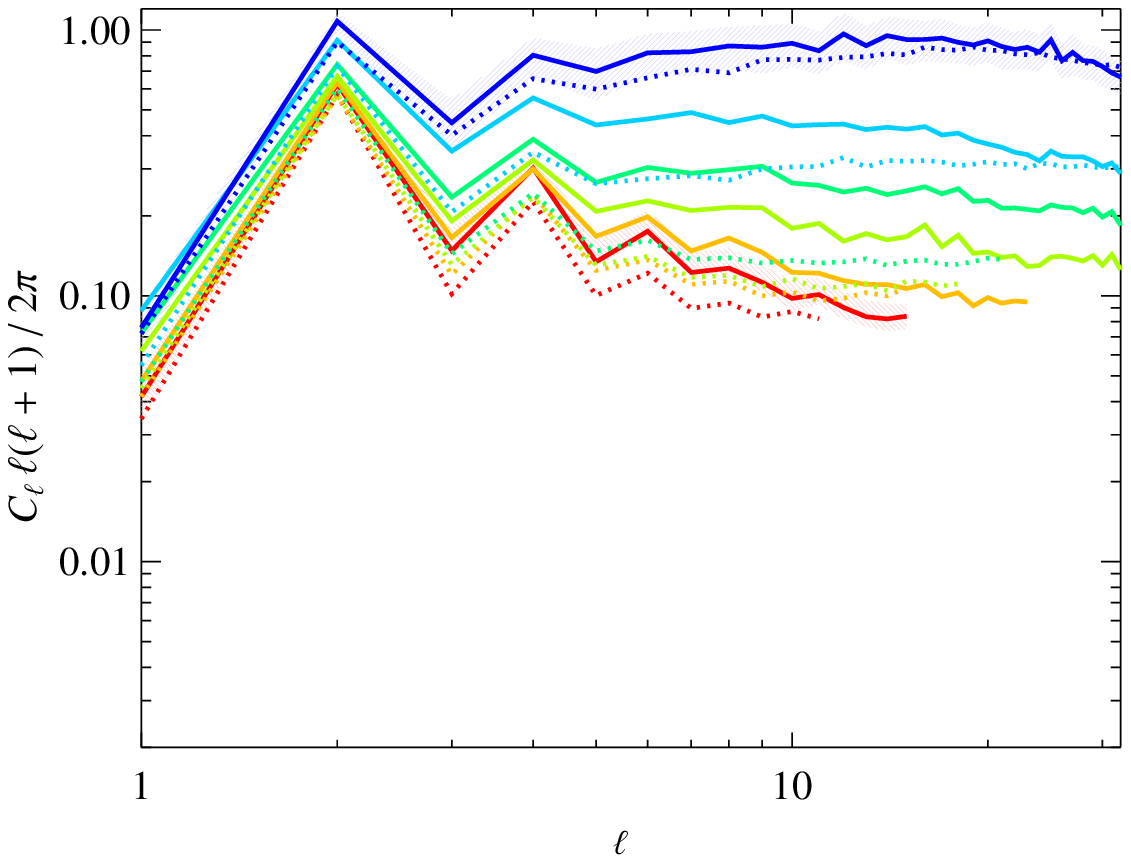}}\\    
  \caption{The average angular power spectra of the fluctuations in thermal
    pressure ($P$, left panels), gas density ($\rho$, middle panels), and the DM
    density ($\rho_\rmn{DM}$, right panels) at various radii (shown with
    different colors) for clusters within the mass ranges provided in the
    legend. The top panels show the power spectra of the {\em AGN feedback}
    simulations and the bottom panels show the {\em shock heating-only}
    simulations. Here the colored bands represent the error on the mean power
    spectrum for the smallest and largest radii, respectively. The quadrupole
    (mode with $\ell = 2$) dominates the normalized power spectrum, which shows
    that cluster ellipticity is an important source of clumping.  The general
    similarities between the average power spectra across the panels suggest
    that these spectra are largely independent of the simulated physics and
    cluster mass. The radial trend results from the different physical sizes
    associated with the angular scale at a given radius and changes in the
    background fields.}
  \label{fig:avgcl}
\end{figure*}

\begin{figure*}
  \begin{minipage}[t]{0.33\hsize}
    \centering{\small $\delta P$}
  \end{minipage}
  \begin{minipage}[t]{0.33\hsize}
    \centering{\small $\delta\rho$}
  \end{minipage}
  \begin{minipage}[t]{0.33\hsize}
    \centering{\small $\delta\rho_\rmn{DM}$}
  \end{minipage}
  \resizebox{0.33\hsize}{!}{\includegraphics{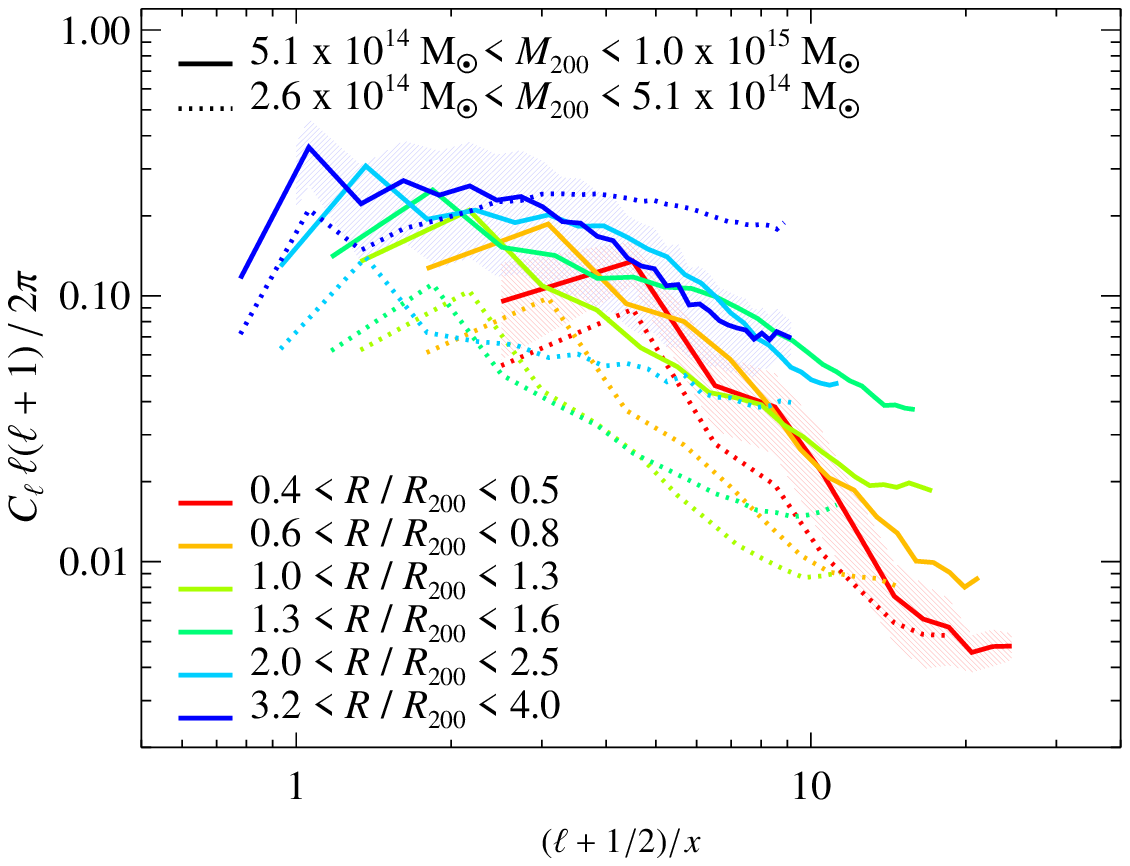}}%
  \resizebox{0.33\hsize}{!}{\includegraphics{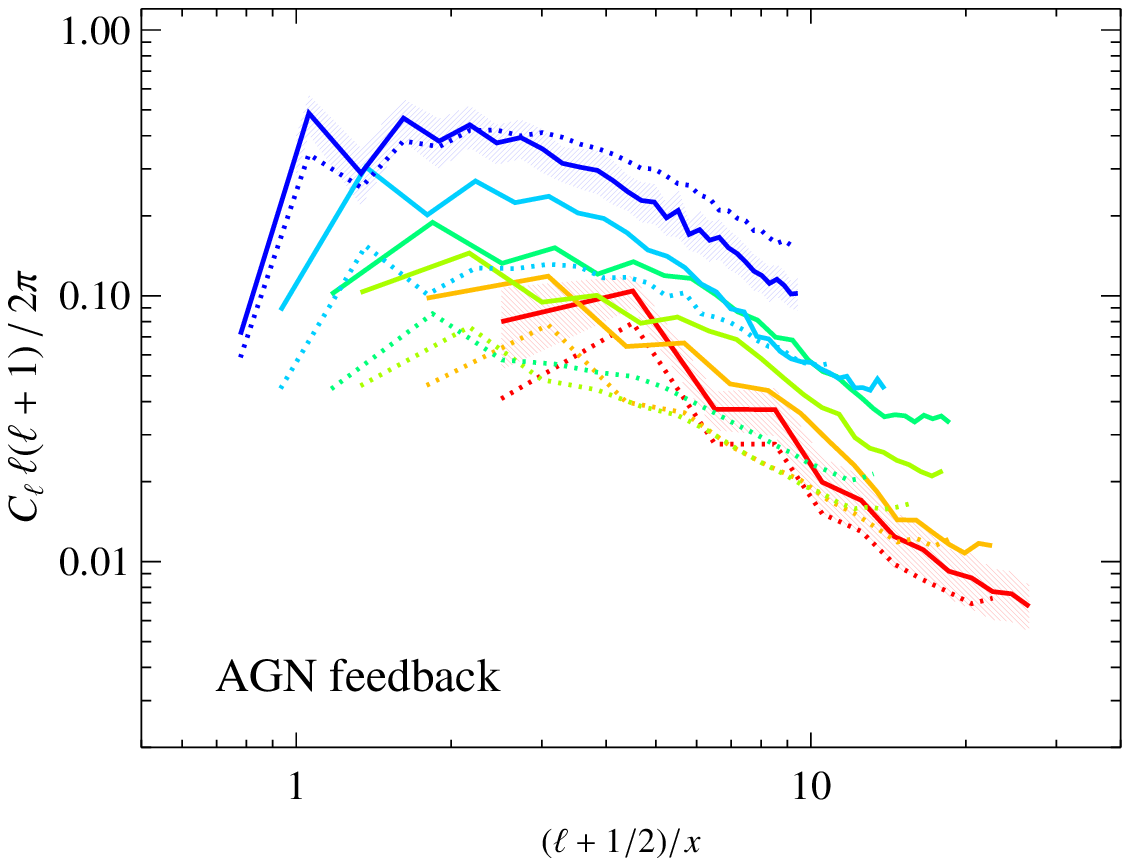}}%
  \resizebox{0.33\hsize}{!}{\includegraphics{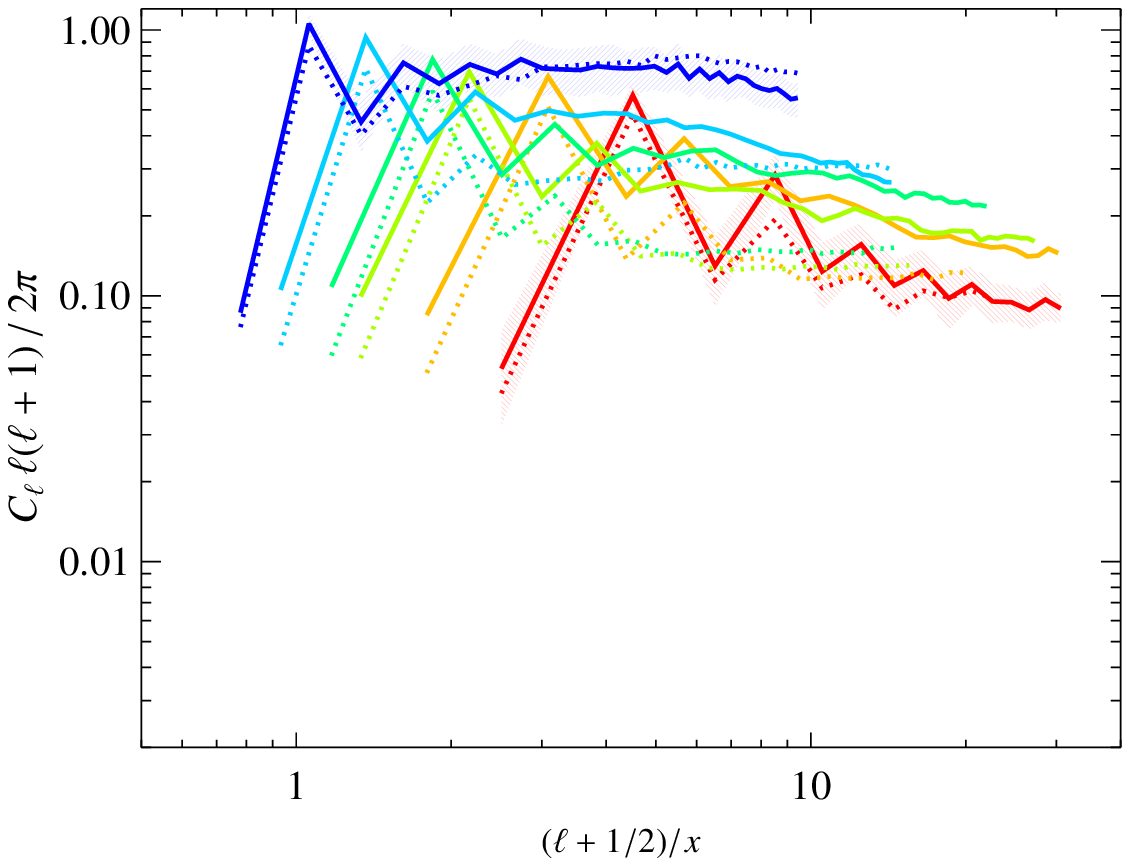}}\\
  \caption{The average angular power spectra as a function of
    $k_\perp\simeq(\ell+1/2) / x$, where $x = R/R_{200}$ (the equality sign in
    the equation for $k_\perp$ assumes the small-angle limit).  We show the
    fluctuations in thermal pressure ($P$, left), gas density ($\rho$, middle),
    and the DM density ($\rho_\rmn{DM}$, right) at various radii (shown with
    different colors) for clusters within the mass ranges provided in the
    legend. Here the colored bands represent the error on the mean power
    spectrum for the smallest and largest radii, respectively. After scaling the
    abscissa to the same physical scale, the gas density and pressure power
    spectra at different radii delineate the same spectral bump (albeit with
    different spatial resolution), indicating the presence of
    ``super-clumping''. In contrast, the DM density power spectrum remains
    comparably flat toward smaller scales, which implies a continuous
    distribution of subhalos.}
  \label{fig:avgclk}
\end{figure*}

In Figure~\ref{fig:avgcl} we show the average auto power spectra for $\delta
\rho$, $\delta P$, and $\delta \rho_\rmn{DM}$ at various radii. Here we averaged
the spectra over massive clusters ($ 5.1 \times 10^{14} M_{\sun}\, < M_{200} <
1.0 \times 10^{14} M _{\sun}$) and intermediate-mass clusters ($ 2.6 \times
10^{14} M_{\sun}\, < M_{200} < 5.1 \times 10^{14} M _{\sun}$) at $z = 0$. We
find that the average power spectra have similar features in comparison to the
individual power spectra shown above. The spectra show a dominant quadrupole
moment and the subsequent ringing in the even harmonics, albeit to a lesser
extent. More massive cluster have more angular power than lighter
clusters. However, the differences in angular power between different mass
ranges is close to the cluster-to-cluster scatter.

The scaled angular power spectra of gas density and pressure show an evolution
of the spectral shape from a broad bump at $R_{200}/2$ to an almost flat
distribution at $4\,R_{200}$. To understand which part of this evolution is
driven by the decrease in angular scale when moving an object of fixed physical
scale toward larger radii, we plot $C_\ell$ as a function of $(\ell+1/2)/x$,
where $x = R/R_{200}$, in Figure~\ref{fig:avgclk}. In the small-angle limit,
$k_\perp=(\ell+1/2)/x$, where $k_\perp$ denotes the physical (perpendicular)
wave number. The resulting power spectra of $\delta P$ at different radii match
very closely in shape while that of $\delta \rho$ shows a residual offset with a
slightly enhanced amplitude of $C_\ell$ at larger radii. Most importantly, the
gas density and pressure power spectra exhibit a broad bump, signaling the
presence of ``super-clumping'' \citep{2014arXiv1404.6250M}. This demonstrates
that gas density and pressure clumping are dominated by comparably large
(sub-)structures with scales $L_\perp\gtrsim \pi R_{200}/k_\perp \sim
R_{200}/5$, which do not get broken up through Kelvin-Helmholtz instabilities or
transformed at shocks due to anisotropic stresses. We emphasize that this
``super-clumping'' is qualitatively different from possible signatures of
``clumping'' inferred in nearby clusters \citep[e.g.,][]{2011Sci...331.1576S,
  2014MNRAS.437.3939U}, which has been interpreted as a distribution of subhalos
accreting onto clusters, whereas group-sized halos would be directly detected
and masked.

\begin{figure*}
  \begin{minipage}[t]{0.33\hsize}
    \centering{\small $\ell = 2$}
  \end{minipage}
  \begin{minipage}[t]{0.33\hsize}
    \centering{\small $\ell =4$}
  \end{minipage}
  \begin{minipage}[t]{0.33\hsize}
    \centering{\small $\ell = 8$}
  \end{minipage}
  \resizebox{0.33\hsize}{!}{\includegraphics{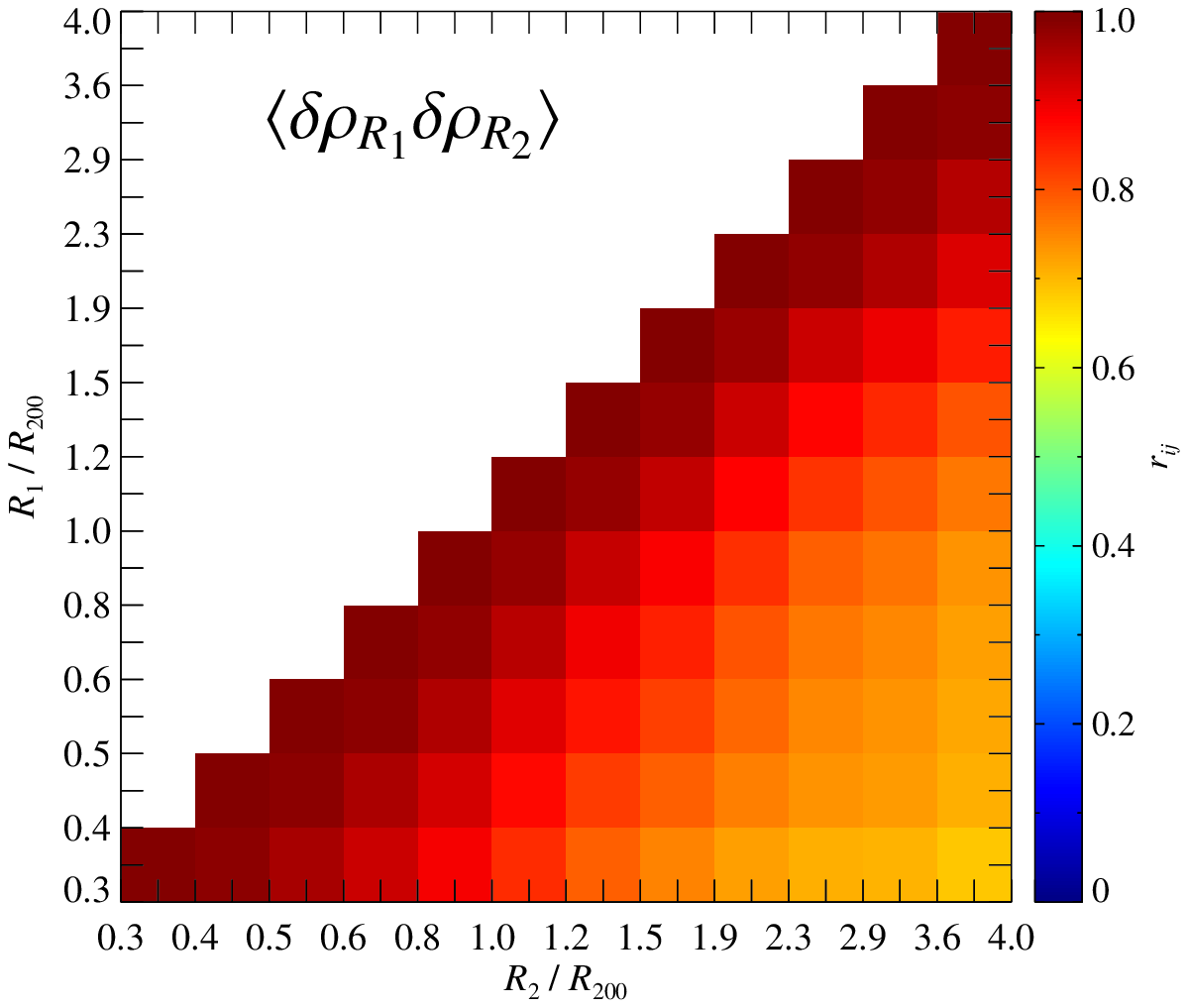}}%
  \resizebox{0.33\hsize}{!}{\includegraphics{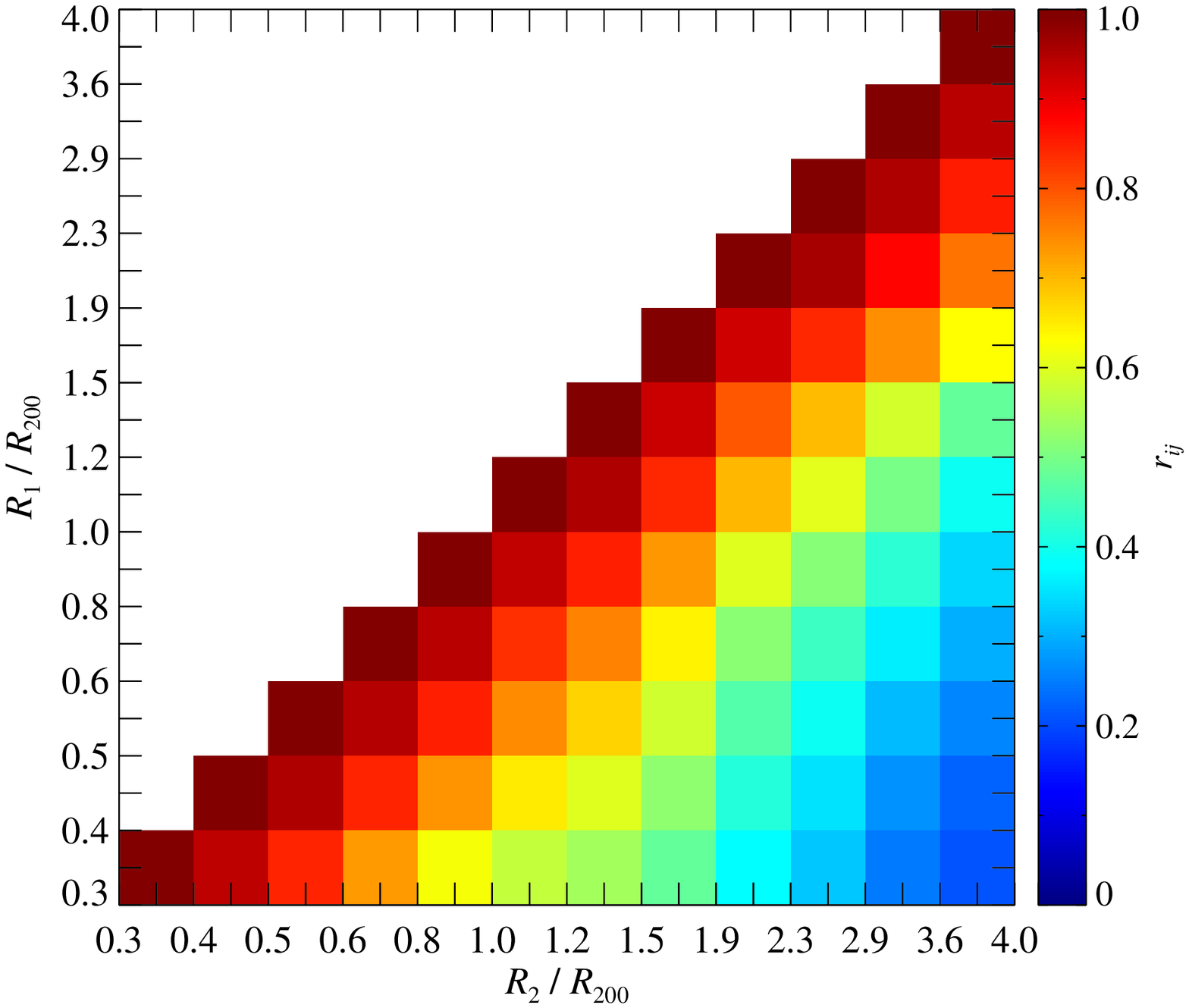}}%
  \resizebox{0.33\hsize}{!}{\includegraphics{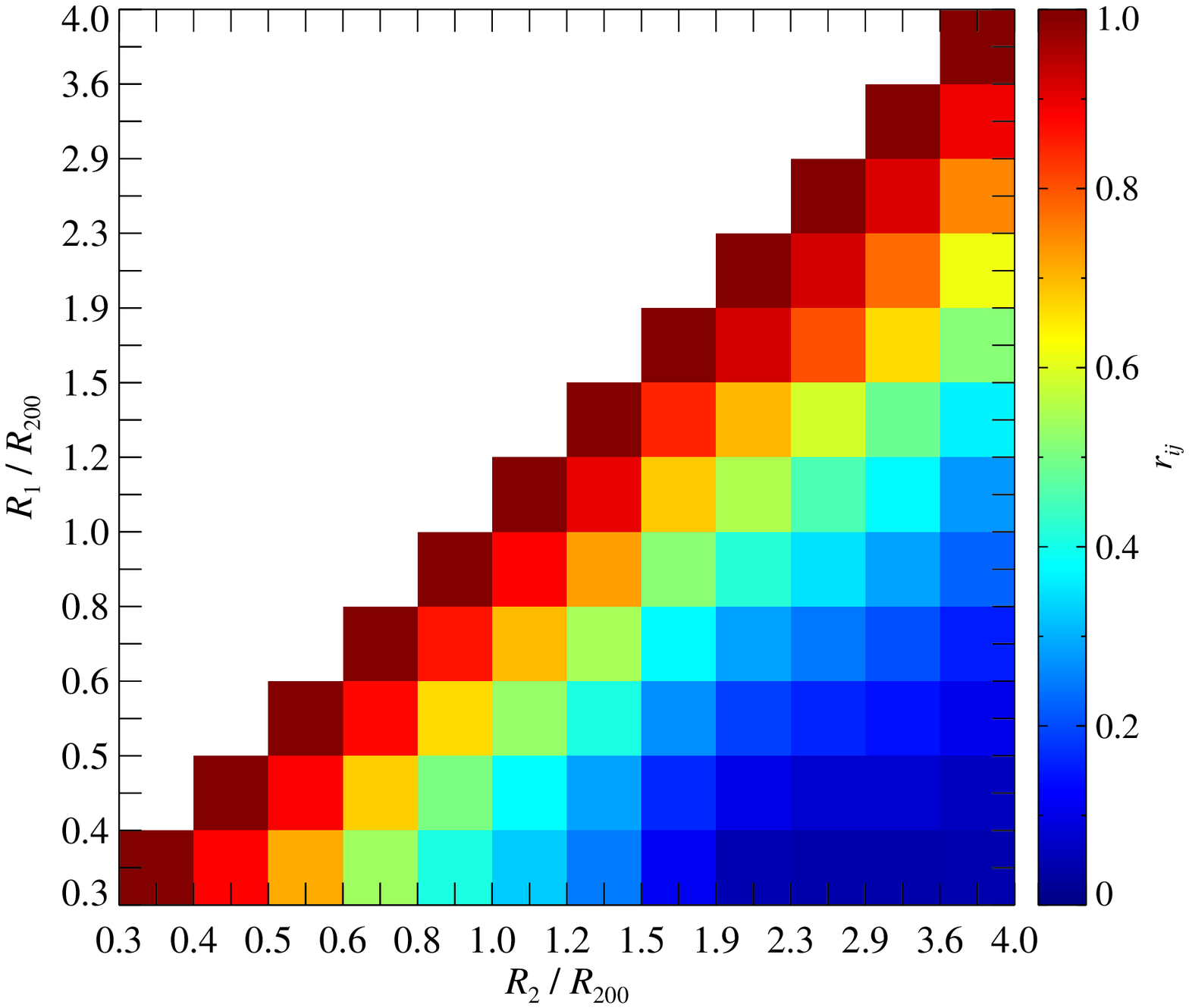}}\\  
  \resizebox{0.33\hsize}{!}{\includegraphics{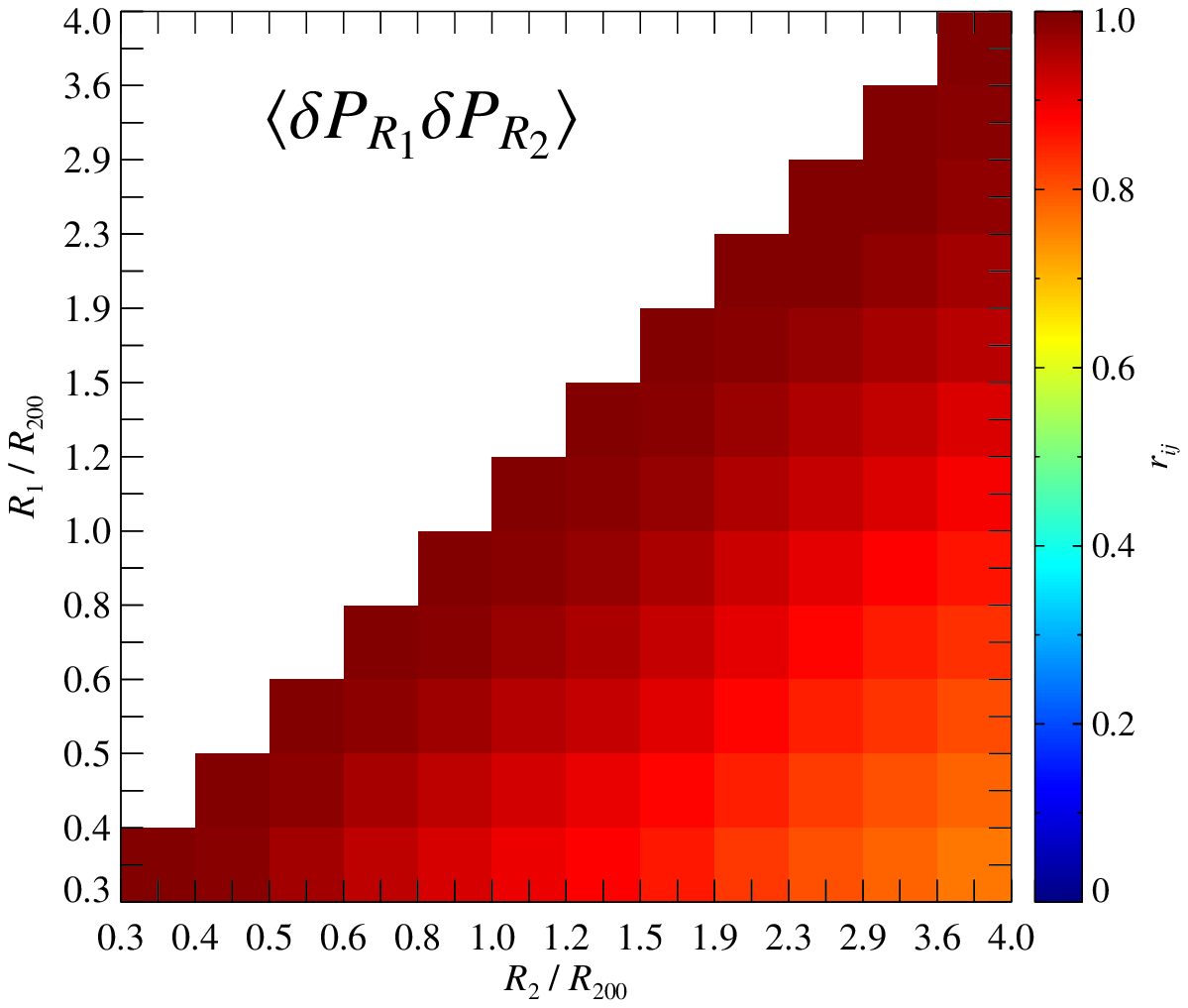}}%
  \resizebox{0.33\hsize}{!}{\includegraphics{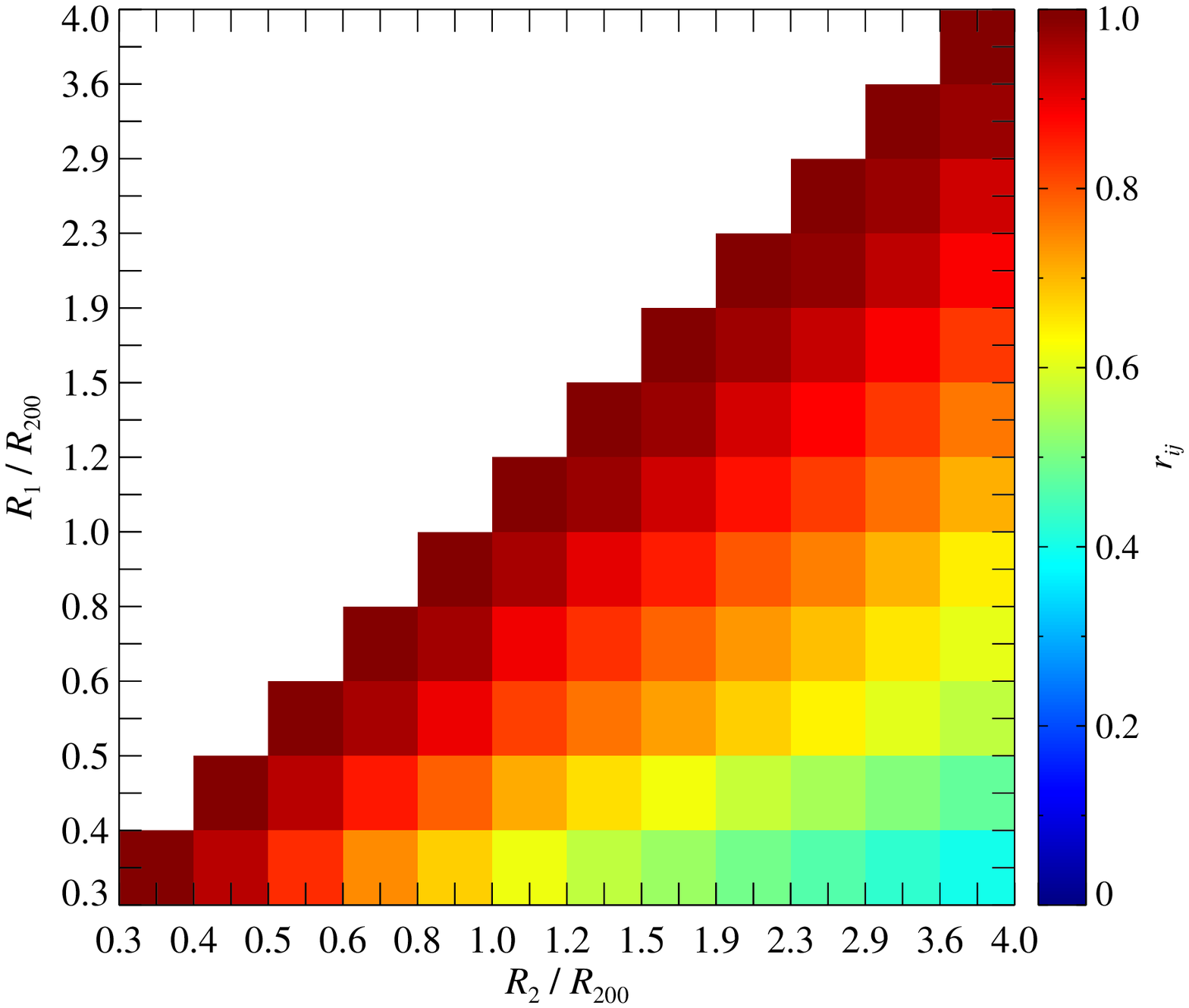}}%
  \resizebox{0.33\hsize}{!}{\includegraphics{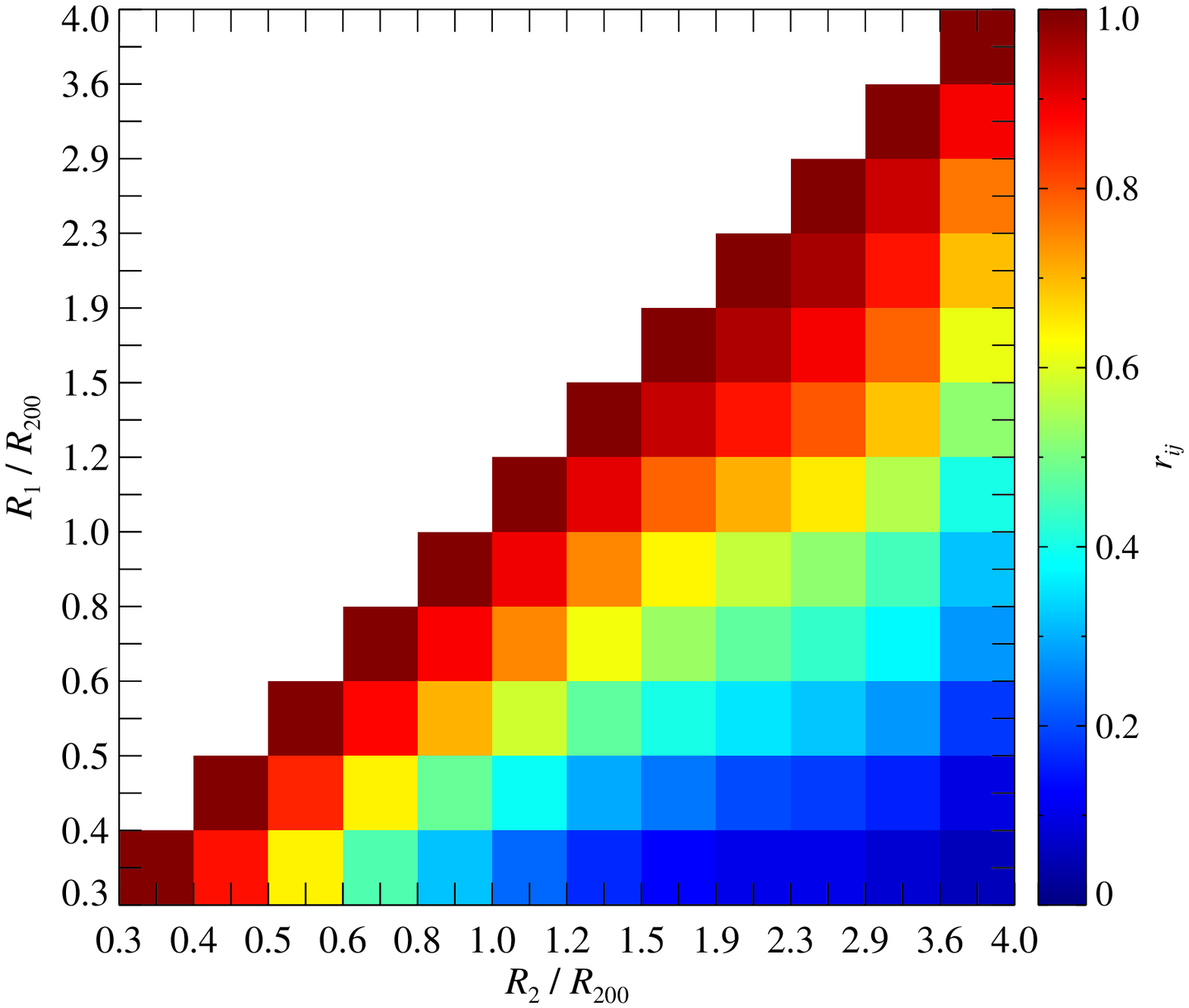}}\\  
  \caption{Matrix of the average radial cross correlation for clusters with
    masses $ 5.1 \times 10^{14} M_{\sun}\,< M_{200} < 1 \times 10^{15} M
    _{\sun}$ at a given multipole moment ($\ell$) that scales inversely with
    angular scale. From the top to bottom, we show the radial cross correlation
    of gas density ($\bra\delta\rho_{R_1}\delta \rho_{R_2}\ket$) and pressure
    ($\bra\delta P_{R_1}\delta P_{R_2}\ket$) at different multipoles $\ell =2$,
    4, and 8 (left to right). The color scale ranges from red (highly
    correlated) to blue (uncorrelated). From the inner to the outer radii the
    $\ell = 2$ mode is highly correlated, so the quadrupole moment at the inner
    radii correlates with the moment at the outer radii. This radial cross
    correlation does not hold for a large radial range for higher multipole
    moments and becomes successively weaker as $\ell$ increases.}
  \label{fig:xgrid}
\end{figure*}

In contrast, the angular power spectrum of the DM density (right panel,
Figure~\ref{fig:avgclk}) shows an almost uniform size distribution (at radii
$>R_{200}/2$) due to subhalos within clusters that are not subject to
hydrodynamic forces. We note that tidal effects are only expected to modify the
DM substructure distribution at smaller radii \citep{2012MNRAS.425.2169G}.  The
auto power spectrum of $\delta \rho_\rmn{DM}$ has the largest amplitude among
the three fields, since the DM only interacts gravitationally whereas gas
additionally feels the hydrodynamic force. This force leads to transient
phenomena such as shocks, contact discontinuities, and hydrodynamic
instabilities, which effectively smooth the gas density and pressure
distributions. This results in smaller contrasts across the sky and less
power. Among the three fields, the pressure shows the smallest amplitude of the
auto power spectrum. While contact discontinuities have density jumps across,
which are in part responsible for the angular power, the pressure is constant
across contact discontinuities. Hence, the pressure field exhibits less
discontinuities and appears generally smoother which causes a smaller angular
power. There is indirect evidence for such an interpretation of our findings
from X-ray observations of the ICM. These find abundant ``cold fronts'', which
are identified with sloshing contact discontinuities \citep[e.g.,][and
references therein]{2007PhR...443....1M,Zuho2010}.
 
The power spectrum amplitude increases with radius. This is in line with our
findings for the clumping factor and expected since the background pressure and
density distributions (gas and DM) are decreasing with radius. Thus, clumps of
similar amplitude have a greater contrast with these backgrounds at larger
radii. Power spectra are measurements of variances, so a larger contrast with
the background implies a greater power spectrum amplitude.

As shown in Figure~\ref{fig:avgcl}, there are no significant differences of the
mean angular power spectra between the various simulated physics models. This
result is similar to the results found in Section~\ref{sec:clumping} and
furthermore confirms that clumping in our simulations is mainly due to
gravitationally dominated processes, which are only modulated by baryonic
physics.

When we discretise the distributions of gas mass and thermal energy into
individual cones, we use the SPH kernel in the radial direction but not for
projection into angular pixels in order not to introduce an artificial smoothing
that is associated with the finite resolution. If we had distributed the
fraction of SPH mass and energy within the smoothing kernel that is subtended by
the solid angle of the HEALPix cone, this would have caused an artificial
decrease of the amplitude of the power spectrum toward small scales, which
scales as the square of the Fourier transform of the SPH smoothing kernel and
reveals the inability to resolve structures on scales below the smoothing
kernel. Instead, we decided to cut the power spectra at the multipole moment
$\ell_\rmn{max}$ at which the SPH particle shot noise term dominates the power,
thereby demarcating their angular resolution. The particle shot noise is smaller
for more massive clusters and at larger radii, implying an increasing
$\ell_\rmn{max}$ with cluster mass and radius. This can be understood with a
simple analytical model that relates the solid angle subtended by the SPH
particle's smoothing radius to the cluster centric radius of that particle,
yielding
\begin{equation}
\ell_\rmn{max}\simeq \frac{\pi}{\theta_\rmn{max}} = 
\left( \frac{\pi^3  x^3\, \rho\,\sqrt{\mathcal{C}_{2,\rho}}\, M_{200}}
  {200 \rho_\rmn{cr} N_\rmn{gas} m_\rmn{gas}}\right)^{1/3},
\end{equation}
where $x=R/R_{200}$ and $\rho_\rmn{cr}$ is the critical mass density of the
universe. Note that the inclusion of the density clumping factor
($\mathcal{C}_{2,\rho}$) is important to account for the increase of
$\ell_\rmn{max}$ at radii much larger than the scale radius of the density
profiles.

\subsection{Cross-power spectra}

We calculate the cross correlation coefficient ($r_{ij}$,
cf. Equation~(\ref{eq:xcor})) for projected quantities between different radial
shells as well as between projected quantities. Values of $r_{ij}$ close to
unity imply strongly correlated quantities and those close to zero show that the
quantities are uncorrelated.

\begin{figure*}
  \begin{minipage}[t]{0.33\hsize}
    \centering{\small $\ell = 2$}
  \end{minipage}
  \begin{minipage}[t]{0.33\hsize}
    \centering{\small $\ell =4$}
  \end{minipage}
  \begin{minipage}[t]{0.33\hsize}
    \centering{\small $\ell = 8$}
  \end{minipage}
  \resizebox{0.33\hsize}{!}{\includegraphics{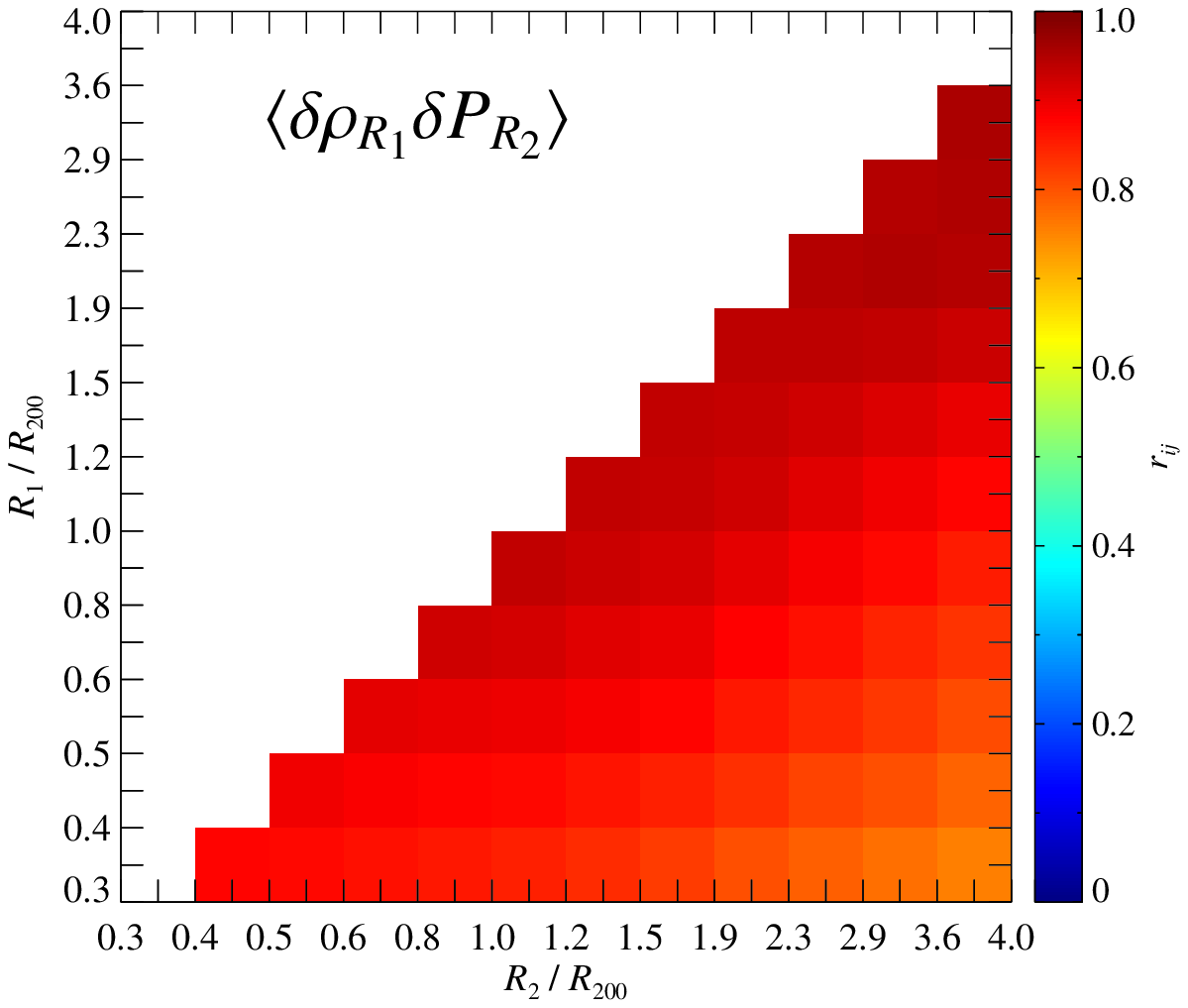}}%
  \resizebox{0.33\hsize}{!}{\includegraphics{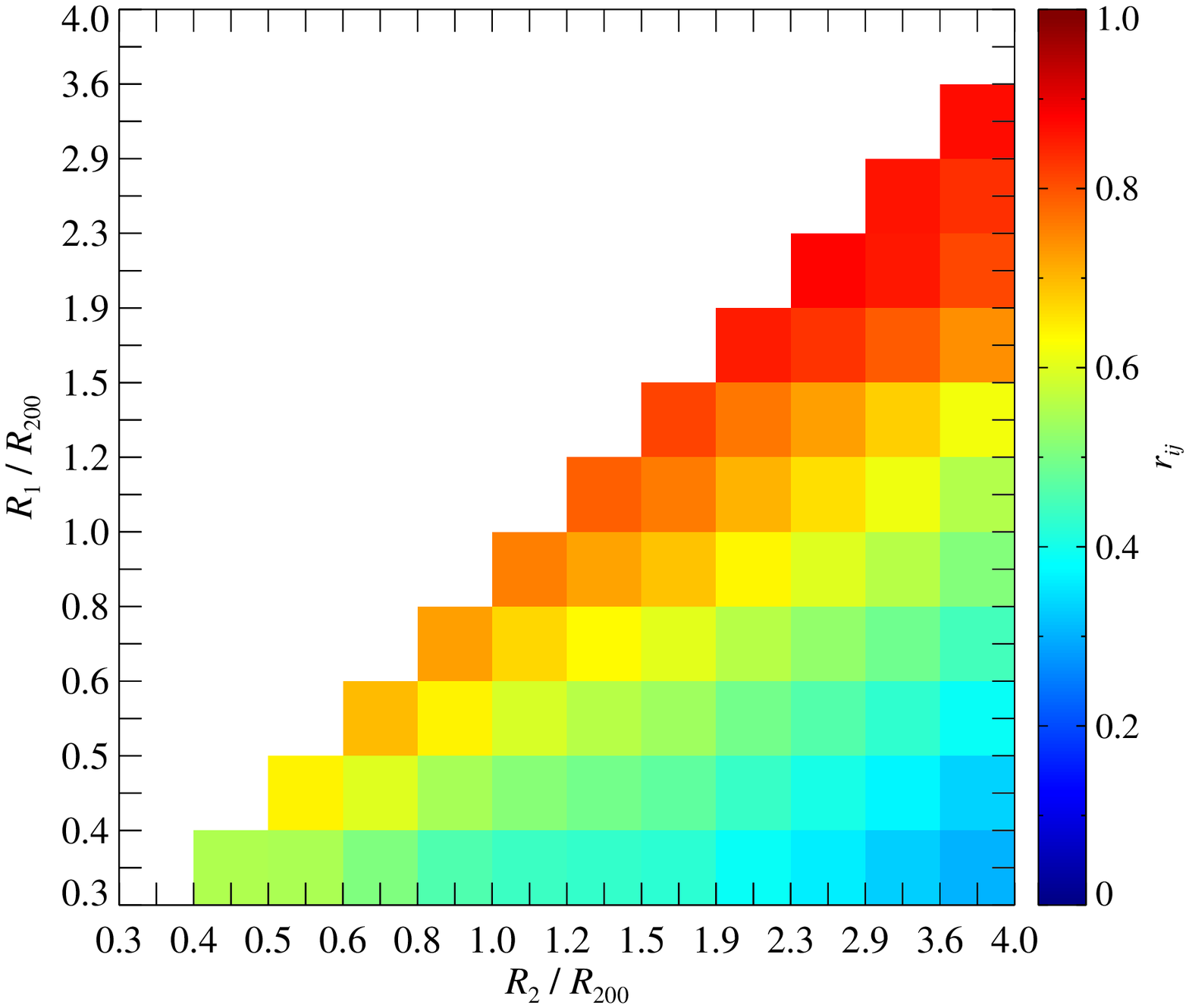}}%
  \resizebox{0.33\hsize}{!}{\includegraphics{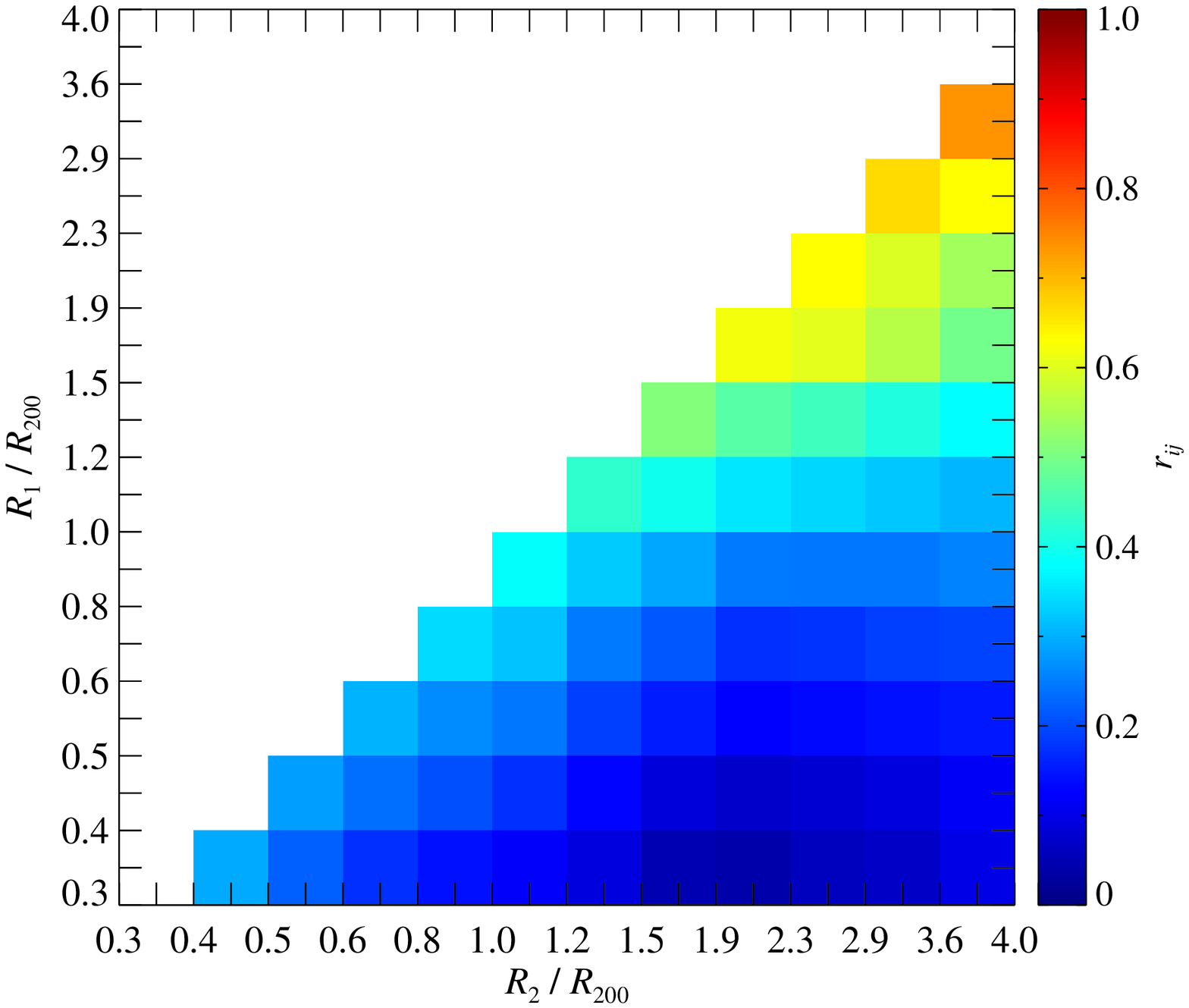}}\\  
  \resizebox{0.33\hsize}{!}{\includegraphics{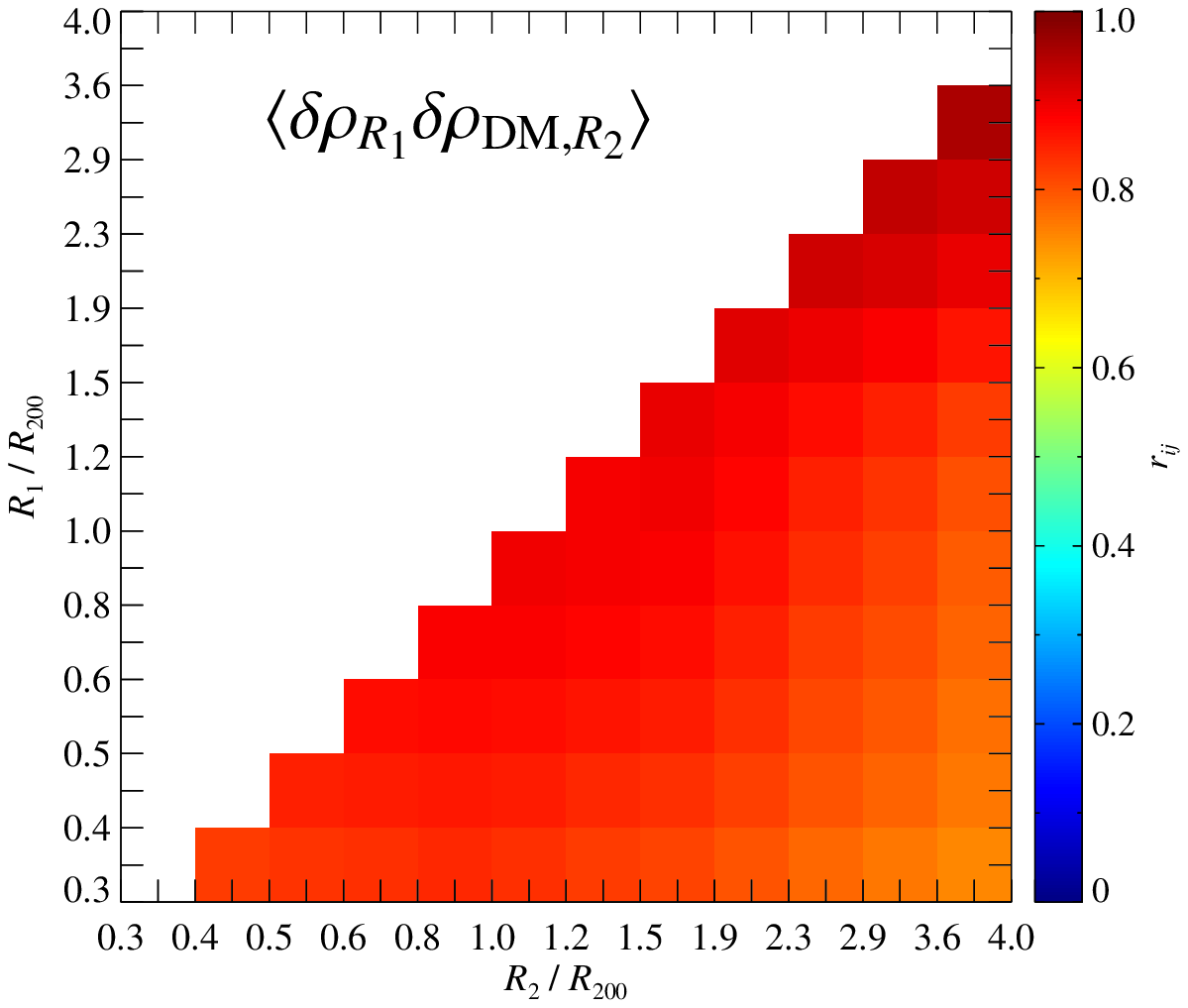}}%
  \resizebox{0.33\hsize}{!}{\includegraphics{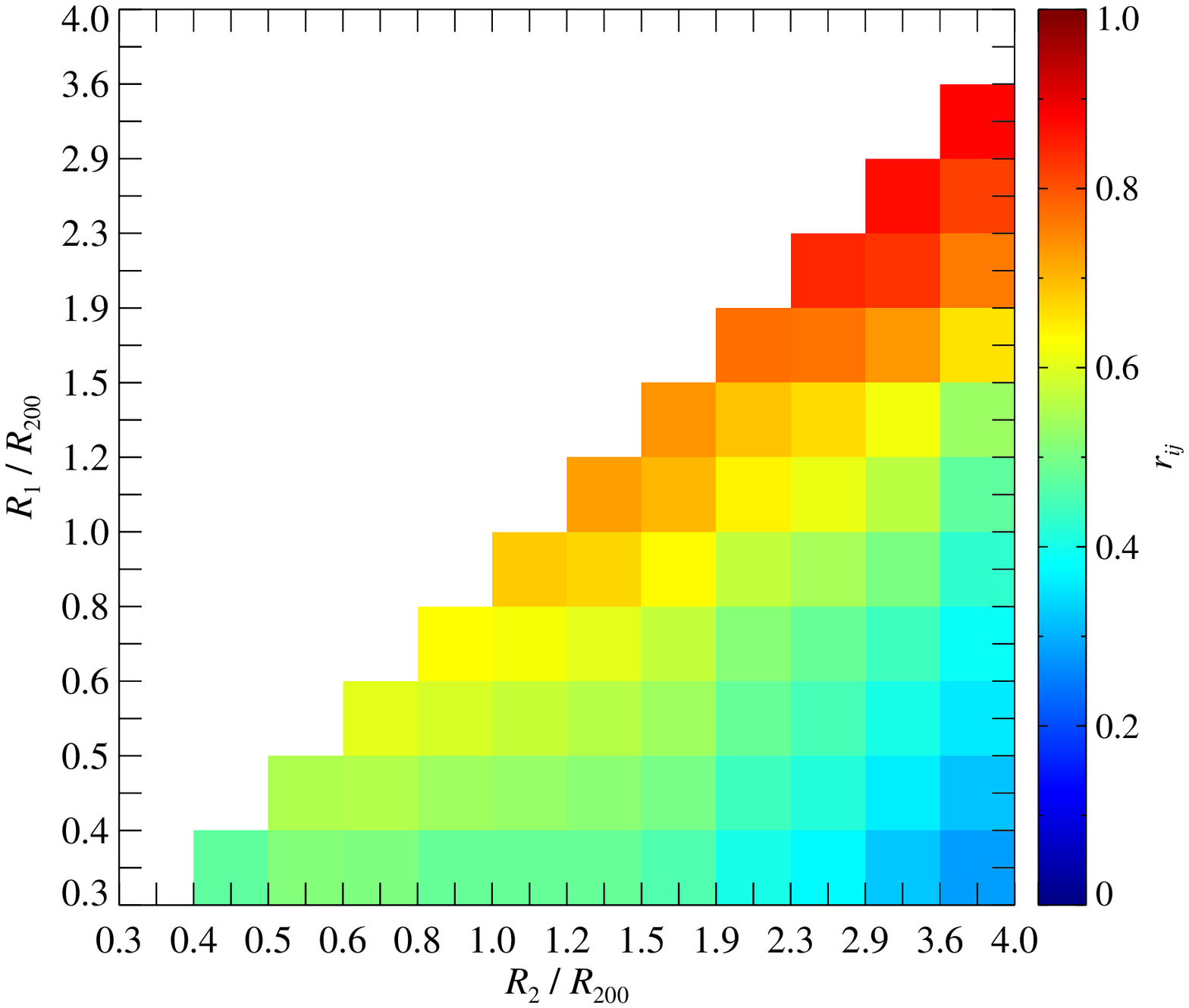}}%
  \resizebox{0.33\hsize}{!}{\includegraphics{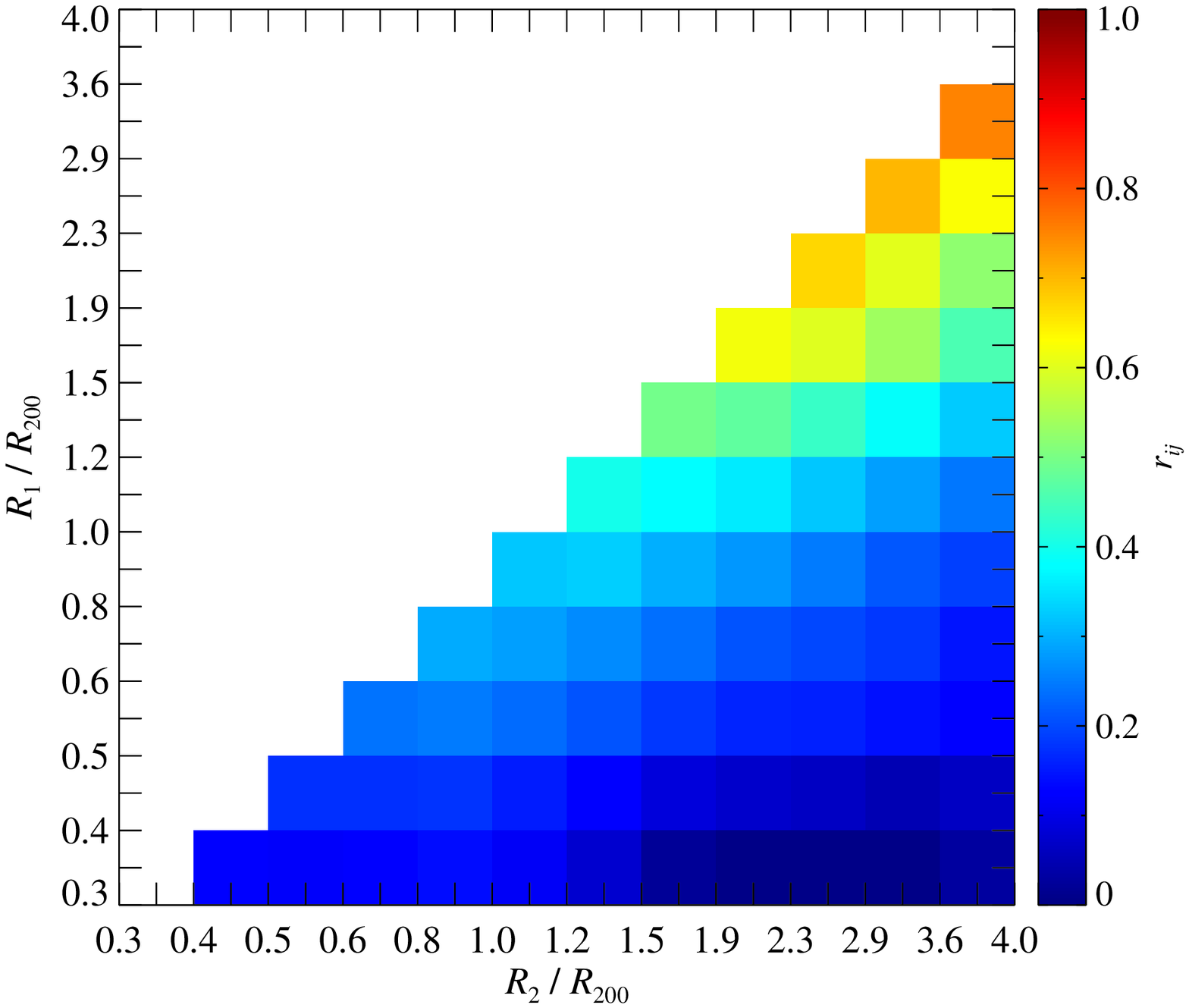}}\\  
  \caption{Same as in Figure~\ref{fig:xgrid}, but for the radial cross
    correlation of $\rho$ and $P$ ($\bra\delta\rho_{R_1}\delta P_{R_2}\ket$,
    top) and $\rho$ and $\rho_\rmn{DM}$ ($\bra\delta\rho_{R_1}\delta
    P_{R_2}\ket$, bottom).  With increasing multipole moments, the correlations
    are confined to successively smaller radial ranges.  These inter-quantity
    cross correlations are generally weaker in comparison to the cross
    correlations of a quantity with itself (cf.{\ }Figure~\ref{fig:xgrid}).}
  \label{fig:xgrid2}
\end{figure*}

\begin{figure*}
  \resizebox{0.50\hsize}{!}{\includegraphics{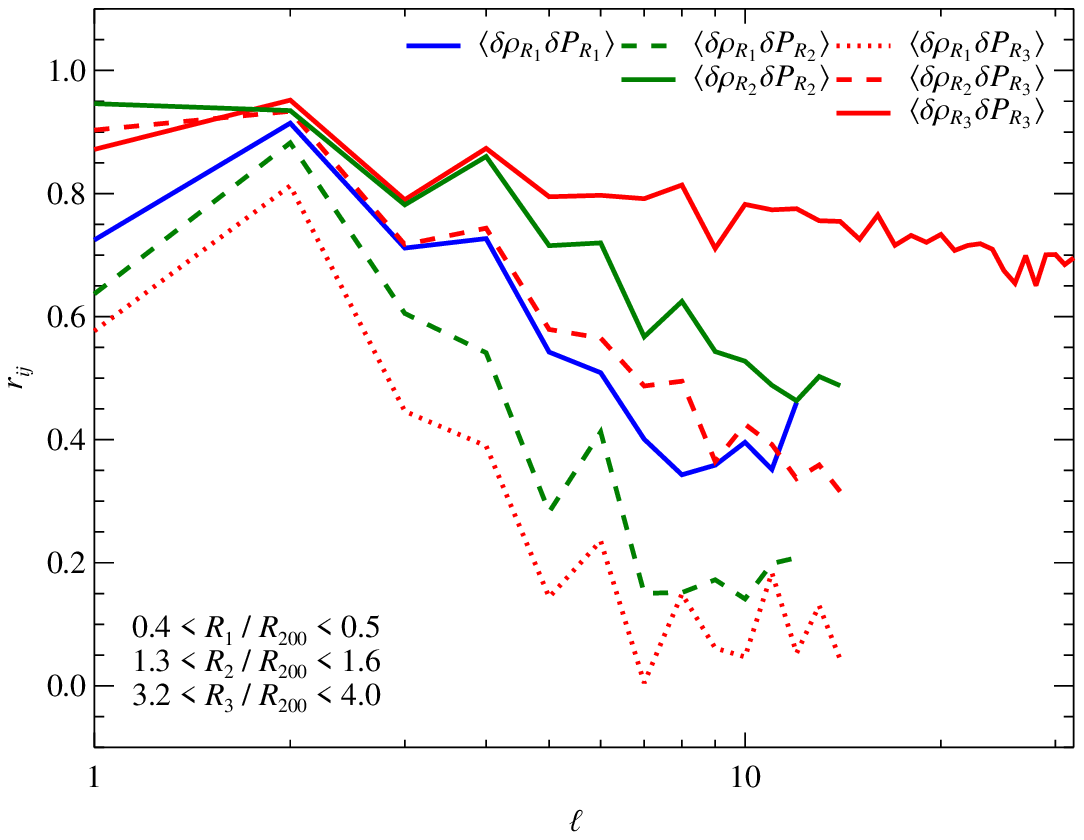}}%
  \resizebox{0.50\hsize}{!}{\includegraphics{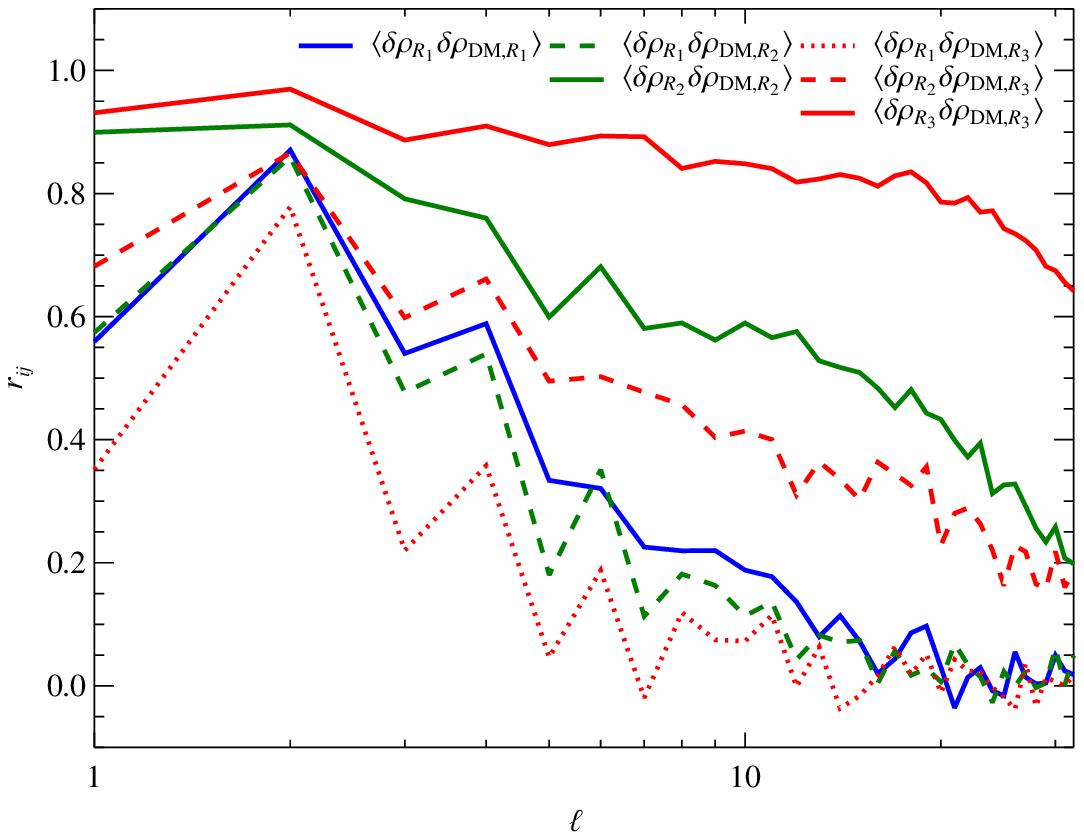}}\\
  \caption{Average radial cross correlation coefficient of $\delta
    \rho$ with $\delta P$ (left) and with $\delta \rho_\rmn{DM}$
    (right) for clusters with masses $ 5.1 \times 10^{14} M_{\sun}\,<
    M_{200} < 1 \times 10^{15} M_{\sun}$ as a function of multipole
    moment $\ell$. The correlation coefficient decreases for
    cluster-centric shells that are further apart and for smaller
    angular scales (as $\ell$ increases). For the greatest shell
    separation (red dotted lines) there is no correlation beyond $\ell
    \gtrsim6$. Gas that is accreted onto clusters passes through the
    cluster accretion shock, which heats it to the clusters' virial
    temperature and starts to separate it from the DM halo that does
    not feel the pressure force.}
  \label{fig:xcorr}
\end{figure*}

In Figures \ref{fig:xgrid} and \ref{fig:xgrid2} we show multiple
panels of cross correlation coefficients $r_{ij}$ between different
radial bins for a fixed angular scale. We present $r_{ij}$ as a matrix
for a given multipole moment $\ell$ that is symmetric if we cross
correlate a quantity with itself.  In the latter case, i.e., when we
cross correlate $\delta\rho$ with $\delta P$ or $\delta\rho_\rmn{DM}$
as a function of radial separation, then the diagonal corresponds to
the cross correlation amplitude at that radius, which we show as a
function of $\ell$ in left panel of Figure~\ref{fig:xcorr}.

For the quadrupole moment ($\ell = 2$), we find cross correlation coefficients
very close to unity for all radial correlations of $\delta \rho$, $\delta P$,
$\delta\rho \times \delta P$ and $\delta\rho \times \delta \rho_\rmn{DM}$. Thus,
the orientation of the gas and pressure ellipticities in the inner regions of
clusters are spatially aligned with the outer regions \citepalias{BBPS1}. We
show that density and pressure enhancements correlate along the cones for small
$\ell$ (i.e., on large angular scales). This suggests that cosmological
filaments, which connect to clusters, have a large influence on the cluster's
quadrupole moment deep into the cluster interior and that clumping is in part
due to substructure that is infalling along filaments into cluster
cores. Therefore, the direction of the most over-dense cones will coincide with
the direction of the major axis of the moment-of-inertia tensor.

The higher angular moments $\ell = 4$ and 8 show less correlation between radial
shells. However, radial shells beyond the virial radius show a stronger
correlation to shells that are further apart in comparison to scales within the
cluster that lose coherence to adjacent radial shells. For $\ell = 4$ the
correlation is still strong for radial bins that are $\sim 3 - 4$ bins
away. This coherence drops to adjacent bins when we look at $\ell = 8$.  Overall
the correlations are weaker for the gas density than for the pressure. 

We find that inter-quantity cross correlations are generally weaker in
comparison to the cross correlations of a quantity with itself.  With increasing
multipole moments, the correlations are confined to successively smaller
radial ranges at any radius. For multipole moments $\ell\geq 4$, the
inter-quantity correlations is weak at large radii and becomes uncorrelated
inside of $R/R_{200} \sim 1.5$, which coincides with the approximate location of
the virial shock.

In Figure \ref{fig:xcorr} we show the radial cross correlation of $\delta \rho$
with $\delta P$ and with $\delta\rho_{\rmn{DM}}$ as a function of $\ell$. As we
increase the separation between the radial shells, the correlations between the
gas density and pressure decreases. The same happens with the correlation between
gas and DM density. Thus, the gas at small radius has lost the memory of its
previous state before infall, when it more closely correlated to DM.  Not
surprisingly this is the result of dissipational virialization. However, this
loss of information is not instantaneous since the intermediate radial bins are
still marginally correlated with the inner most radial bin.

\section{The physics of cluster outskirts}
\label{sec:synthesis}

We now attempt a comprehensive view of the physics of cluster outskirts and
provide a synopsis of various non-equilibrium measures that all become
increasingly important beyond $R_{200}$. Those include the kinetic pressure
support of the gas (see Equation~(\ref{eq:Pkin})), the gas ellipticity
(characterized by the axis ratio of the moment-of-intertia tensor), density and
pressure clumping as a function of cluster radius. First we review the
definition of ellipticity \citepalias[for further details, please refer
to][]{BBPS1}.

We estimate the deviation from spherical symmetry of the cluster gas and dark
matter distribution through the moment-of-inertia tensor,
\begin{equation}
I_{ij} (r<R) = 
\frac{\sum_{\alpha} m_{\alpha}(x_{i,\alpha} - \bar{x})(x_{j,\alpha} - \bar{x})}
{\sum_{\alpha} m_{\alpha}} ,
\end{equation}
where $\alpha$ labels all particles within a given radius $R$, $x_i$ is the
$i$th coordinate of particle $\alpha$, and $m$ is its mass. We quantify the
ellipticity of cluster gas and DM density using the eigenvalues $\lambda_i$,
which are computed for each inertial tensor at a given radius. We adopt the
convention $\lambda_1 < \lambda_2 < \lambda_3$ for the eigenvalues. The
ellipsoid associated with the tensor has half-axis lengths $a =
\sqrt{\lambda_1}$, $b = \sqrt{\lambda_2}$, and $c = \sqrt{\lambda_3}$.

In Figure~\ref{fig:all}, we show the radial profiles of the kinetic-to-thermal
pressure ratio $P_{\mathrm{kin}}/P_{\mathrm{th}}$, the axis ratio of the
moment-of-inertia tensor $1-c/a$, and the density and pressure clumping factors
(corrected for SPH volume bias). Inside $R_{200}$, all these dissipative
non-equilibrium statistics show fairly small values, which increase dramatically
for $R\gtrsim R_{200}$ (within a radial range of a factor of 2). At any radius,
heavier clusters and/or clusters at higher redshift show an increased level of
these non-equilibrium statistics since these probe on average dynamically
younger objects. 

Starting from the inside out, we encounter first a rising ratio of
$P_{\mathrm{kin}}/P_{\mathrm{th}}$, followed by density and pressure clumping,
and last by the gas ellipticities. Those different transition radii provide
evidence for the different timescales on which these non-equilibrium statistics
approach sphericalization/virialization as structures get accreted by a cluster,
transforms into substructures, and eventually resolve within the ICM.

The smooth transition of the kinetic-to-thermal pressure ratio demonstrates that
cluster accretion shocks do not thermalize all kinetic energy
immediately. Instead the curvature of their shock surfaces induces vorticity
\citep{2011ApJ...730...22P} which drives a turbulent cascade that dissipates the
energy contained within a few eddy-turnover times in the comoving frame of the
accreting gas \citep{2014arXiv1401.7657S}.

In contrast, these post-shock shear flows destroy the peripheral parts of
accreted substructures on a faster time scale through Kelvin-Helmholtz
instabilities, causing the density and pressure clumping factors to drop shortly
after the accretion shocks, the location of which apparently moves in as a
function of redshift in our radial representation $R/R_{200}$. However, as we
show in \citetalias{BBPS1} (and later on confirmed by
\citet{2014arXiv1404.4636N}), this is an artifact of the definition of the
virial radius $R_{200}$ (of taking a sphere enclosing a mean density that is 200
times the {\em critical} density of the universe) and almost stays fixed when
changing the definition to $R_{200m}$ (i.e., evaluating the mean density within
a sphere that is 200 times the {\em mean} density of the universe).

Finally, the gas ellipticities show the transition to a smaller value at the
largest radius, as this statistic responds to the overall potential shape that
has been established through accretion along a few filaments connecting to the
cluster from different angular directions. Contrarily, the dissipationless DM
component shows an elevated level of ellipticity with only modest decline
towards smaller radii, presumably owing to gravitational exchange with the
dissipational gas component.


\section{Conclusions}
\label{sec:conclusions}

Quantifying and characterizing density and pressure clumping is essential for
eliminating observational biases in determining various ICM properties. In X-ray
measurements, density clumping will bias the estimates of the ICM density, which
has observational consequences such as an apparent flattening in the entropy
profile and apparent baryon fractions that exceed the cosmic mean. The
consequence of pressure clumping are more subtle in comparison to density
clumping. Variations about mean pressure profiles will increase the amplitude of
the thermal SZ auto spectrum as a function of $\ell$ and increase the scatter in
the $Y-M_{200}$ scaling relation.

\begin{figure*}
  \begin{minipage}[t]{0.5\hsize}
    \centering{\small Redshift $z=0$:}
  \end{minipage}
  \begin{minipage}[t]{0.5\hsize}
    \centering{\small Redshift $z=1$:}
  \end{minipage}
  \resizebox{0.5\hsize}{!}{\includegraphics{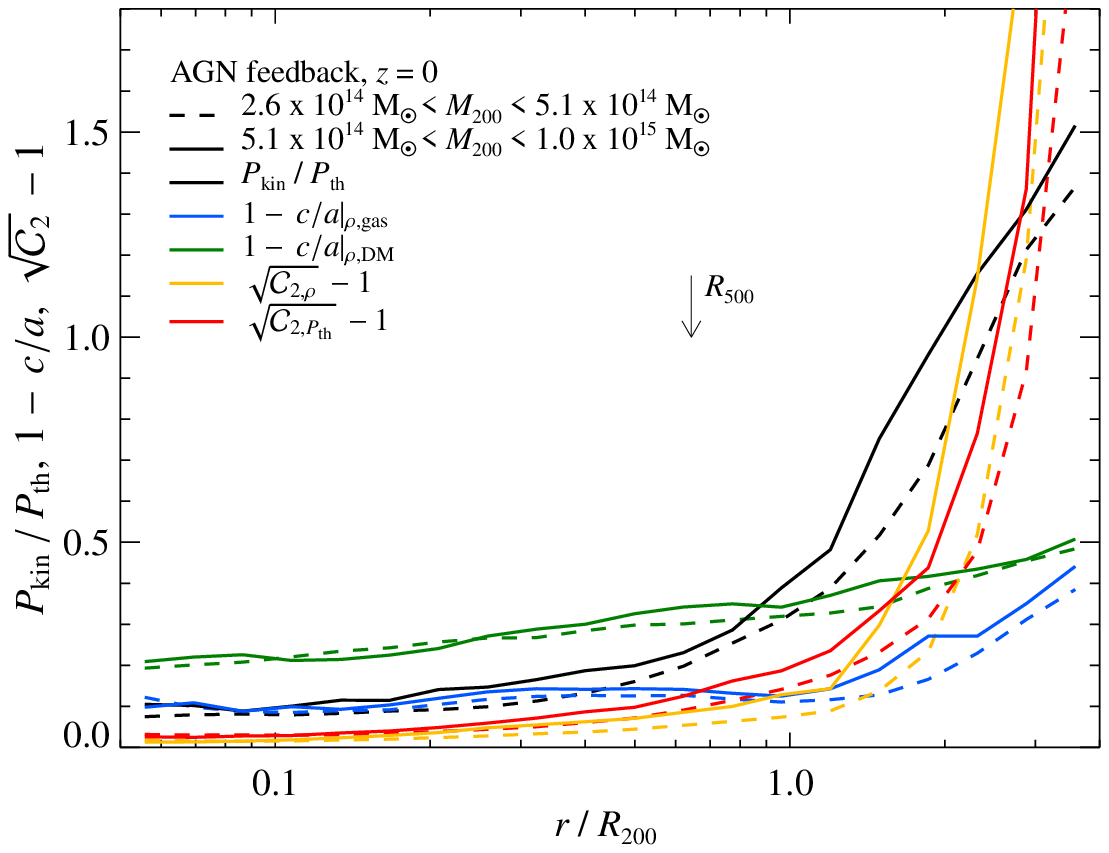}}%
  \resizebox{0.5\hsize}{!}{\includegraphics{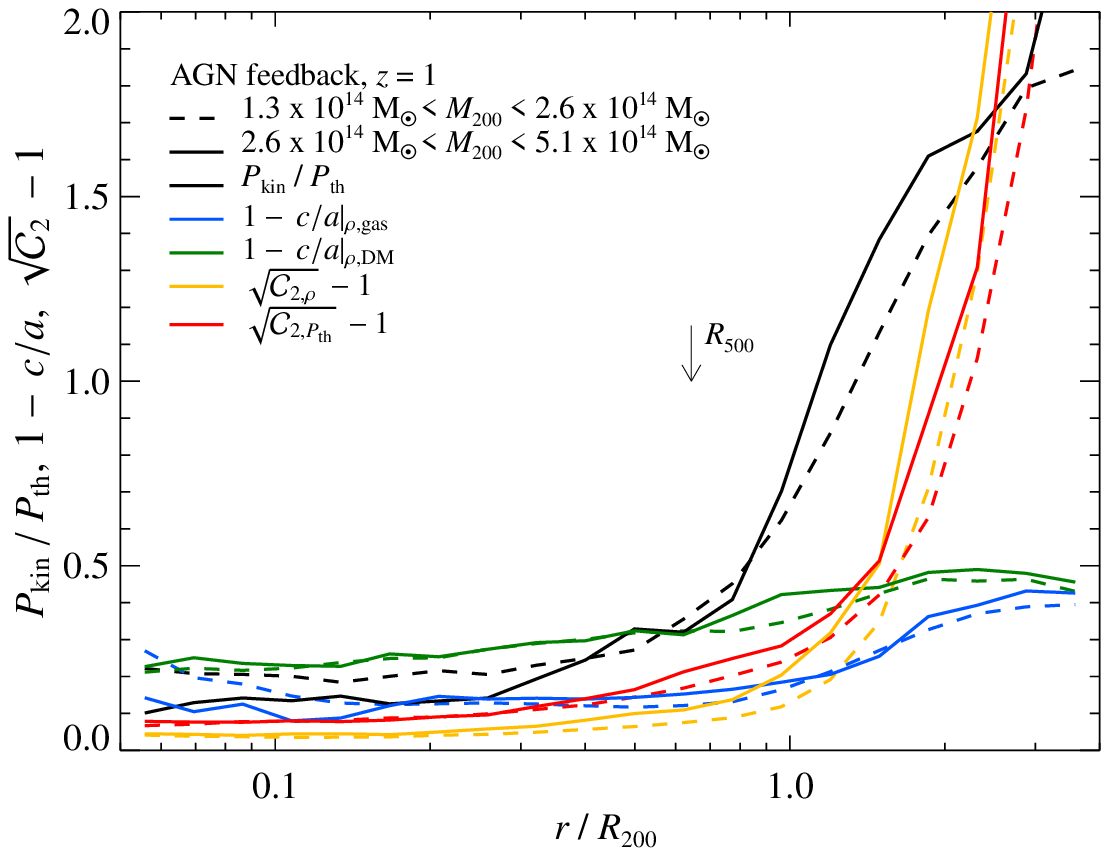}}\\
  \caption{The outskirts of clusters become increasingly complicated since
    traditionally simplifying assumptions dramatically break down around and
    beyond the virial radius. This is demonstrated by comparing the median
    profiles of the non-thermal pressure contribution, $P_\rmn{kin}/P_\rmn{th}$,
    the asphericity estimate for DM and gas, $1-c/a$ (where $c/a$ is the ratio
    of the largest-to-smallest main axis of the moment-of-inertia tensor), as
    well as gas density and pressure clumping factor, $\mathcal{C}_2$, which all
    increase dramatically with radius. We contrast low- and high-mass clusters
    (dashed and solid) at $z=0$ (left) and $z=1$ (right). Note the different
    range of the ordinate values in both panels.  Those simplifying assumptions
    break down more significantly for larger clusters (at the same $z$) and
    earlier times (i.e., larger $z$ for fixed cluster mass) since both criteria
    probe on average dynamically younger objects in a hierarchically growing
    universe. }
\label{fig:all}
\end{figure*}

To address these biases, we presented an analysis of ICM inhomogeneities
(clumping) using cosmological hydrodynamic simulations with different variants
of subgrid physics included. Our analysis can be categorized into two
approaches: (1) quantifying the so-called {\it clumping factor} (i.e., the
dimensionless second-order moment of density or pressure) as a function of
radius and (2) characterizing the angular structure of these inhomogeneities
with the goal of understanding its nature.  Our results can be summarized as
follows.

\begin{itemize} 

\item Similar to the previous work \cite[e.g.,][]{2011ApJ...731L..10N,Ronc2013}
  we find that the clumping factor increases with radius and is independent of
  the sub-grid physics model used (see Figures~\ref{fig:clumping_linear} and
  \ref{fig:clumping_physics}). This indicates that the process giving rise to
  clumping is gravitational in nature and only modulated by baryonic physics.
  However, our AGN feedback model at higher redshifts ($z\gtrsim 1$) shows a
  moderate suppression of clumping outside $R_{200}$ due to the smoothing effect
  of the feedback energy.

\item Density and pressure clumping are increasing functions of cluster mass and
  redshift (see Figures~\ref{fig:clumping_mass} and
  \ref{fig:clumping_redshift}), which probe on average dynamically younger
  objects that are still in the process of assembling. Hence this is a
  consequence of hierarchical structure growth.
 
\item We find a small trend with the dynamical state of a cluster, with relaxed
  clusters exhibiting lower clumping factors for both density and pressure (see
  Figure~\ref{fig:clumping_redshift}). However, the differences are within the
  25$^{\rmn{th}}$ and 75$^{\rmn{th}}$ percentiles of the distributions of
  clumping factors.

\item The inhomogeneities that are responsible for density and pressure clumping
  can be clearly identified in the angular projections of radial shells (see
  Figures~\ref{fig:mole1} and \ref{fig:mole2} for examples of a massive and
  intermediate-mass cluster). The angular power spectra of these inhomogeneities
  have a prominent quadrupole feature and increase with amplitude as the
  spatially smoothed background field decreases (see Figure~\ref{fig:avgcl}).

\item When the angular power spectra of gas density and pressure are plotted as
  a function of physical (perpendicular) wavenumber, they show a universal broad
  peak at large scales (see Figure~\ref{fig:avgclk}), signaling the presence of
  gravitationally-driven ``super-clumping'' \citep{2014arXiv1404.6250M}. Hence,
  the gas density and pressure clumping signal is dominated by comparably large
  substructures with scales $\gtrsim R_{200}/5$. Smaller structures apparently
  get broken up through Kelvin-Helmholtz instabilities or are transformed at
  shocks due to anisotropic stresses. In contrast, the angular power spectrum of
  the DM density shows an almost uniform size distribution at all radii due to
  unimpeded distributions of subhalos within clusters (see
  Figure~\ref{fig:avgclk}), which are not subject to hydrodynamic forces nor are
  they affected by tidal effects at cluster-centric radii $>R_{200}/3$.

\item The quadrupole correlates across all radial shells (for radii $>
  R_{200}/3$), indicating that the direction of clumping is determined by
  infalling substructures into clusters (see Figure~\ref{fig:xgrid}). At smaller
  angular scales (for multipole moments $\ell\geq 4$), the cross correlation
  becomes weaker with increasing radial distance between the two shells, in
  particular for smaller radii and when we correlate different quantities such
  as gas density and pressure (see Figure~\ref{fig:xgrid2}).

\item Combining our results from clumping, angular power spectra, and cross
  correlation coefficients, we find that these inhomogeneities are sourced by
  cosmic filaments that are channeling baryonic and dark matter onto clusters.

\item We finally provide a synopsis of the clusters' non-equilibrium and
  non-spherical measures such as kinetic pressure support of the gas, gas
  ellipticity (characterized by the axis ratio of the moment-of-intertia
  tensor), density and pressure clumping as a function of cluster radius. All
  these measures increase sharply beyond $R_{200}$ (see Figure~\ref{fig:all}).
\end{itemize}

\acknowledgments 

We thank Chris Thompson for useful discussions. Research in Canada is supported
by NSERC and CIFAR. Simulations were run on SCINET and CITA's Sunnyvale
high-performance computing clusters. SCINET is funded and supported by CFI,
NSERC, Ontario, ORF-RE and UofT deans.  C.P. gratefully acknowledges financial
support of the Klaus Tschira Foundation. We also thank KITP for their
hospitality during the 2011 galaxy cluster workshop. KITP is supported by
National Science Foundation under Grant No. NSF PHY05-51164.

\bibliography{bibtex/nab}
\bibliographystyle{apj}
\vspace{1em}

\begin{appendix}

\section{Insights into corrections for SPH volume bias}
\label{sec:volume_bias}

\begin{figure*}[thbp]
  \begin{minipage}[t]{0.5\hsize}
    \centering{\small Density moments {\em corrected} for SPH volume bias}
  \end{minipage}
  \begin{minipage}[t]{0.5\hsize}
    \centering{\small Density moments {\em uncorrected} for SPH volume bias}
  \end{minipage}
  \resizebox{0.5\hsize}{!}{\includegraphics{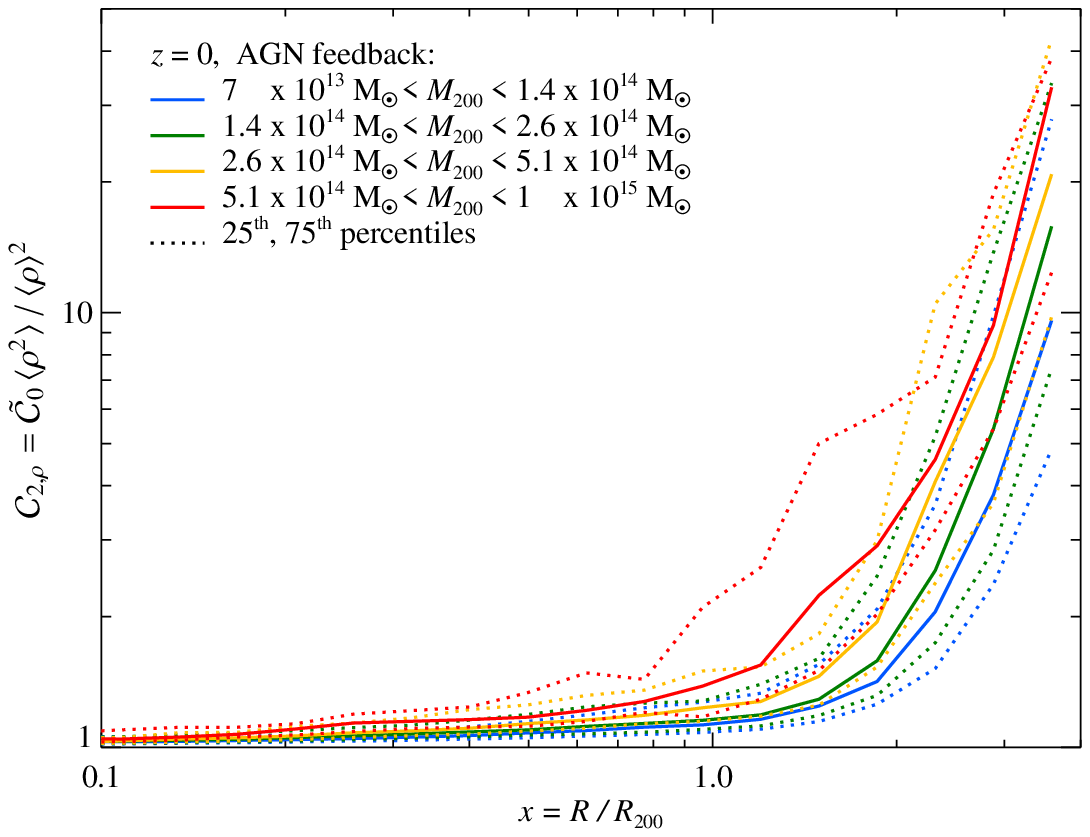}}%
  \resizebox{0.5\hsize}{!}{\includegraphics{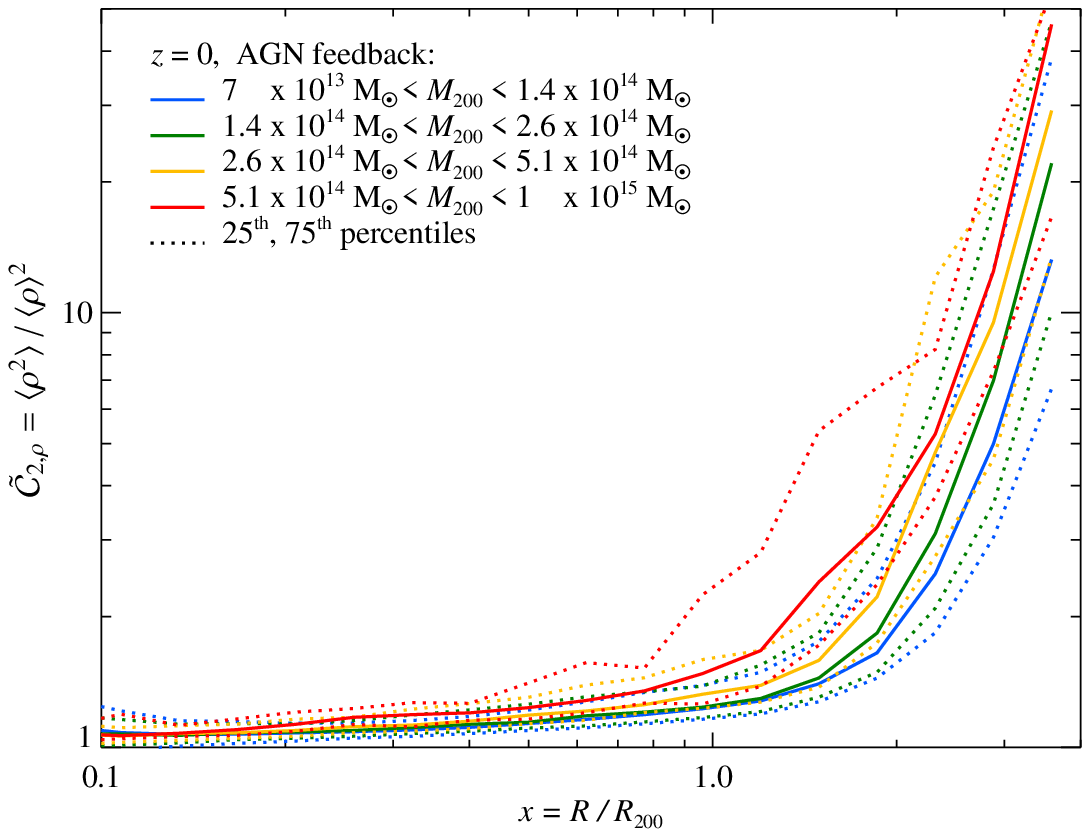}}\\
  \resizebox{0.5\hsize}{!}{\includegraphics{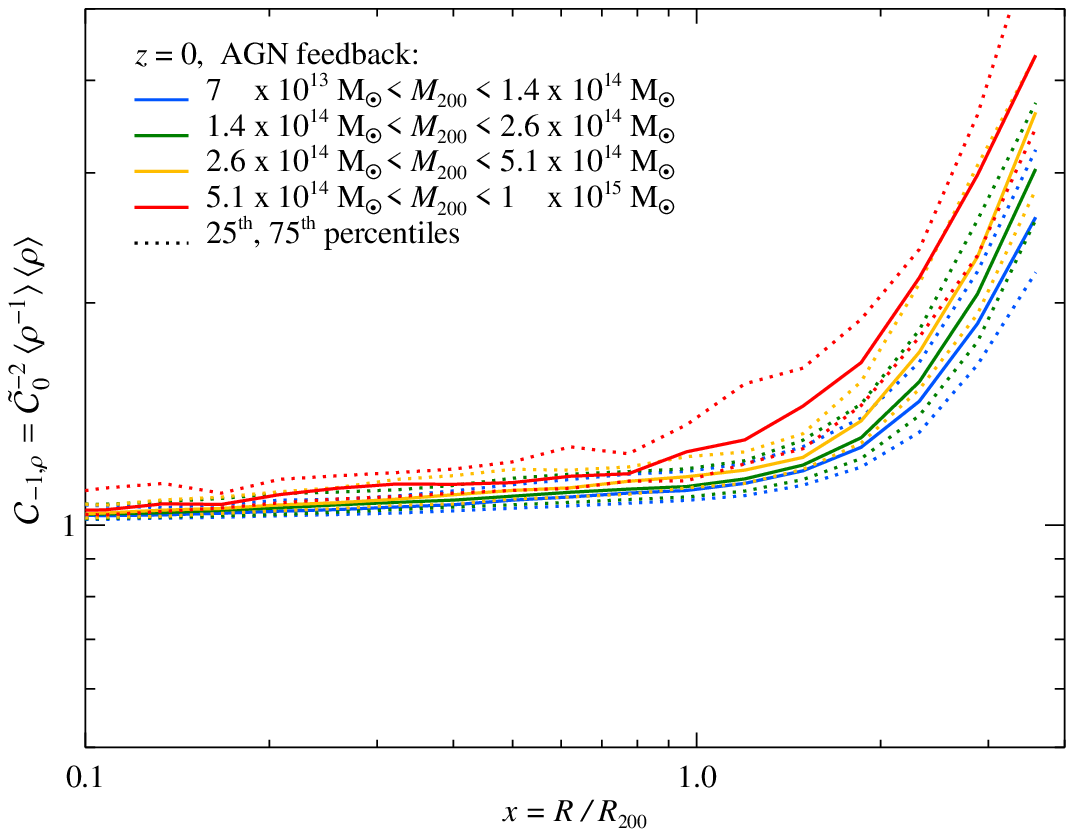}}%
  \resizebox{0.5\hsize}{!}{\includegraphics{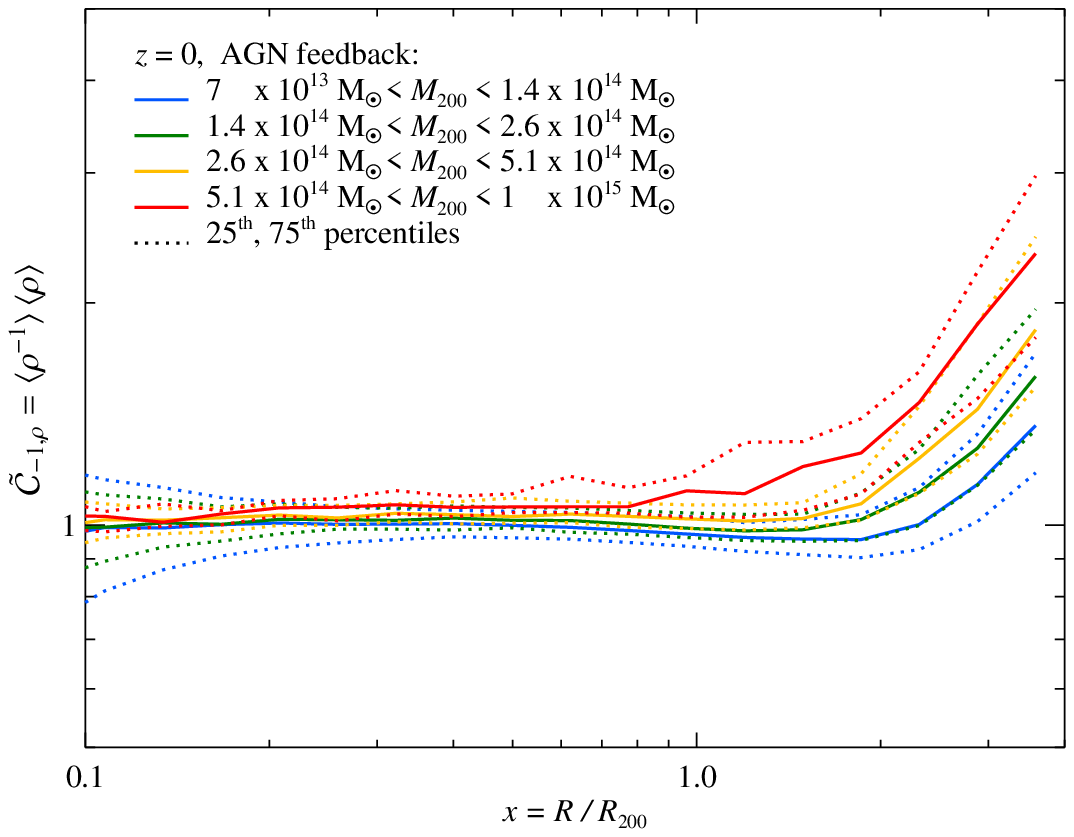}}\\
  \caption{We compare density moments that are {\em corrected} for the SPH
    volume bias (left) with those that are {\em uncorrected} (right). Shown are
    median (solid) and $25^\rmn{th}$, $75^\rmn{th}$ percentiles (dotted) for
    different cluster mass ranges (indicated with different colors). Upper
    panels show clumping factors $\mathcal{C}_{2,\rho}$, and bottom panels the
    negative first order density moment $\mathcal{C}_{-1,\rho}$. In the latter
    case, the SPH volume bias correction removes unphysical density moment
    estimates smaller than unity.}
\label{fig:bias_correction}
\end{figure*}

In Figure~\ref{fig:bias_correction}, we demonstrate how SPH volume correction
affect the density moments (to second order and to first negative order). For the
latter case, the SPH volume bias correction removes unphysical density moment
estimates smaller than unity. The temperature cut in radiative simulations
(i.e., considering only SPH particles with $T>10^6$~K) is only important for
positive density moments (see Figure~\ref{fig:temp_cuts}), which are biased by
dense gas clumps. 

The influence of a temperature cut is negligible for zeroth or $n$th order
negative density moments that intentionally down-weight the dense and cold
clumps. This is explicitly shown in Figure~\ref{fig:sph_volume_bias}
demonstrating that the SPH volume bias is almost independent of the simulated
physics. In contrast, applying a temperature cut to $\tilde{C}_0$ or $n$th order
negative density moments would artificially cut the colder pre-shock regions of
the warm-hot intergalactic medium $T<10^6$~K, introducing a progressively larger
bias to smaller systems (see Figure~\ref{fig:temp_cuts}).

\end{appendix}

\end{document}